\documentclass[11pt]{article}
\usepackage[utf8]{inputenc}
\usepackage{import}
\usepackage{amsmath}
\usepackage{slashed}
\usepackage{mathtools}
\usepackage{caption}
\usepackage{authblk}
\usepackage{blindtext}
\usepackage[titletoc]{appendix}
\usepackage{subcaption}
\usepackage{color}
\usepackage{setspace}

\DeclarePairedDelimiterX\braket[2]{\langle}{\rangle}{#1 \delimsize\vert #2}
\usepackage{abstract}
\usepackage{cite}

\usepackage{graphicx}
\usepackage{multirow}
\title{\textbf{$B\to X_{c(u)}\ell\bar{\nu}_{\ell}\gamma$ and determination of non-perturbative parameters}}
\author[1]{Namit Mahajan\thanks{ nmahajan@prl.res.in}}
\author[1,2]{Dayanand Mishra\thanks{dayanand@prl.res.in}}
\affil[1]{Theoretical Physics Division, Physical Research Laboratory, Ahmedabad, 380009, India.}
\affil[2]{Indian Institute of Technology, Gandhinagar, 382424, India.}
\date{}

\begin{document} 
\maketitle
\doublespacing
This article explores the calculation of non-perturbative parameters present in the matrix element of semileptonic inclusive $B$ decays. These are important for the inclusive determination of the Cabbibo-Kobayashi-Maskawa (CKM) parameters in the standard model. We focus on calculating the rate for radiative inclusive decay, $B\to X_c\ell \bar{\nu}_{\ell}\gamma$, where $\gamma$ is hard. Both the radiative and non-radiative ($B\to X_c \ell \bar{\nu}_{\ell}$) modes are found to require the same non-perturbative parameters. We propose forming ratios from $B\to X_c \ell \bar{\nu}_{\ell}\gamma$ and $B\to X_c \ell \bar{\nu}_{\ell}$ widths in different lepton energy ranges. Our results provide an efficient way to unambiguously calculate the non-perturbative parameters, especially when combined with total width calculations of $B\to X_c \ell \bar{\nu}_{\ell}$. This is shown explicitly by working to $\mathcal{O}(1/m_b)$ and determining $\lambda_1$ and $\lambda_2$.

\section{Introduction} 
$B$ meson decays are complex processes that provide an ideal platform for precision studies of the Standard Model (SM) as they involve various physical scales. Inclusive decay modes (where the final state mesons are summed over) are considered to be theoretically cleaner than exclusive ones (where the final state particles are explicitly identified) as the latter involve transition form factors, which are rather difficult to evaluate. The inclusive $B$ meson decays are theoretically cleaner as they take advantage of hard scale, $m_b$, which is much larger than $\Lambda_{QCD}$. The presence of the hard scale enables the use of the Heavy Quark Expansion (HQE) \cite{Isgur:1988gb,Nussinov:1986hw} and local Operator Product Expansion (OPE) \cite{Khoze:1986fa,Blok:1993va}, and hence a systematic theoretical treatment of the decay rates.

HQE is a fundamental tool in the study of heavy quark physics. It provides a series expansion in $\frac{1}{m_Q}$, where $Q$ is the heavy quark \cite{Shifman:1987rj}. 
Though the mass of $b$ quark is heavy enough to use the expansion in $\frac{1}{m_b}$, yet corrections to the heavy quark limit are significant for achieving high precision.   
OPE is a powerful tool that enables a systematic treatment of non-perturbative Quantum Chromodynamics (QCD) effects by separating the effects originating at large and small distances \cite{Wilson:1969zs,Symanzik:1971vw}. For a heavy quark system, it turns out to be a series expansion in powers of $\frac{\Lambda_{\text{QCD}}}{m_b}$ \cite{Eichten:1989zv,Georgi:1990um} when combined with the theory of heavy quark expansion.

The corrections to the leading term ($m_b\to \infty$) in the heavy quark expansion are expected to be small in the end region of the phase space. This part of the phase space allows the contribution of several hadronic states which satisfy the condition: $m_X^2\to m_q^2 + \#\Lambda_{\text{QCD}}m_b$. An observable, like the decay rate, which averages over these hadronic states, can then be predicted reliably.
  To $\mathcal{O}(\frac{1}{m_b})$ in HQE, the major uncertainty in these predictions arises from the values of non-perturbative parameters, $\lambda_1$ and $\lambda_2$. Throughout the article, we consistently work to $\mathcal{O}(\frac{1}{m_b})$ in HQE. These parameters are defined as \cite{Neubert:1993mb,Manohar:2000dt}
  \begin{eqnarray}
      \lambda_1=\frac{1}{2m_{H_Q}}\langle H_Q|\bar{Q} (i{\bf D})^2 Q|H_Q\rangle, \hspace{0.1cm}  \hspace{0.2cm}\text{and} \hspace{0.2cm} 3\lambda_2=\frac{1}{2m_{H_Q}}\langle H_Q|\bar{Q} \frac{i}{2} \sigma_{\mu\nu}G^{\mu\nu} Q|H_Q\rangle, 
      \label{def_lambdas}
  \end{eqnarray}
  where $D$ is the covariant derivative and $|H_Q\rangle$ represents the states of hadrons containing one heavy quark and light cloud (for $b$ hadrons $H_Q=B,\Lambda_b$, etc.). 
  Physically, $\lambda_1$ provides information about the average spatial momentum squared of the heavy quark, while $\lambda_2$ represents the amount of color magnetic field produced by the light cloud at the position of the heavy quark \cite{Bigi:1993ex,Uraltsev:2000qw}.
  
  In the past, the parameters $\lambda_1$ and $\lambda_2$ have been determined using QCD models \cite{Bagan:1991sg,Neubert:1992fk,Ball:1993xv,Bigi:1994ga}, and also have been extracted by fitting to the experimental data \cite{Ligeti:1994qi,Kapustin:1995nr,Falk:1995kn,Gremm:1996yn,Gremm:1996df}. 
   Ref. \cite{JLQCD:2003chf} determines $\mu_{\pi}^2$ and $\mu_{G}^2$\footnote{In this study, we are specifically considering the $\lambda_{1(2)}$ parameters rather than $\mu_{\pi(G)}^2$.
} in quenched lattice QCD using the NRQCD action including $\mathcal{O}(\frac{1}{m_Q})$ terms for the heavy quark. They compute $\lambda_2$ explicitly but not $\lambda_1$, and rather calculate the difference of matrix elements using two different methods. Also, these two different methods of computation have followed different central values.
   Consequently, having an unambiguous prediction for these parameters becomes important.

It should be noted that both $B\to X_c \ell\bar{\nu}_\ell$ and $B\to X_u \ell\bar{\nu}_\ell$ exhibit similar decay signatures characterized by a high-momentum lepton, a hadronic system, and undetected neutrino energy. Thus, distinguishing between these two processes is challenging. Further, close to the lepton energy end point regions, non-perturbative shape functions enter the description of the decay kinematics, making predictions for the decay rates dependent on the precise modeling of these shape functions \footnote{ We briefly touch upon the shape functions in Section-\ref{discussion}.}. Instead, we are exploring the possibility of determining the non-perturbative parameters in an efficient manner. Considering modes with similar cuts such as $B\to X_{c(u)} \ell \bar{\nu}_\ell$ together with $B\to X_{c(u)} \ell \bar{\nu}_\ell \gamma$ may provide complementary information allowing to extract the non-perturbative parameters. Further, the inclusion of a hard photon in the decay process introduces additional degrees of freedom, such as the angle between the lepton and the photon. As an example, the forward-backward symmetry has been calculated (for details, see section- \ref{result}). The complete angular analysis is left for future work. 

In this article, we explore how the experimental determination of the decay rate for the $B\to X_{c(u)} \ell\bar{\nu}_\ell\gamma$ mode, in conjunction with the $B\to X_{c(u)} \ell\bar{\nu}_\ell$ mode.
It is worth reiterating that the emitted photon is a hard photon, and the process should not be thought of as soft photon correction to $B\to X_{c(u)}\ell\bar{\nu}_{\ell}$. To this order in $\frac{1}{m_b}$, the decay widths of both these modes, i.e. $B\to X_{c(u)}\ell\bar{\nu}_{\ell}$ and $B\to X_{c(u)}\ell\bar{\nu}_{\ell}\gamma$, have linear dependence on $\lambda_1$ and $\lambda_2$:
\begin{eqnarray}
   \text{decay width} \sim A+B\lambda_1+C\lambda_2.
    \label{totgrep}
\end{eqnarray}
for different values of ${A,B,C}$ for the two modes. Therefore, knowing (or experimentally measuring) one of the sides of these equations will provide a simultaneous set of linear equations, which can then be solved to get unambiguous determinations of $\lambda_1$ and $\lambda_2$. While we are including terms $\mathcal{O}(\frac{1}{m_b})$ in HQE, it is straightforward to include higher order terms. To avoid the uncertainties which may arise due to the presence of the Cabibbo-Kobayashi-Maskawa (CKM) element ($V_{ub}$ or $V_{cb}$), we propose to consider the ratio of the decay widths in different ranges of the leptonic energy instead of directly working with the decay width (see Section-\ref{result} for details). Such ratios are defined as (for radiative width, $\Gamma_\gamma$, and non-radiative, $\Gamma$)
\begin{eqnarray}
     R_1=\frac{\int_{0}^{0.2} dy \frac{d\Gamma_{\gamma}}{dy}}{\int_{0}^{0.5}dy \frac{d\Gamma_{\gamma}}{dy}}
      \ \ \ \ \ \text{ and}
\hspace{2cm}
     R_2=\frac{\int_{0}^{0.5} dy \frac{d\Gamma}{dy}}{\int_{0}^{1}dy \frac{d\Gamma}{dy}},
     \label{btoxur}
 \end{eqnarray}
 where $y$ is the lepton energy expressed in dimensionless units. Further, the considered cuts are representative cuts and can be chosen in accordance with the experimentally suitable ones. Here, we are exploring an avenue which could provide complementary information in a simple way.
 
The rest of the article is organised as follows: in Section- \ref{nonrad}, as a warm-up, we first calculate the decay width for the non-radiative process $B\to X_{c(u)}\ell\bar{\nu}_{\ell}$ mode. This is achieved by directly using the Cutkosky method applied to the amplitude in contrast with the usual approach of writing down the hadronic tensor in terms of invariant quantities and employing analytical properties. Both approaches are actually equivalent, and we explicitly verify that the result matches with \cite{Blok:1993va,Manohar:2000dt} as it should. In Section- \ref{radiative}, we provide the details of the calculation of the decay width for $B\to X_{c(u)}\ell\bar{\nu}_{\ell}\gamma$ mode. In Section- \ref{result}, we discuss our results for the differential decay rate for $B\to X_{c(u)}\ell\bar{\nu}_{\ell}\gamma$ and its sensitivity to the energy of the hard photon. Also, we provide a complimentary method to extract non-perturbative parameters $\lambda_1$ and $\lambda_2$. Finally, we conclude in Section-\ref{discussion} and discuss the implications of the decay rate of $B\to X_{c}\ell\bar{\nu}_{\ell} \gamma$ for calculation of $\lambda_1$ and $\lambda_2$.

\section{ Decay rate of $B\to X_{q}\ell\bar{\nu}_{\ell}$ $(q=c/u)$} 
\label{nonrad}
 The weak Hamiltonian density for the inclusive semi-leptonic $B$ meson decays to final state containing a $c$ or $u$ quark ($B\to X_q\ell\bar{\nu}_{\ell}$) is given by
 \begin{eqnarray}
     \mathcal{H}_{weak}=\frac{4G_F}{\sqrt{2}}V_{qb} (\bar{q}\gamma_{\mu}P_L b) (\bar{\ell}\gamma^{\mu}P_L \nu_{\ell}),
     \label{weakh}
 \end{eqnarray}
 where, $G_F$ and $V_{qb}$ are Fermi constant and CKM element, respectively. 
 \begin{figure}[h]
	   \centering
	    \includegraphics[width=0.4\linewidth]{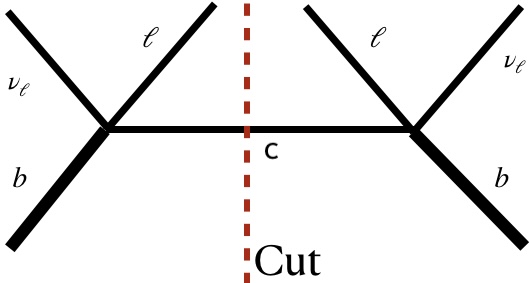}
	   \caption{Representative Diagram of forward scattering for $b\to c \ell \nu_{\ell}$}
	    \label{forml}
	\end{figure}
The process of $B\to X_q\ell\bar{\nu}_{\ell}$ involves the $b\to q$ transition and sum over the final state mesons containing the $q$ quark. 
To calculate the decay width for the inclusive process, the forward scattering matrix element is a useful quantity. Fig.(\ref{forml}) shows the Feynman diagram for the forward scattering matrix element of the inclusive semi-leptonic $B \to X_c \ell \bar{\nu}_\ell$ decay. 
 The imaginary part of the forward scattering matrix element called the transition matrix element, $\langle B|\hat{T} |B\rangle$, is related to the decay to an inclusive final state through the optical theorem. It is given by
 \begin{eqnarray}
     \Gamma\propto \frac{1}{2m_B} Im\big\lbrace \langle B|\hat{T}(b\to q \to b) |B\rangle\big\rbrace,
 \end{eqnarray}
 where the operator sandwiched between the hadronic states is called forward scattering operator (or transition operator). The explicit form of the transition operator reads  
 \begin{eqnarray}
     \hat{T}^{\mu\nu}(b\to q \to b)=i\int d^4x e^{-iq.x}\frac{1}{2m_B} T\lbrace J^{\mu\dagger}_w(x)J^{\nu}_w(0)\rbrace,
     \label{transop}
 \end{eqnarray}
 where $T$ denotes the time ordered product and $J^{\mu}_w\ (=\bar{q}\gamma^{\mu}P_L b)$ is the weak current. 
 Further, the differential decay rate for $B\to X_q\ell\bar{\nu}_{\ell}$ process is given by
 \begin{eqnarray}
     \frac{d^3\Gamma}{dq^2dE_{\ell}dE_{\nu}}&=&\frac{1}{4}\sum_{X_q}\sum_{\text{spins}} \frac{1}{2m_B} |\langle X_q \ell \bar{\nu}_\ell|\mathcal{H}_{weak}|B\rangle|^2\delta^4(p_B-p_X-p_{\ell}-p_{\nu}),\nonumber\\
     &=&2G_F^2 |V_{ub}|^2 \mathcal{M}_{\mu\nu}\mathcal{L}^{\mu\nu},
 \end{eqnarray}
 where $q^2=(p_{\ell}+p_{\nu})^2$, $E_{\ell}$ is the lepton energy, and $E_{\nu}$ is the neutrino energy. Sum over spins indicates sum over lepton spins. Furthermore, the leptonic tensor ($\mathcal{L}_{\mu\nu}$) is defined directly from the electroweak Lagrangian,
 \begin{eqnarray}
     \mathcal{L}_{\mu\nu}=\sum_{\text{spins}}(\bar{\nu}_{\ell}\gamma_{\mu}(1-\gamma_5)\ell)(\bar{\ell}\gamma_{\nu}(1-\gamma_5)\nu),
 \end{eqnarray}
 whereas the hadronic tensor, $\mathcal{M}_{\mu\nu}$, is expressed in terms of matrix elements of electroweak currents and is given by 
 \begin{eqnarray}
     \mathcal{M}_{\mu\nu}=\frac{1}{2m_B}\sum_{X_q} \langle B|J_{\mu}|X_q\rangle\langle|X_q|J_{\nu}|B\rangle \delta^4(p_B-p_X-p_{\ell}-p_{\nu}).
 \end{eqnarray}
  The hadronic tensor ($\mathcal{M}_{\mu\nu}$) is defined as the absorptive part of the matrix element of the transition operator (i.e. Eqn. (\ref{transop})). Explicitly,
 \begin{eqnarray}
     \mathcal{M}_{\mu\nu}=-i \hspace{0.1cm}Disc.(\langle B|\hat{T}_{\mu\nu}|B\rangle). 
 \end{eqnarray}
Instead of using the general parametrization for the hadronic tensor and relying on the analytical properties, we directly evaluate the transition operator ($\hat{T}_{\mu\nu}$) along with the leptonic tensor ($\mathcal{L}_{\mu\nu}$) to obtain the matrix element involved in the inclusive semi-leptonic $B\to X_q \ell\bar{\nu}_{\ell}$ decay. At the lowest order in perturbation theory (i.e. in $\alpha_s$ expansion), the quark level amplitude ($\mathcal{M}_{NR}$) for the Fig. (\ref{forml}) is given by 
 \begin{eqnarray}
     \mathcal{M}_{NR}=\frac{i}{(p_b+\Pi-q)^2-m_q^2}\bar{b} \gamma_{\nu}P_L (\slashed{p_b}+\slashed{\Pi}-\slashed{q}+m_b)\gamma_{\mu}P_L b  \mathcal{L}^{\mu\nu}.
    \label{mlnr}
 \end{eqnarray}
 The subscript `NR' refers to a non-radiative process, i.e., the process without a photon in the final state. Further, $p_b +\Pi$ is the effective momentum of $b$ quark, where $p_b=m_b v$ is heavy quark momentum while $\Pi$ is residual momentum of heavy quark. Also, $q\ (=p_\ell+p_\nu)\sim \mathcal{O}(m_b)$ while $\Pi$ $\sim\mathcal{O}(\Lambda_{\text{QCD}})$. Hence, the expansion of $ \mathcal{M}_{NR}$ in powers of $\Pi$ produces an expansion in powers of $\frac{\Lambda_{\text{QCD}}}{m_b}$. Expanding the denominator to $(\frac{\Lambda_{\text{QCD}}}{m_b})^2$,
 {\small\begin{eqnarray}
     \frac{1}{(p_b+\Pi-q)^2-m_q^2}&=&  \frac{1}{ ((p_b-q)^2-m_q^2)}\Big[ 1-  \frac{2(p_b-q).\Pi+\Pi^2}{((p_b-q)^2-m_q^2)}+\frac{2((p_b-\Pi).\Pi)^2}{((p_b-q)^2-m_q^2))^2}\Big].
     \label{inr}
 \end{eqnarray}}
 The hadronic part of $\mathcal{M}_{NR}$ is then sandwiched between the $B$ meson states. It is to be noted that since the OPE involved here is in the expansion of inverse power of $m_b$, therefore the $b$-quark field in QCD is converted to those in HQET at order $1/m_b$ through the relation: 
 \begin{eqnarray}
     b(x)=e^{-im_b\,v.x} \left(1+\frac{i\slashed{D}}{2m_b}\right)b_v(x).
 \end{eqnarray}
 Now, the obtained matrix elements are simplified as \cite{Manohar:2000dt}
 \begin{eqnarray}
     \langle B(v)|\bar{b}_v\gamma^{\mu} b_v|B(v) \rangle &=& 2p_B^\mu,\nonumber\\
     \langle B(v)|\bar{b}_v\gamma_{\mu}D_{\tau} b_v|B(v) \rangle &=& \frac{\lambda_1+3\lambda_2}{3m_b}(2g_{\mu\tau} -5v_{\mu}v_{\tau}),\nonumber\\
     \langle B(v)|\bar{b}_v\gamma_{\mu}D_{(\alpha}D_{\beta)} b_v|B(v) \rangle &=& \frac{2\lambda_1}{3m_b}(g_{\alpha\beta} -v_{\alpha}v_{\beta})v_{\mu},\nonumber\\
      \langle B(v)|\bar{b}_v\gamma_{\mu}D^2 b_v|B(v) \rangle &=& \frac{2\lambda_1}{m_b} v_{\mu}.
      \label{ope}
 \end{eqnarray}
 Next, we calculate the imaginary part of the denominator,
  \begin{eqnarray}
     \frac{1}{((p_b-q)^2-m_q^2)}\to (-2\pi i) \delta((p_b-q)^2-m_q^2)\Theta((p_{b}^0-q^0)).
 \end{eqnarray}
Integrating over the neutrino energy, the double differential decay rate for $B\to X_c\ell\bar{\nu}_{\ell}$ mode is calculated as
{\small\begin{eqnarray}
    \frac{d^2\Gamma}{dy d\hat{q^2}}&=&\frac{G_F^2 |V_{ub}|^2 m_b^5}{96\pi^3}y\Big[6(1-\frac{\hat{q^2}}{y})(1-\rho-y+\hat{q^2})+\lambda_1\Big(-3+3\rho+4\frac{\hat{q}^2}{y}-4\rho\frac{\hat{q^2}}{y}-6\hat{q^2}+4\frac{(\hat{q^2})^2}{y}\nonumber\\&-& \delta(z)\Big(1-2\rho+\rho^2-3y(1-\rho)-3\frac{\hat{q^2}}{y}+2\rho \frac{\hat{q^2}}{y}+\rho^2 \frac{\hat{q^2}}{y}+11\hat{q^2}-3\rho \bar{q^2}-3x\hat{q^2}-6\frac{(\hat{q^2})^2}{y}\nonumber\\&-&2\rho\frac{(\hat{q^2})^2}{y}+2(\hat{q^2})^2+\frac{(\hat{q^2})^3}{y}\Big)+\delta'(z)(1-\frac{\hat{q^2}}{y})(1-\rho-y+\hat{q^2})(1-2\rho+\rho^2-2\hat{q^2}-2\rho \hat{q^2}+(\hat{q^2})^2)\Big)\nonumber\\&+&3\lambda_2\Big(1-5\rho+2\frac{\hat{q^2}}{y}+10\rho\frac{\hat{q^2}}{y}+10\hat{q^2}(1-\frac{\hat{q^2}}{y})-\delta(z)\Big(-1+6\rho-5\rho^2+y(1-5\rho)+\frac{\hat{q^2}}{y}(1-2\rho)\nonumber\\&+&5\rho^2\frac{\hat{q^2}}{y}+\hat{q^2}(1+15\rho)+5y\hat{q^2}-2\frac{(\hat{q^2})^2}{y}-10(1+y)\frac{(\hat{q^2})^2}{y}+\frac{(\hat{q^2})^3}{y}\Big)\Big)\Big],
    \label{nrdiffdecay}
\end{eqnarray}}
 where $y=\dfrac{2E_{\ell}}{m_b}$, $q^2=(p_{\ell}+p_{\nu})^2$, $\hat{q^2}=\dfrac{q^2}{m_b^2}$, $\rho=\dfrac{m_c^2}{m_b^2}$ and $z=1-y-\frac{\hat{q^2}}{y}+\hat{q^2}-\rho$. Eqn. (\ref{nrdiffdecay}) is in perfect agreement with \cite{Blok:1993va,Manohar:2000dt}. Integrating over $\hat{q}^2$, the lepton spectrum for $B\to X_c\ell\bar{\nu}_{\ell}$ is obtained as 
 \begin{eqnarray}
  \frac{d\Gamma}{dy}&=&2\Gamma_0 y^2\Big[ (3-2y)-3\rho -\frac{3\rho^2}{(1-y)^2}+\frac{(3-y)\rho^3}{(1-y)^3}-\frac{\lambda_1}{m_b^2}\Big(-\frac{5}{3}y-\frac{y(5-2y)\rho^2}{(1-y)^4}+\frac{2y(10-5y+y^2)\rho^3}{3(1-y)^5}\big)\nonumber\\&&-\frac{3\lambda_2}{m_b^2}\Big(-\frac{(6+5y)}{3}+\frac{2(3-2y)\rho}{(1-y)^2}+\frac{3(2-y)\rho^2}{(1-y)^3}-\frac{5(6-4y+y^2)\rho^3}{3(1-y)^4}\Big)\Big].   
 \end{eqnarray}
 Further,in the limit $\rho\to 0$ the lepton spectrum for $B\to X_u\ell\bar{\nu}_{\ell}$ can be obtained as
 {\small\begin{eqnarray}
     \frac{d\Gamma}{dy}&=&2\Gamma_0 \Big[ y^2(3-2y)-\frac{\lambda_1}{m_b^2}\Big(-\frac{5}{3}y^3+\frac{1}{6}\delta(1-y)+\frac{1}{6}\delta'(1-y)\big)-\frac{\lambda_2}{m_b^2}\Big(-(6+5y)y^2\nonumber\\&&+\frac{11}{2}\delta(1-y)\Big)\Big],
 \end{eqnarray}}
 where, $\Gamma_0=\frac{G_F^2 |V_{c(u)b}|^2 m_b^5}{192\pi^3}$. It is important to note that the contribution from the parton model, which is proportional to $2y^2(3-2y)$, does not vanish at the endpoint. This leads to the presence of delta functions and their derivatives in the lepton spectrum. After integrating over the lepton energy, the total decay rate for $B\to X_c\ell\bar{\nu}_\ell$ is obtained as
 \begin{eqnarray}
     \Gamma=\Gamma_0\Big(1+\frac{\lambda_1}{2m_b^2}+\frac{3\lambda_2}{2m_b^2}\big(2\rho\frac{d}{d\rho}-3 \big)\Big)(1-8\rho+8\rho^3-\rho^4-12\rho^2\ln{\rho}),
     \label{nrtd}
 \end{eqnarray}
 which has the same form as shown in Eqn. (\ref{totgrep}), and matches the result obtained in\cite{Blok:1993va,Manohar:2000dt}. 

\section{Differential rate of $B\to X_{c}\ell \bar{\nu}_{\ell}\gamma$ }
   \label{radiative}
\begin{figure}[h]
    \centering
    \begin{subfigure}{0.32\textwidth}
    \includegraphics[width=0.9\textwidth]{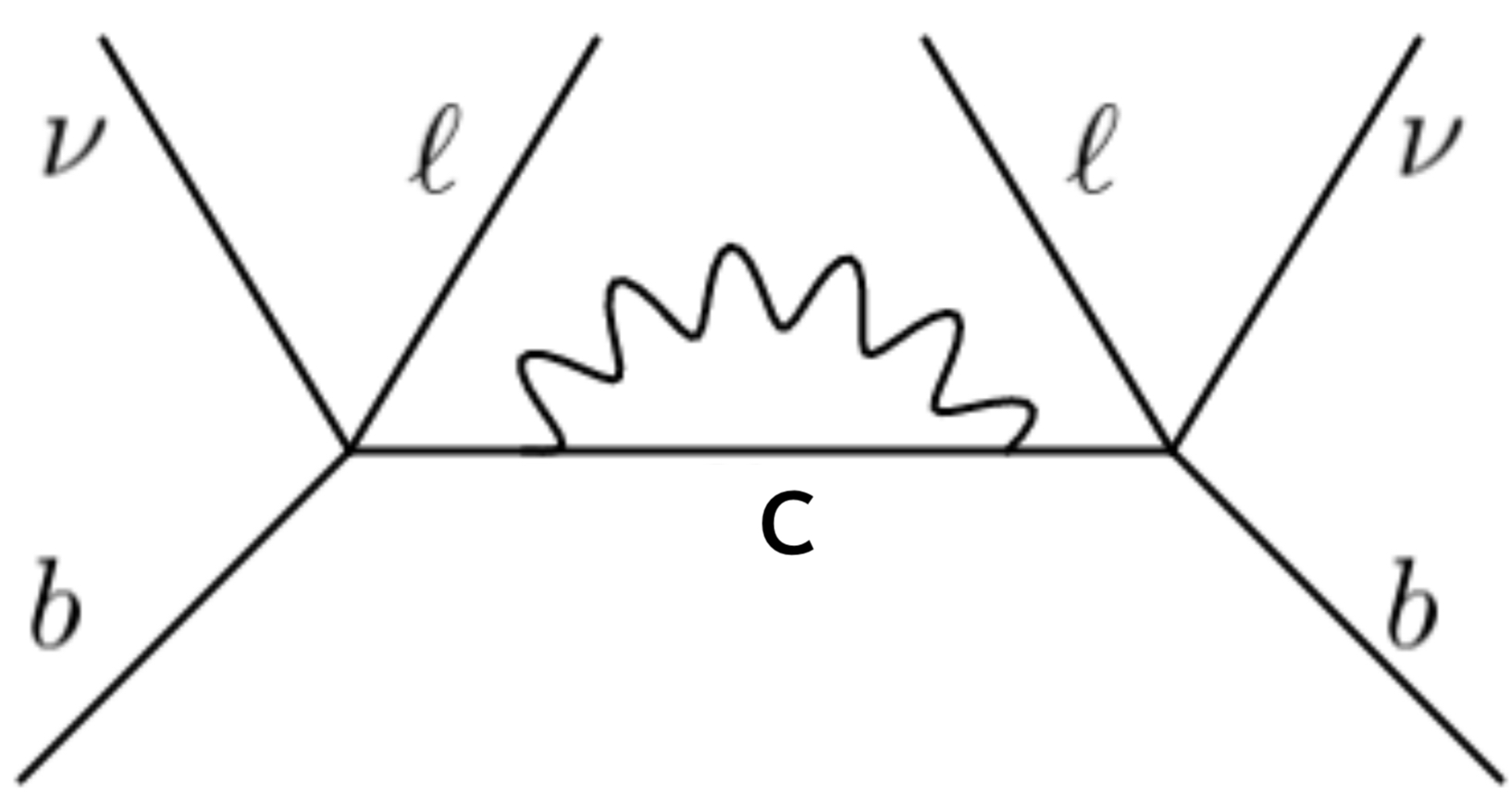}
    \caption{}
    \end{subfigure}
    \begin{subfigure}{0.32\textwidth}
    \includegraphics[width=0.9\textwidth]{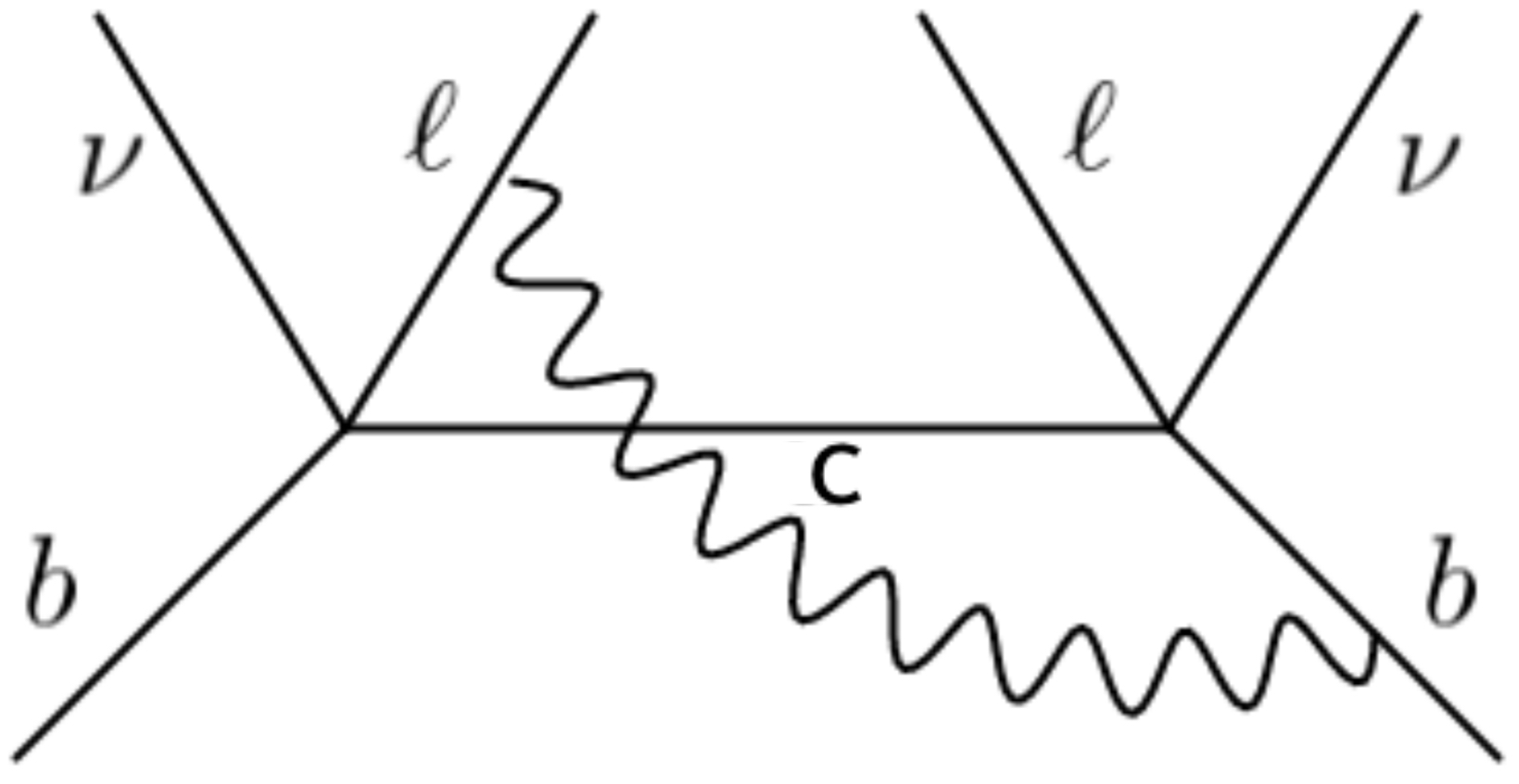}
      \caption{}
    \end{subfigure}
    \begin{subfigure}{0.32\textwidth}
    \includegraphics[width=0.9\textwidth]{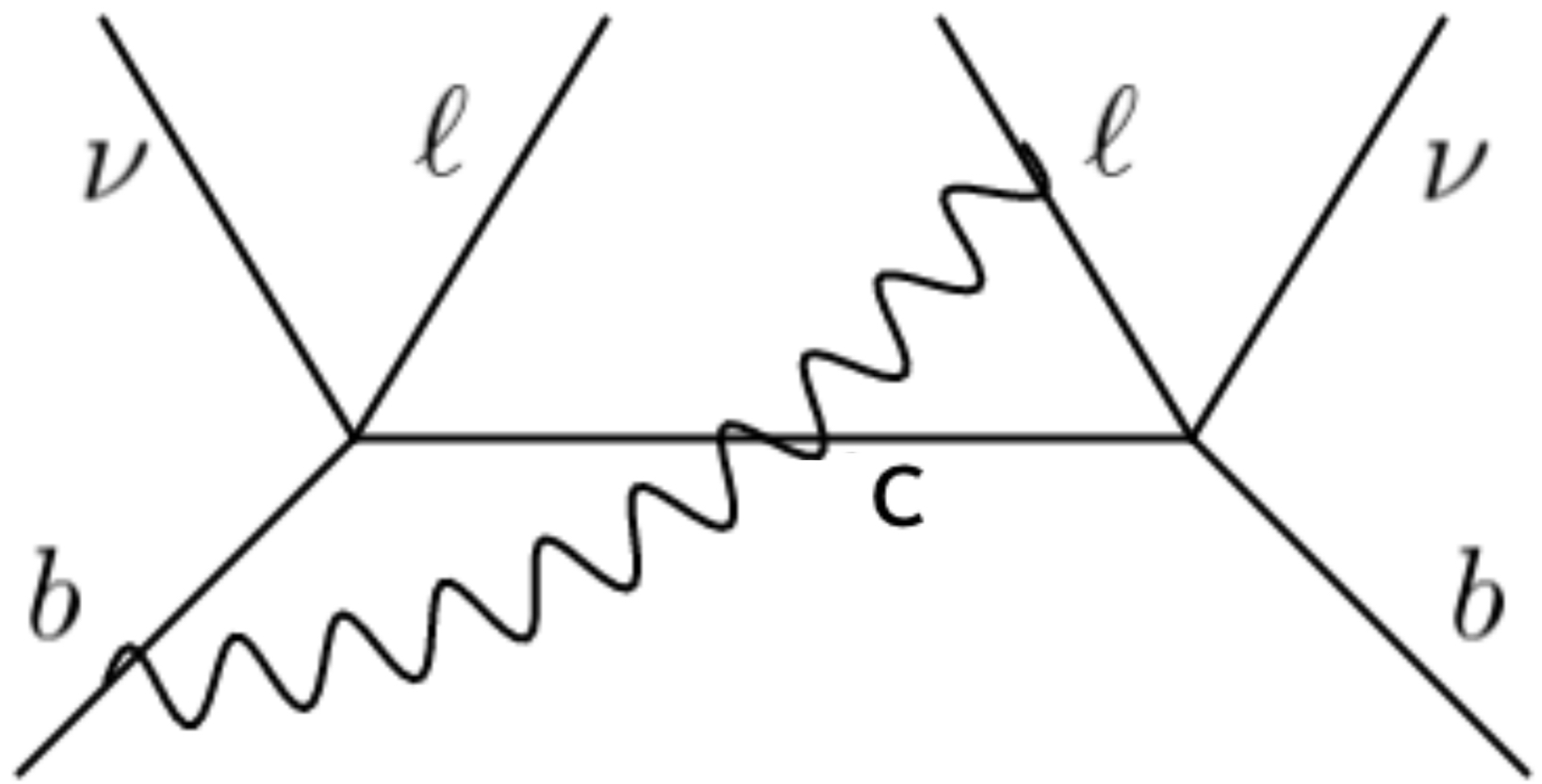}
      \caption{}
    \end{subfigure}\\
     \begin{subfigure}{0.32\textwidth}
    \includegraphics[width=0.9\textwidth]{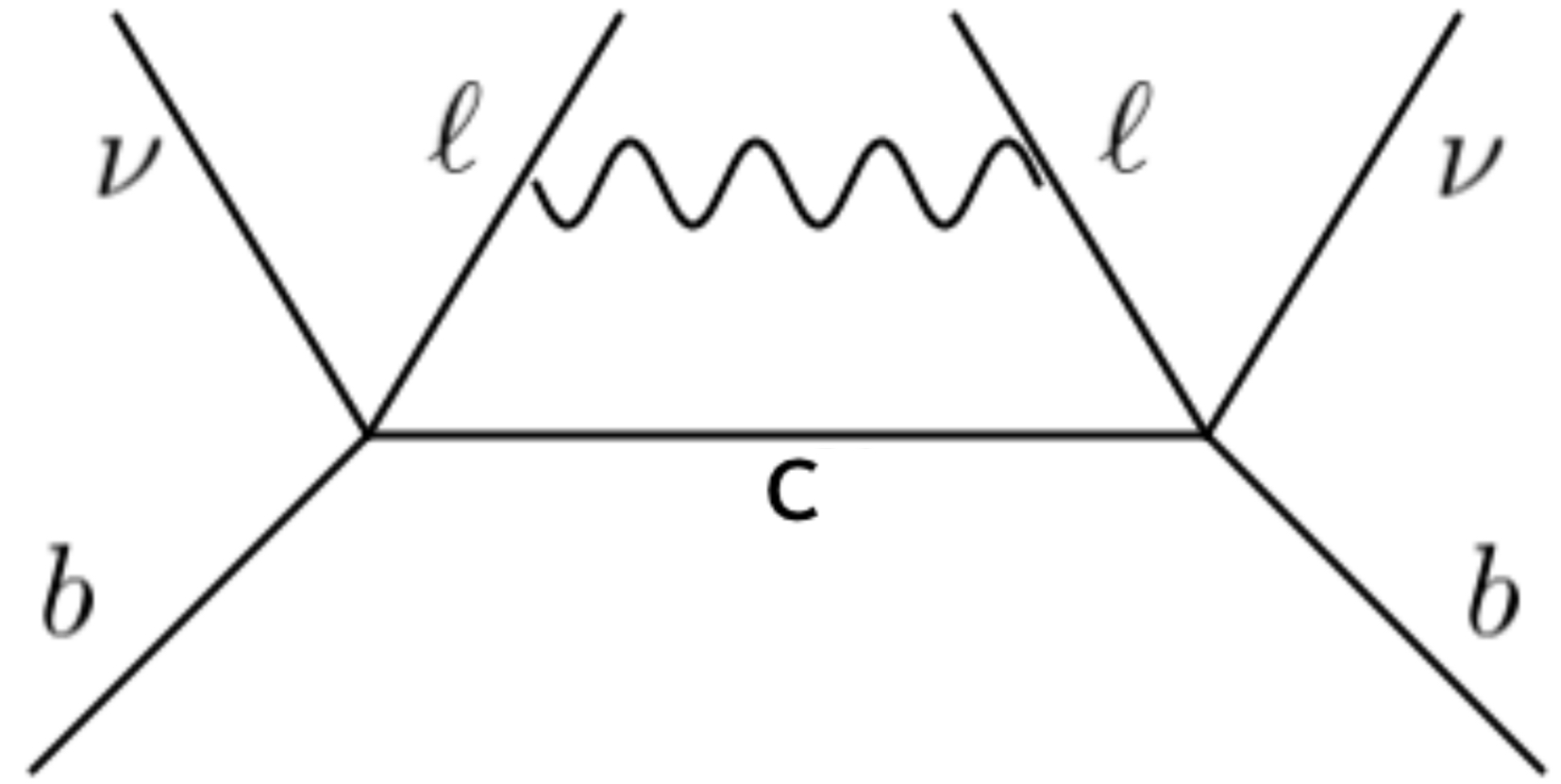}
      \caption{}
    \end{subfigure}
     \begin{subfigure}{0.32\textwidth}
    \includegraphics[width=0.9\textwidth]{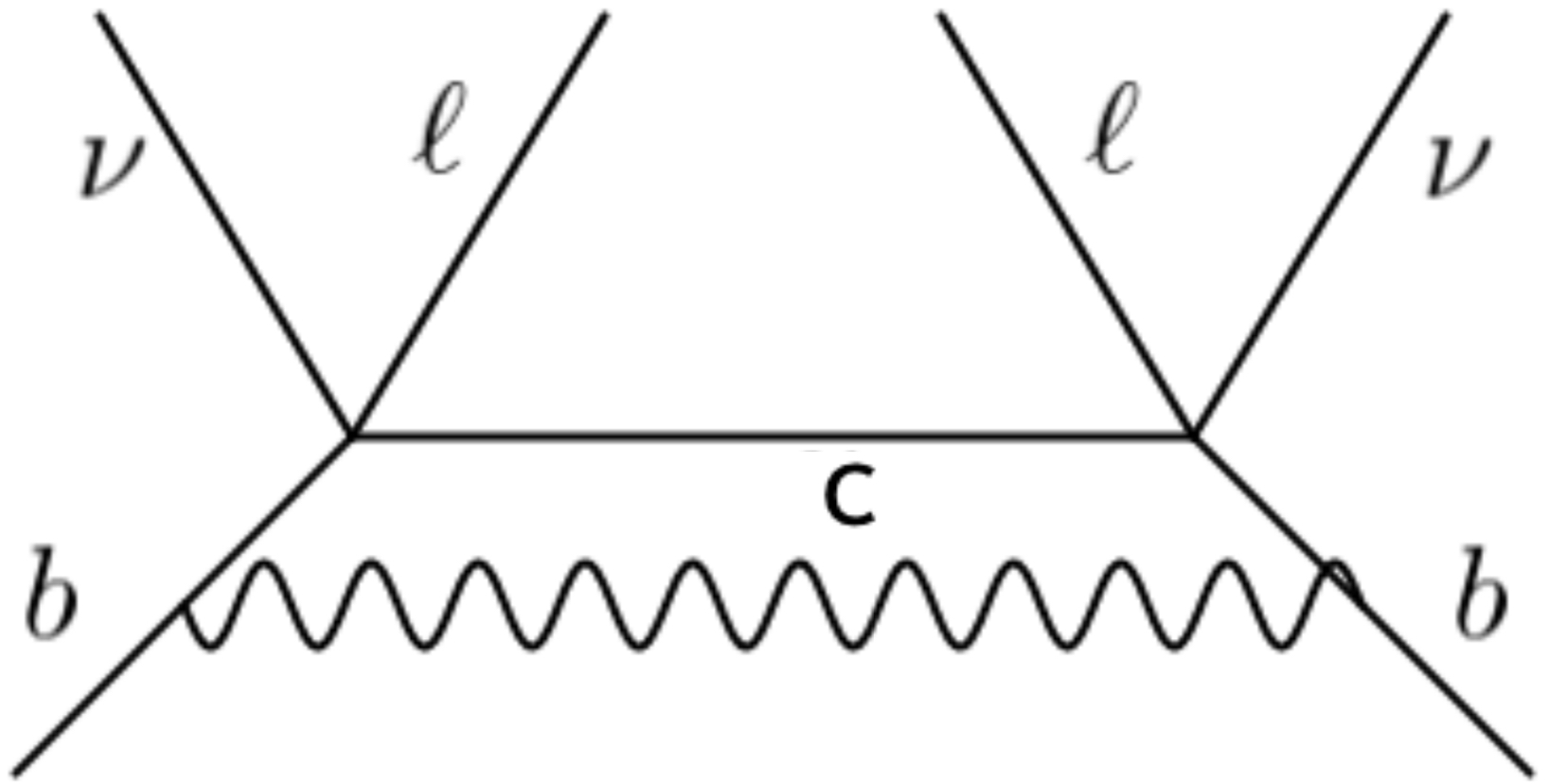}
      \caption{}
    \end{subfigure}
     \begin{subfigure}{0.32\textwidth}
    \includegraphics[width=0.9\textwidth]{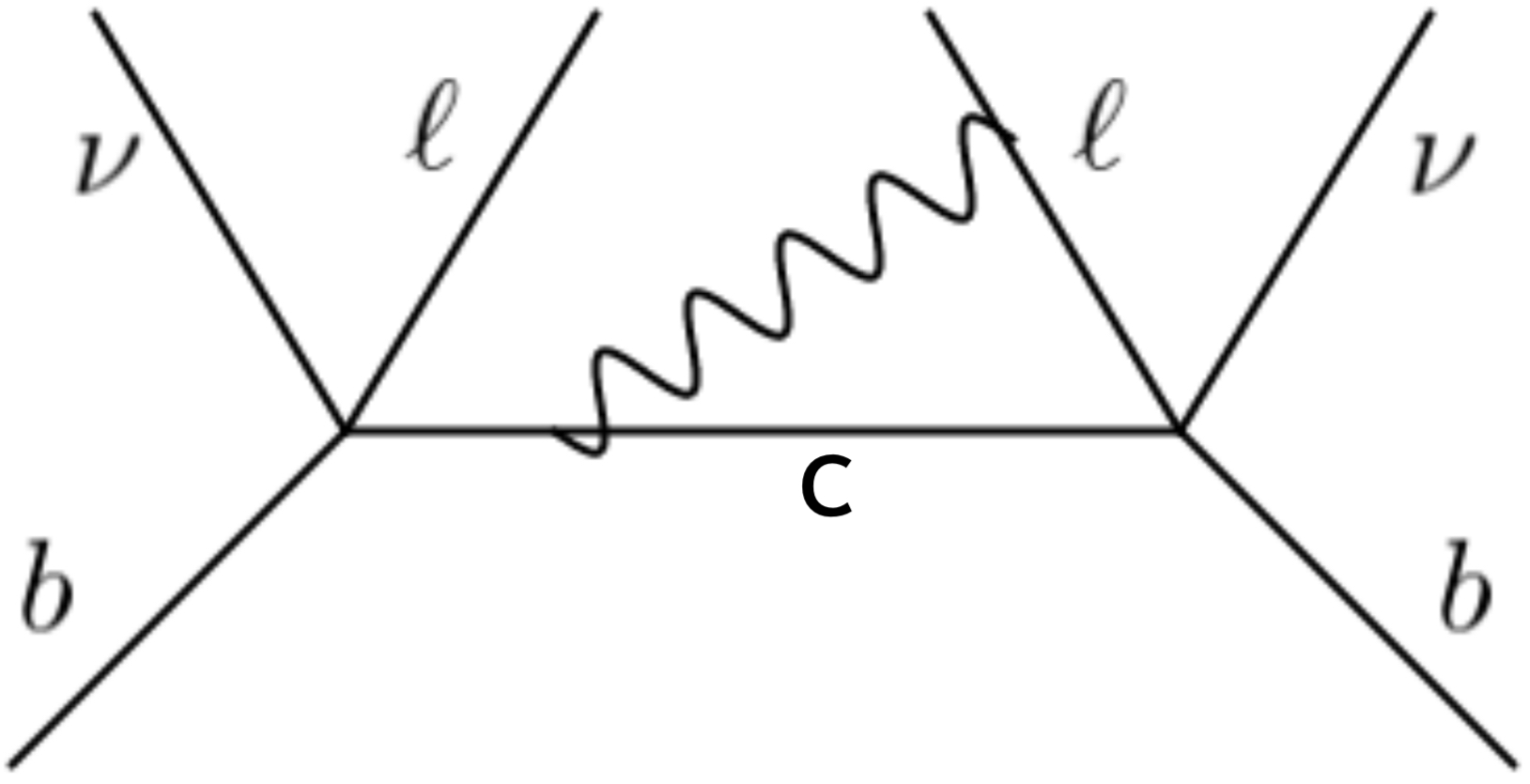}
      \caption{}
    \end{subfigure}\\
     \begin{subfigure}{0.32\textwidth}
    \includegraphics[width=0.9\textwidth]{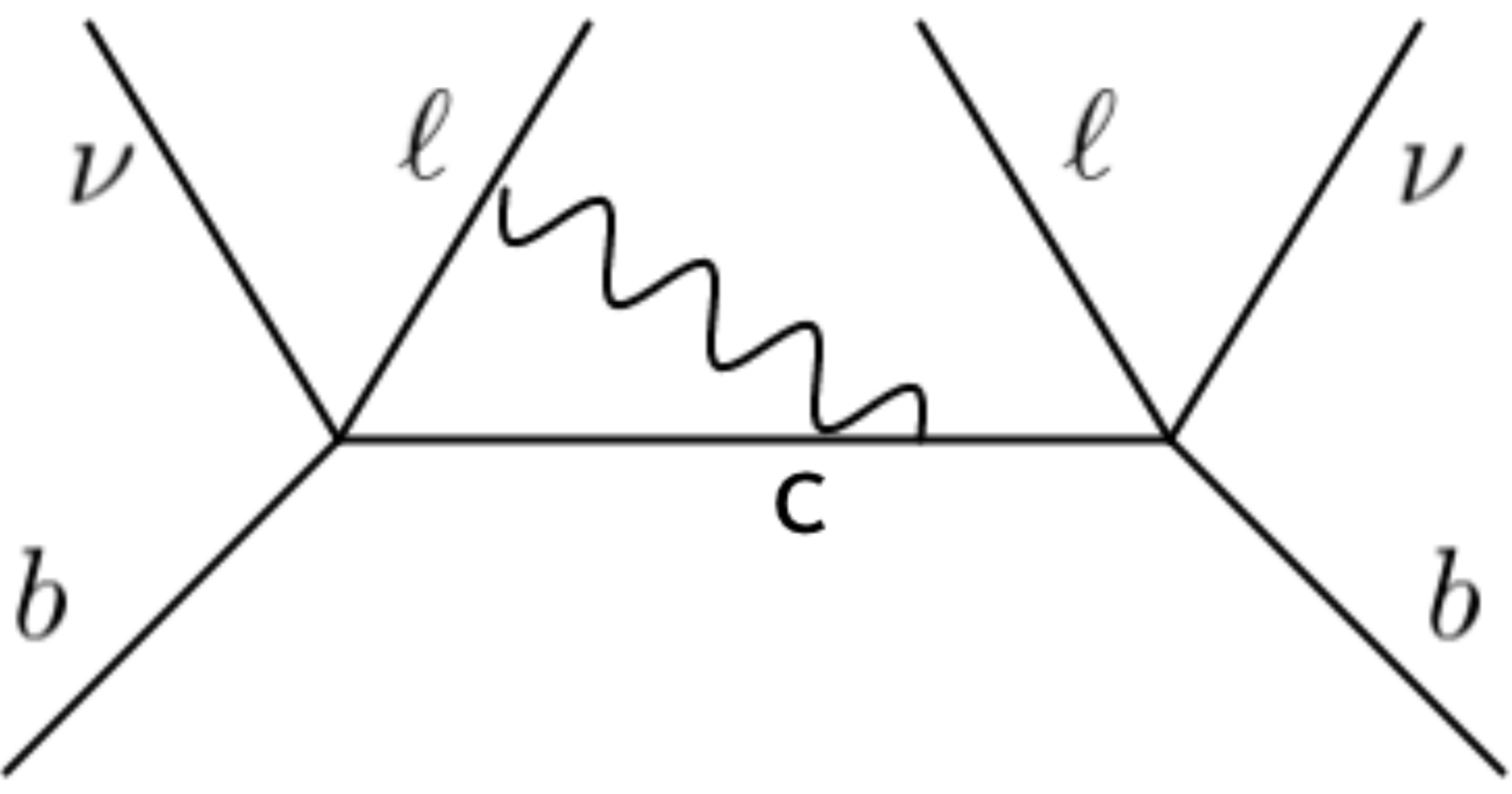}
      \caption{}
    \end{subfigure}
     \begin{subfigure}{0.32\textwidth}
    \includegraphics[width=0.9\textwidth]{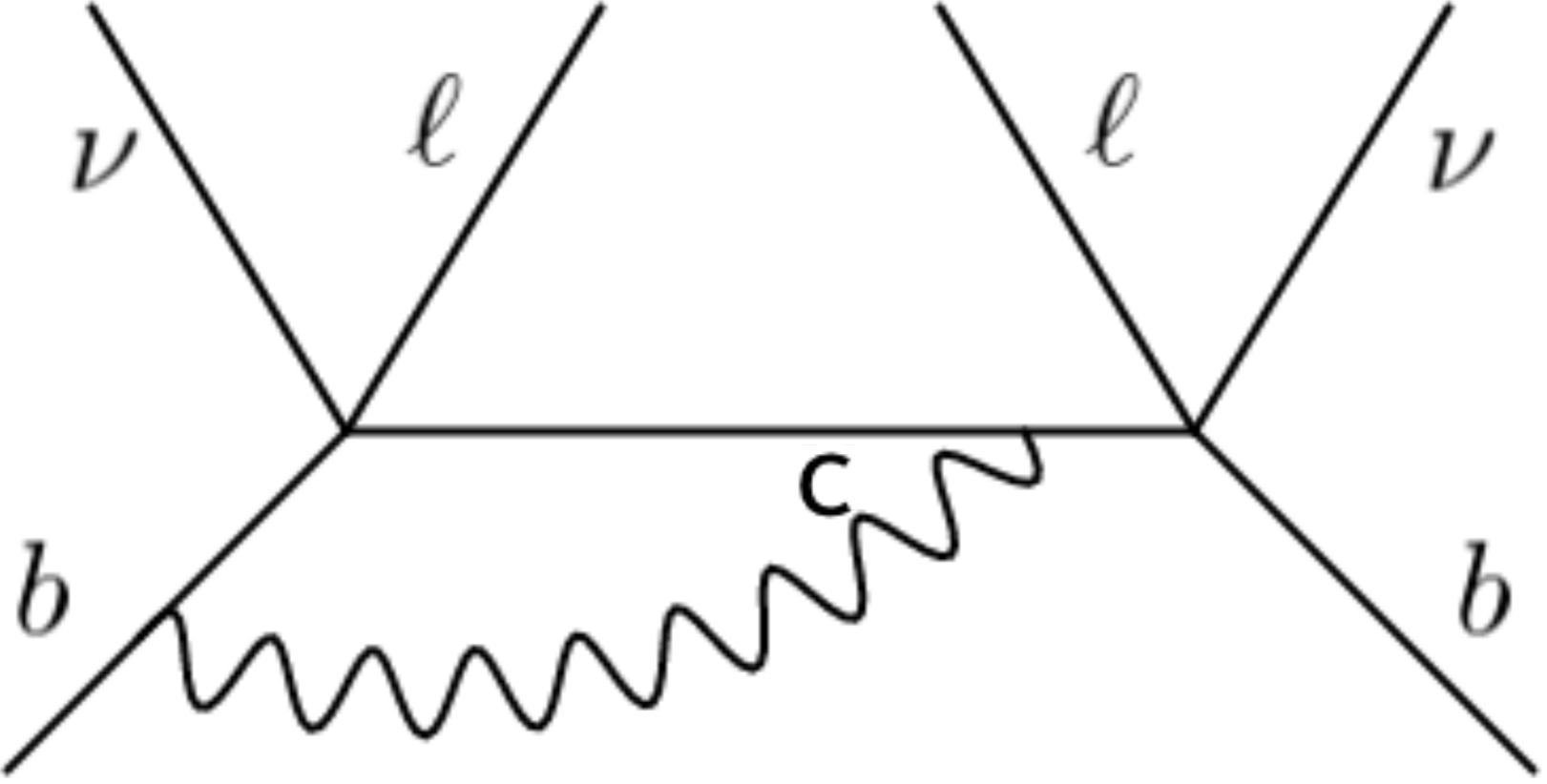}
      \caption{}
    \end{subfigure}
     \begin{subfigure}{0.32\textwidth}
    \includegraphics[width=0.9\textwidth]{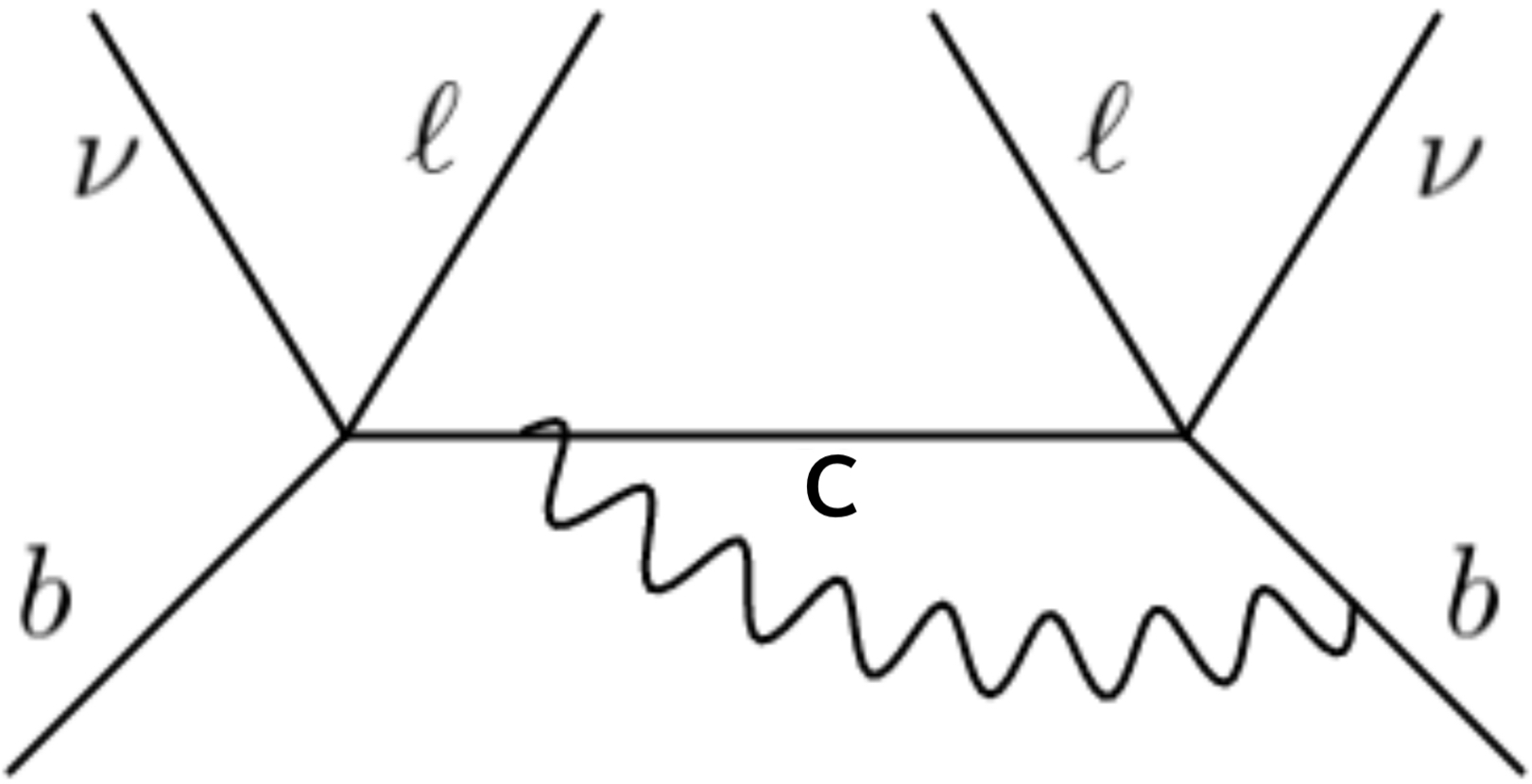}
      \caption{}
    \end{subfigure}
    \caption{Feynman diagrams for $B\to X_c \ell \bar{\nu}_{\ell}\gamma$.}
    \label{fdtot}
\end{figure}

\begin{figure}[h]
	     \centering
	     \includegraphics[width=0.4\linewidth]{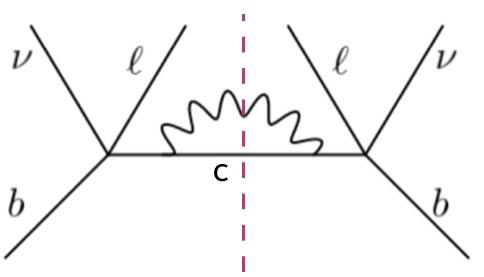}
	     \caption{Representative diagram with explicit cut}
	     \label{fircut}
	 \end{figure}

 	 In this section, we calculate the differential rate for $B\to X_c\ell \bar{\nu}_{\ell}\gamma$ mode. This decay mode is more likely to be measured, compared to the case of the
$B\to X_u$ mode which is additionally strongly suppressed by $|V_{ub}/V_{cb}|^2$. Fig. (\ref{fdtot}) shows all the Feynman diagrams\footnote{We have considered only those diagrams which after cutting the photon and $c$-quark lines lead to $B\to X_c \ell \bar{\nu}_{\ell} \gamma$.} contributing to the decay width of $B\to X_c \ell \bar{\nu}_{\ell} \gamma$ at leading order in perturbation theory. At the leading order ($m_b\to \infty$), the decay width for the process $B\to X_c\ell \bar{\nu}_{\ell}\gamma$ is obtained by the partonic result and further the preasymptotic effects i.e. the sub-leading contributions in heavy quark expansion are obtained in the powers of $\frac{\Lambda_{\text{QCD}}}{m_b}$. 
Similar to Section-\ref{nonrad}, our focus is on the $B\to B$ forward scattering matrix element instead of the amplitude for $B\to X_c \ell \bar{\nu}_{\ell}\gamma$ itself. The imaginary part of the amplitude shown in Fig.(\ref{fircut}) is related to the inclusive rate for the $B\to X_c \ell \bar{\nu}_{\ell}\gamma$ transition, as dictated by the optical theorem. However, the process $B\to X_c \ell \bar{\nu}_{\ell}\gamma$ is highly non-trivial compared to the $B\to X_c \ell \bar{\nu}_{\ell}$ mode due to the presence of a photon line between the charged quarks and leptons, as shown in Fig.(\ref{fdtot}). This means that some diagrams, such as Figs.2(b) and 2(c), do not easily separate into leptonic and hadronic parts like in the $B\to X_c \ell \bar{\nu}_{\ell}$ mode. Since the hadronic part in some of the diagrams communicates with the leptonic part through the photon, the calculation of the matrix element is more complex when expressed in terms of invariant tensors and using analytic properties of the transition operator. Also, in the present case, the transition tensor will be a four index object, two for the weak currents and two for the electromagnetic currents representing the photon emission. Therefore, it is more straightforward to use the Cutkosky method directly to compute the matrix element. As we verified in the last section, this method reproduces correct results for the decay rate of $B\to X_q\ell\bar{\nu}_{\ell}$ mode.
 
 Now, the decay rate for the semi-leptonic inclusive process $B\to X_c \ell \bar{\nu}_{\ell}\gamma$ is given by 
\begin{eqnarray}
    \Gamma_{\gamma}=\Big(\frac{4G_F}{\sqrt{2}}\Big)^2 |V_{ub}|^2 \frac{1}{2m_B}\int \frac{d^3p_l}{(2\pi)^3 2E_l}\int \frac{d^3p_{\nu}}{(2\pi)^3 2E_{\nu}}\int \frac{d^4 k}{(2\pi)^4} Im\Big[\langle B | I \mathcal{M}_{\mu\nu}\mathcal{L}^{\mu\nu} |B\rangle\Big],
    \label{totdwg}
\end{eqnarray}
with
\begin{eqnarray}
I\mathcal{M}_{\mu\nu}\mathcal{L}^{\mu\nu}=\sum_{m=1}^9 I_m \mathcal{M}_{\mu\nu}^{(m)}\mathcal{L}^{\mu\nu(m)},
\end{eqnarray}
where, $\mathcal{M}_{\mu\nu}^{(m)}$ and $\mathcal{L}_{\mu\nu}^{(m)}$ contain the Dirac structure for quark and leptonic part respectively, and, $I_m$ contains denominator part of the propagator. Also, $m=1,...9$ corresponds to Fig. 2 $(a),...(i)$. $k$ is photon four momentum, $p_{\ell(\nu)}$ is lepton (neutrino) four momentum. The explicit calculation of forward scattering operator for Fig. 2 (a) is presented below. All other diagrams can be calculated analogously. Calculating the hadronic and leptonic tensors requires the computation of $\mathcal{M}_{\mu\nu}^{(m)}$, $\mathcal{L}_{\mu\nu}^{(m)}$, and $I_m$.  
At the leading order in $\alpha_s$, the explicit forms of 
 $\mathcal{M}_{\mu\nu}^{(1)}$, $\mathcal{L}_{\mu\nu}^{(1)}$, and $I_1$ are  
 \begin{eqnarray}
     \mathcal{M}_{\mu \nu}^{(1)}&=&2(-ig^{\alpha \beta}) \bar{b}\gamma ^{\nu }\left(1-\gamma ^5\right)i\left(\slashed{ p_b}+\slashed{\Pi}-\slashed{q}+m_c\right)(-ieQ_u)\gamma ^{\alpha }i\left(\slashed{p_b}+\slashed{\Pi}-\slashed{k}-\slashed{q}+m_c\right)(-ieQ_u)\gamma ^{\beta }i\big(\slashed{p_b}\nonumber\\ &+&\slashed{\Pi}-\slashed{q}+m_c\big)\gamma ^{\mu }\left(1-\gamma ^5\right)b,\nonumber\\
     &=&2(e^2Q_u^2g^{\alpha \beta}) \Big[\bar{b}\gamma ^{\nu }\left(1-\gamma ^5\right)\left(\slashed{ p_b}+\slashed{\Pi}-\slashed{q}\right)\gamma ^{\alpha }\left(\slashed{p_b}+\slashed{\Pi}-\slashed{k}-\slashed{q}\right)\gamma ^{\beta }\big(\slashed{p_b}+\slashed{\Pi}-\slashed{q}\big)\gamma ^{\mu }\left(1-\gamma ^5\right)b\nonumber\\ &&+z^2 m_b^2 \big( \bar{b}\gamma ^{\nu }\left(\slashed{ p_b}+\slashed{\Pi}-\slashed{q}\right)\gamma ^{\alpha }\left(\slashed{p_b}+\slashed{\Pi}-\slashed{k}-\slashed{q}\right)\gamma ^{\beta }\big(\slashed{p_b}+\slashed{\Pi}-\slashed{q}\big)\gamma ^{\mu }\left(1-\gamma ^5\right)b\big)\Big],\\
     \mathcal{L}_{\mu \nu}^{(1)}&=&\left(\bar{\ell}\gamma^{\mu }(1-\gamma^5)\nu_{\ell}\right) \left(\bar{\nu}_{\ell} \gamma^{\nu }(1-\gamma^5)\ell\right),
     \label{m1l1}
 \end{eqnarray}
and 
 \begin{eqnarray}
     I_1&=&\frac{1}{(k^2+i\epsilon)((p_b+\Pi-q)^2-m_c^2+i\epsilon)((p_b+\Pi-q-k)^2-m_c^2+i\epsilon)((p_b+\Pi-q)^2-m_c^2+i\epsilon)},
     \label{i0}
 \end{eqnarray}
respectively. 
As before, the effective momentum of $b$ quark is given by $p_b+\Pi$. Expanding $I_1$ in powers of $\Pi$ yields an expansion in terms of $\frac{\Lambda_{\text{QCD}}}{m_b}$, similar to the $B\to X_c\ell\bar{\nu}_{\ell}$ mode. Further, the matrix element is also expanded in the powers of $z$ to order $z^3$, where $z=m_c/m_b$. Now, the explicit form of $I_1$ up to $\mathcal{O}(\Pi^2)$ is obtained from Eqn.(\ref{i0}), and is given by
\begin{eqnarray}
     I_1&=&  \frac{1}{ k^2((p_b-q-k)^2-m_c^2)}\Big[ \frac{1}{(p_c.k)^2}-\frac{2(p_b-q).\Pi}{(p_c.k)^3}-\frac{\Pi^2}{(p_c.k)^3}+\frac{2((p_b-q).\Pi)^2}{(p_c.k)^4}\Big]\nonumber\\&-& \frac{1}{ k^2((p_b-q-k)^2-m_c^2)^2}\Big[ \frac{2p_c.\Pi}{(p_c.k)^2}-\frac{4(p_c.\Pi)(p_b-q).\Pi}{(p_c.k)^3}-\frac{\Pi^2}{(p_c.k)^2}\Big]\nonumber\\&+&\frac{1}{ k^2((p_b-q-k)^2-m_c^2)^3}\Big[\frac{2(p_c.\Pi)^2}{(p_c.k)^2}\Big].
     \label{i1}
 \end{eqnarray}
Similar to Section-\ref{nonrad}, the Cutkosky method is exploited (see Fig.(\ref{fircut})) to calculate the imaginary part of the matrix element.
Mathematically, this essentially replaces the cut propagator with a product of delta function and theta function, enforcing the positive energy condition. For example
 \begin{eqnarray}
     \frac{1}{((p_b-q-k)^2-m_c^2+i\epsilon)}\to (-2\pi i) \delta((p_b-q-k)^2-m_c^2)\Theta((p_{b}^0-q^0-k^0)).
 \end{eqnarray}
More generally, one has the identity 
 \begin{eqnarray}
     -\frac{1}{\pi} Im\Big(\frac{1}{(p_b-q-k)^2 -m_c^2}\Big)^n=\frac{(-1)^{(n-1)}}{(n-1)!}\delta^{(n-1)}((p_b-q-k)^2 -m_c^2),
     \label{imdelta}
 \end{eqnarray}
where the superscript of the delta function denotes $(n-1)$th derivative with respect to its argument itself. 
When working with terms involving the derivatives of the delta function, it is important to handle them with care. The first step is to use integration by parts to remove the derivatives from the delta function and transfer them onto other functions by multiplying it. However, it is crucial to use the theta function carefully during this process because it determines the minimum value of the neutrino energy, denoted by $E_{\nu}$. See Appendix \ref{appB} for details of kinematics.  

Subsequently, we proceed by combining the terms $\mathcal{M}_{\mu\nu}^{(1)}$, $\mathcal{L}^{\mu\nu\,(1)}$, and $I_1$ to compute the imaginary part of the amplitude. Notably, our analysis reveals that no new operators are produced beyond those already present in the decay rate of the $B\to X_c\ell\bar{\nu}_{\ell}$ process. All the pertinent operators, up to dimension five, are listed in Eqn.(\ref{ope}). 
 
 It is evident that only $I_1$ contributes to the imaginary part of the matrix element. Hence, we have presented an explicit expression of the matrix element in accordance with the representation outlined in Eqn. (\ref{i1}). Each square bracket contains terms with expansion in $\Pi$ up to second order. The imaginary parts of the coefficients of these square brackets contribute to the delta function and its derivatives. We designate the forward scattering matrix element as  
 \begin{eqnarray}
     \mathcal{J}_1(n;\alpha)=\langle B(v)|Im\lbrace I_1\mathcal{M}_{\mu\nu}^{(1)}\mathcal{L}^{\mu\nu\,(1)}\rbrace|B(v)\rangle (n\alpha).
 \end{eqnarray}
 Where, $n=0,1,2$ denotes expansion powers of $\Pi$ and $\alpha=a,c,d$ represents square brackets of Eqn.(\ref{i1}) in order. For example in $\mathcal{J}(0;a)$, $'0'$ denotes the expansion in $\Pi$ to $\mathcal{O}(\Pi^0)$ and $'a'$ tells that elements of first square bracket of Eqn.(\ref{i1}) are chosen. We now explicitly list all the expressions for each of the terms of $I_1$ without the delta function or its derivatives.
\begin{eqnarray}
\mathcal{O}(\Pi^0): \nonumber\\
      \mathcal{J}_1(0;a)=&\frac{-1}{m_b(p_c.k)^2}16(p_b.p_{\ell})\Big( (q^2+m_b^2-2(p_b.q))(p_{\nu}.(p_b+k-q))-2((p_b-q).k)\nonumber\\&(q-p_b).p_{\nu}+4 z^2 m_b^2 \left(3 p_b . p_\nu+  k . p_\nu -3 q . p_\nu\right)\Big).
    \label{pi0}
\end{eqnarray}
{\small
\begin{eqnarray}
  \mathcal{O}(\Pi):\nonumber\\ \mathcal{J}_1(1;a)&=&\frac{-1}{3m_b^3(p_c.k)^3}64(\lambda_1+3\lambda_2)\Big( 2m_b^2(p_{\ell}.q)+(p_b.p_{\ell})(3m_b^2-5(p_b.q))\Big) \Big( (q^2+m_b^2-2(p_b.q))\nonumber\\&&(p_{\nu}.(p_b+k-q))-2((p_b-q).k)(p_{\nu}.q-p_b.p_{\nu})+ z^2 m_b^2\Big(2 \left(3  p_b+ k -3 q\right). p_n \big(2 m_b^2 \left(q. p_l\right)\nonumber\\&&+\left( p_b. p_l\right) \left(3 m_b^2-5 \left( q .p_b\right)\right)\big)+3 \left( k .p_c\right) \left(2 m_b^2 \left( p_l. p_n\right)-5 \left( p_b .p_l\right) \left( p_b. p_n\right)\right)\Big)\Big).
    \label{pi1}
    \end{eqnarray}}
    {\small
    \begin{eqnarray}
    \mathcal{O}(\Pi^2): \nonumber\\
    \mathcal{J}_1(2;a)&=&\frac{1}{3m_b^3(p_c.k)^4}32\lambda_1 \Big((p_b.p_{\ell})\Big( -2(p_c.k)^2(m_b^2(p_{\nu}.k-5(p_{\nu}.q))+(p_b.p_{\nu})(2(p_b.(q+k))+3m_b^2))\nonumber\\&+&((q^2+m_b^2-2(p_b.q))(p_{\nu}.(p_b+k-q))-2(p_b-q).k((p_b-q).p_{\nu}))((p_b.q)^2+m_b^2(3(p_c.k)-q^2))\nonumber\\&+& 4(p_c.k)\Big(m_b^2((p_{\ell}.p_{\nu})(2p_b.(q+k)-4(q.k)-3q^2)+(p_b.p_{\nu})(p_b.q+2(q.k+q^2))+2q^2(p_{\nu}.k))\nonumber\\&+&(p_b.q)(2(p_{\nu}.q)(p_b.(q+k))-2(p_b.q)(p_{\nu}.k)+(p_b.p_{\nu})(q^2+2(q.k)-4p_b.(q+k)))\nonumber\\&&-m_b^4(p_{\nu}.q)\Big)\Big)+4  z^2 m_b^2\Big(- p_b. p_l \left(3  p_b+ k -3 q\right) .p_n \left(\left( q. p_b\right)^2+m_b^2 \left(3 k. p_c-q^2\right)\right)-6 \left( k. p_c\right) \nonumber\\&&\left(m_b^2 \left(\left( p_b .p_l\right) \left( \bar{q} .p_n\right)+\left(p_b p_n\right) \left( \bar{q} .p_l\right)\right)-2 \left( p_b .p_l\right) \left(p_b. p_n\right) \left( q.p_b\right)\right)\Big)\Big).
    \label{pi2}
\end{eqnarray}}
 The sum, $\mathcal{J}_1(0;a)+\mathcal{J}_1(1;a)+\mathcal{J}_1(2;a)$, is multiplied with $\delta(k^2)\theta(k^0)\delta(((p_b-q-k)^2-m_c^2))\theta((p_b-q-k)^0)$. 
 Similarly, the second square brackets of Eqn. (\ref{i1}) with hadronic and leptonic tensors (Eqn. (\ref{m1l1})) produce the terms expanded in powers of $\Pi$. Since this set of terms comes multiplied with $\frac{1}{((p_b-q-k)^2-m_c^2)^2}$ i.e. square of the propagator, the sum of these terms has an overall factor of $\delta(k^2)\theta(k^0)\delta'((p_b-q-k)^2-m_c^2)\theta((p_b-q-k)^0)$.
{\small
\begin{eqnarray}
    \mathcal{O}(\Pi): \nonumber\\ \mathcal{J}_1(1;c)&=&\frac{-1}{3m_b^3(p_c.k)^2}128(\lambda_1+3\lambda_2)\Big(\Big((m_b^2+q^2-2p_b.q)((p_b-q+k).p_{\nu})-2((p_b-q).k)((p_b-q).p_{\nu})\Big)\nonumber\\&& \Big( (p_b.p_{\ell})(5p_b.(q+k)-3m_b^2)-2m_b^2(p_{\ell}.(q+k))\Big)+z^2 m_b^2 \left(3  p_b .p_\nu+  k .p_\nu-3 \left( q. p_\nu\right)\right) \big(2 m_b^2 \nonumber\\&&\left(  k. p_l+ q. p_l\right)+\left( p_b. p_l\right) \left(3 m_b^2-5 \left(  k. p_b+  q. p_b\right)\right)\big))\Big).\\
    \label{pi1d}
     \mathcal{O}(\Pi^2): \nonumber\\ \mathcal{J}_1(2;c)&=&\frac{-1}{3m_b^3(p_c.k)^3}128\Big(\lambda_1 (p_b.p_{\ell})\Big( 8((p_c.k+p_b.q)(p_{\nu}.q) -(2(p_c.k)+p_b.q)(p_b.p_{\nu}))(p_b.k)^2\nonumber\\&+&2\Big((4(p_{\nu}.k)(p_c.k)+(7p_c.k-4k.q-4q^2)(p_{\nu}.q)+(4(q.k)-3(p_c.k)+4q^2)(p_b.p_{\nu}))m_b^2\nonumber\\&-& 2(p_b.q)((p_{\nu}.k)(2(p_c.k)-q^2+2(p_b.q)-m_b^2)+(p_{\nu}.q)(2k.q -4(p_c.k)+q^2-4(p_b.q)+m_b^2))\nonumber\\&+& 2(2(q.k)(p_c.k+p_b.q)+(p_c.k)(q^2-8(p_b.q)+m_b^2)+(p_b.q)(q^2-4(p_b.q)+m_b^2))(p_b.p_{\nu}))(p_b.k)\nonumber\\&+& \Big(8(p_b-p_{\nu}).q(k.q)^2+2(2(3q^2-2(q.p_b)+m_b^2)(p_{\nu}.q-p_b.p_{\nu})+p_c.k(11(p_b.p_{\nu})-15(q.p_{\nu})))(k.q)\nonumber\\& +&4q^2(q^2-2(p_b.q)-m_b^2)(p_b-q).p_{\nu}+ p_c.k((2(p_b.q)-m_b^2)(7q-3p_b).p_{\nu}+q^2(11p_b-15q).p_{\nu}) \Big) m_b^2\nonumber\\&+& 4 (p_b.q)\Big(p_b.q p_{\nu}.q(2(p_c-q).k-q^2+2p_b.q-m_b^2)+(2(q.k)(p_c.k+p_b.q)+p_c.k(q^2-4p_b.q+m_b^2)\nonumber\\&+&p_b.q(q^2-2p_b.q+m_b^2))(p_b.p_{\nu}) \Big)+(p_{\nu}.k)\Big(4(q^2-2p_b.q+m_b^2)((p_b.q)^2-(q^2+k.q)m_b^2)\nonumber\\&+&(p_c.k)((7q^2+2p_b.q-m_b^2)m_b^2-8(p_b.q)^2)\Big)+4  z^2 m_b^2 \left(p_b .p_l\right) \big(m_b^2 \big(3 \left(  p_b .p_\nu\right) \left(4 q^2-3 \left(  k .p_c\right)\right)+4 \left( k .q\right) \nonumber\\&&\left(3 p_b. p_\nu+ k .p_n-3 \left( q .p_\nu\right)\right)+\left(k .p_\nu\right) \left(3  k. p_c+4 q^2\right)+3 \left(\bar{q}. p_\nu\right) \left(5 \left( k .p_c\right)-4 q^2\right)\big)-2 \left(k .p_b+  q. p_b\right)\nonumber\\&& \left(3 \left( p_b. p_\nu\right) \left(2 q .p_b + k. p_c\right)+2 \left(  q. p_b\right) \left( k .p_\nu-3 \left( q. p_\nu\right)\right)\right)\big)\Big)\Big).
    \label{pi2d}
\end{eqnarray}

In a similar way, we then consider the imaginary part of the third square bracket of Eqn. (\ref{i1}), and combine with Eqn. (\ref{m1l1}). Explicitly, the amplitude in the expanded in powers of $\Pi$ is
\begin{eqnarray}
   \mathcal{O}(\Pi^2): \nonumber\\ \mathcal{J}_1(2;d)&=&\frac{1}{3m_b^3(p_c.k)^2}256\lambda_1 (p_b.p_{\ell})\Big(\Big((-2p_b.q+m_b^2+q^2)(p_b+k-q).p_{\nu}-2(p_b-q).k(q-p_b).p_{\nu}\Big)\nonumber\\&&\Big(((q+k).p_b)^2-m_b^2(2q.k+q^2)\Big)-4z^2 m_b^2 \left(\left( k .p_b+  q. p_b\right)^2-m_b^2 \left(2  k. q+q^2\right)\right) \left(3  p_b+  k -3 q\right). p_\nu\Big). 
    \label{pi2dd}
\end{eqnarray}
Here, $`d'$ refers to the elements of the third square bracket of Eqn. (\ref{i1}). Moreover, the $\mathcal{J}_1(2;d)$ has a multiplicative factor of $\delta(k^2)\theta(k^0)\delta''(((p_b-q-k)^2-m_c^2))\theta((p_b-q-k)^0)$. 

Next, combining all the amplitudes, the total forward matrix element for Fig.2(a) is given by
\begin{eqnarray}
\langle B|Im\lbrace I_1\mathcal{M}_{\mu\nu}^{(1)}\mathcal{L}_{\mu\nu}^{(1)}|B\rbrace\rangle&=&\delta(k^2)\theta(k^0)\Big[(\mathcal{J}_1(0;a)+\mathcal{J}_1(1;a)+\mathcal{J}_1(2;a))\delta((p_b-q-k)^2-m_c^2)\nonumber\\&&+(\mathcal{J}_1(1;c)+\mathcal{J}_1 (2;c))\delta'((p_b-q-k)^2-m_c^2)+(\mathcal{J}_1(2;d))\delta''((p_b-q-k)^2-m_c^2)\Big]\nonumber\\&&\theta((p_b-q-k)^0).
\end{eqnarray}
Integration by parts is then used to simplify such expressions;
\begin{eqnarray}
     \delta'(x)\theta(x)f(x)&=&-\delta(x) \delta(x)f(x)-\delta(x) \theta(x)f'(x),\ \text{and} \\
     \delta''(x)\theta(x)f(x)&=&\delta(x)\delta'(x)f(x)+ 2\delta(x)\delta(x)f'(x)+\delta(x)\theta(x) f''(x).
\end{eqnarray}
These relations will be used to carry out integrals over phase space.
In a similar way, the forward matrix elements for the remaining eight Feynman diagrams, as shown in Fig. (\ref{fdtot}), are calculated. Relevant expressions for $\mathcal{M}_{\mu\nu}^{(m)}$, $\mathcal{L}_{\mu\nu}^{(m)}$, and $I_m$  $m=2,...9$ for all the Feynman diagrams are provided in Appendix \ref{appA}.

\section{Differential rate for $B\to X_u \ell \bar{\nu} \gamma$}
\label{m_diffdecay}
In the limit $m_u\to 0$ the above method can be carried to $B\to X_u$ decay. The matrix element for Fig. 3(a) replacing $c$ quark to $u$ quark is given by 
\begin{eqnarray}
     \mathcal{M}_{\mu \nu}^{(1)}(u)&=&2(-ig^{\alpha \beta}) \bar{b}\gamma ^{\nu }\left(1-\gamma ^5\right)i\left(\slashed{ p_b}+\slashed{\Pi}-\slashed{q}\right)(-ieQ_u)\gamma ^{\alpha }i\left(\slashed{p_b}+\slashed{\Pi}-\slashed{k}-\slashed{q}\right)(-ieQ_u)\gamma ^{\beta }i\big(\slashed{p_b}\nonumber\\ &+&\slashed{\Pi}-\slashed{q}\big)\gamma ^{\mu }\left(1-\gamma ^5\right)b,\nonumber\\
     \mathcal{L}_{\mu \nu}^{(1)}&=&\left(\bar{\ell}\gamma^{\mu }(1-\gamma^5)\nu_{\ell}\right) \left(\bar{\nu}_{\ell} \gamma^{\nu }(1-\gamma^5)\ell\right),
     \label{m1l1u}
 \end{eqnarray}
and 
 \begin{eqnarray}
     I_1(u)&=&\frac{1}{(k^2+i\epsilon)((p_b+\Pi-q)^2+i\epsilon)((p_b+\Pi-q-k)^2+i\epsilon)((p_b+\Pi-q)^2+i\epsilon)}.
     \label{i0u}
 \end{eqnarray}
$I_1(u)$ has a similar expansion as Eqn. (\ref{i1}) replacing $m_c$ with $m_u$. The differential decay rate for $B\to X_u\ell\nu\gamma$ is then calculated following the procedure outlined for $B\to X_c$, but now in the limit $z\to 0$. In the above processes, one difference is the expansion in power of $z$ and another crucial difference is in kinematics\footnote{In Appendix- \ref{appB}, we have retained $m_X$ in corresponding equations which can be put to zero for $u$ quark case.}. 
 
    \section{Results}
\label{result}
As a general rule, the four-body phase space comprises five distinct variables. In the context of inclusive decays, we have an additional variable, namely the invariant mass squared of the final state meson $(p_X^2)$, which we trade for $q'^2\ (=(p_{\ell}+p_{\nu}+k)^2)$. The other independent variables are the lepton energy $y\ (=\frac{2p_B.p_{\ell}}{m_B^2})$, the energy of the hard photon $x\ (=\frac{2p_B.k}{m_B^2})$, the neutrino energy, and three angles. These are defined in the rest frame of the $B$ meson. A detailed description of the kinematics is furnished in Appendix \ref{appB}. All the input parameters required for the evaluation of the differential decay rate are listed in Table- \ref{tab_input}
\begin{center}
  \begin{tabular}{ |p{2cm}|p{4cm}|p{1.5cm}|  }
\hline
Parameter & Numerical value & Refs. \\
\hline
$m_{B(b)}$ & $5.28(4.18)\,GeV$ & \cite{ParticleDataGroup:2020ssz} \\
$m_{c}$ & $1.27 \,GeV$& \cite{ParticleDataGroup:2020ssz} \\
$m_\mu$ & $0.10 \,GeV$ & \cite{ParticleDataGroup:2020ssz} \\
$G_F$ & $1.166\times 10^{-5}\,GeV^{-2}$  & \cite{ParticleDataGroup:2020ssz} \\ 
$\alpha_{em}$ & $1/137$ & \cite{ParticleDataGroup:2020ssz} \\
$\lambda_1$ & $-0.3\,GeV^2$ & \cite{Finauri:2023kte} \\
$\lambda_2$ & $0.117\,GeV^2$ & \cite{Finauri:2023kte} \\
$|V_{ub}|$ & $3.82\times 10^{-3}$ & \cite{ParticleDataGroup:2020ssz} \\
$|V_{cb}|$ & $41\times 10^{-3}$ & \cite{ParticleDataGroup:2020ssz} \\
\hline
\end{tabular} 
\captionof{table}{Numerical inputs used for the decay rate calculation.}
\label{tab_input}
\end{center}
We begin by integrating over all variables except $y$ to determine the spectrum of the charged lepton for different values of $x$. Fig. (\ref{dgvsdytxc}) and (\ref{dgvsdyt}) illustrate the differential decay rate as a function of the lepton energy ($y$) for various photon energy values ($x$) for $B\to X_c\ell\bar{\nu}_\ell\gamma$ and $B\to X_u\ell\bar{\nu}_\ell\gamma$ respectively. We observe that as the photon becomes increasingly soft, the leptonic energy end-point shifts accordingly, which is consistent with the kinematics of the process.
\begin{figure}[h]
    \centering
	    \includegraphics[width=0.5\linewidth]{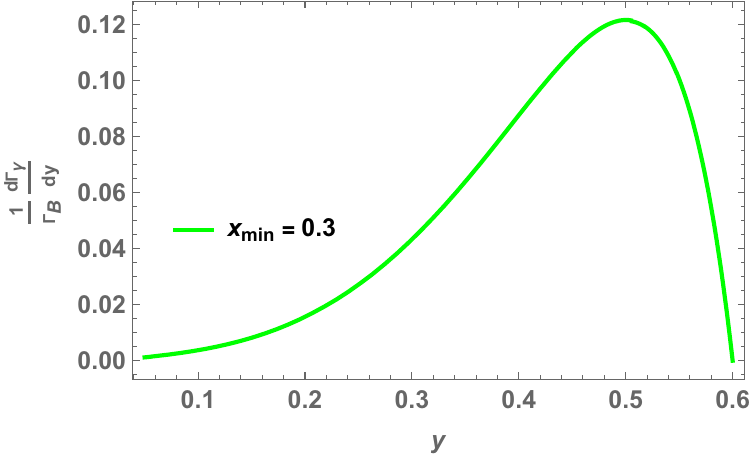}
	   \caption{Differential decay width of $B\to X_c \mu \bar{\nu}_{\mu}\gamma$ for particular normalized photon energy ($x_{min}=0.3$).}
	    \label{dgvsdytxc}
	\end{figure}
\begin{figure}[h]
\begin{subfigure}{0.5\textwidth}
    \centering
	    \includegraphics[width=1\linewidth]{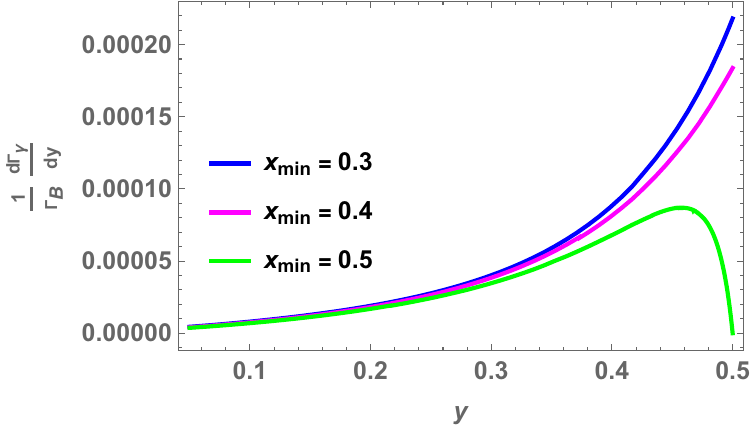}
	   \caption{}
	    \label{dgvsdyt}
\end{subfigure}
\hfill
	   \begin{subfigure}{0.5 \textwidth}
	        \centering
	    \includegraphics[width=1\linewidth]{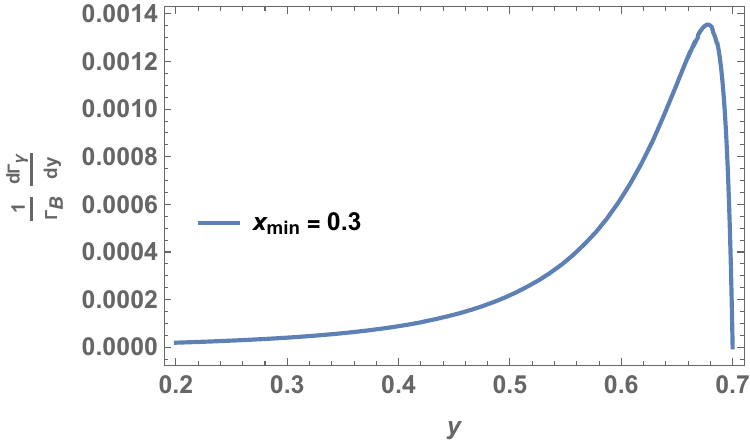}
	   \caption{}
	    \label{dgvsdys}
	   \end{subfigure}
    \caption{(a): Differential decay width of $B\to X_u \mu \bar{\nu}_{\mu}\gamma$ for different values of normalized photon energy ($x$), and (b): Differential decay width of $B\to X_u \mu \bar{\nu}_{\mu}\gamma$ for particular normalized photon energy ($x_{min}=0.3$).}
	\end{figure}
 
To provide a complete picture of the distribution for the differential decay rate for $u-$ quark mode, we present the distribution for $x_{min}=0.3$, which corresponds to $k_{min}\sim 0.8 \ GeV$ in Fig.(\ref{dgvsdys}). The plot shows that the distribution ends at the kinematical boundary, which is, as expected, larger than that for $x_{min}=0.5$, and more towards non-radiative, $B\to X_u \ell \bar{\nu}_{\ell}$, case. Further, it is observed that apart from the difference in CKM elements between the $X_c$ and $X_u$ modes, the decay rate for radiative $B\to X_c$ mode receives a correction of approximately $10\%$ due to mass effects. This finding aligns with the difference of about $12\%$ in the two inclusive rates as per the PDG values\cite{ParticleDataGroup:2020ssz} after correcting for the difference in the CKM factors. 

Further, as an example of possible additional observable, we define the photon (differential) forward-backward asymmetry, $A_{FB}(y)$, with respect to recoiling final state hadron as
 \begin{eqnarray}
     A_{FB}(y)=\frac{\int_{0}^{1}dt \frac{d^2\Gamma_{\gamma}}{dy dt}-\int_{-1}^{0}dt \frac{d^2\Gamma_{\gamma}}{dy dt}}{\int_{-1}^{1}dt \frac{d^2\Gamma_{\gamma}}{dy dt}},
 \end{eqnarray}
 where $t=cos\theta_{X\gamma}$ is the angle between the outgoing photon and recoiling hadron ($X$) in the rest frame of $B$ meson. The forward-backward asymmetry for $B\to X_u \ell\bar{\nu}_\ell\gamma$ is shown in Fig.(\ref{afbvsy}) for $\lambda_1$ and $\lambda_2$ from Table \ref{tab_input}.
 \begin{figure}[h]
	   \centering
	    \includegraphics[width=0.6\linewidth]{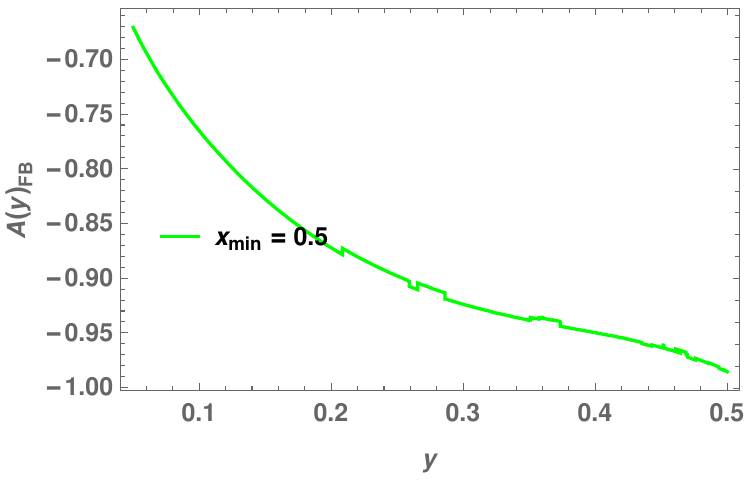}
	   \caption{Forward-backward asymmetry ($A_{FB}$) with lepton energy (y).}
	    \label{afbvsy}
	\end{figure}

Finally, Fig.(\ref{dgvsdx}) depicts the differential decay rate as a function of the normalized photon energy ($x$), which shows that with lowering of the energy of the photon, the decay rate behaves as for the non-radiative mode.

It is worth noting that infrared divergences can be effectively circumvented by assigning sufficient mass to the photon, i.e., the photon is made sufficiently hard. To avoid any contribution from mass singularities, we choose to work with muons in the final state. Furthermore, by choosing a lower cut for the polar angle, we can eliminate any collinear singularities that might arise.
\begin{figure}[h]
	   \centering
	    \includegraphics[width=0.6\linewidth]{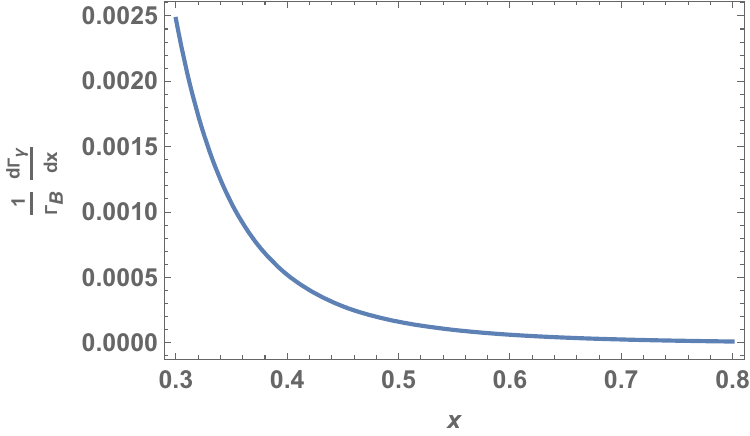}
	   \caption{Differential decay width of $B\to X_u \mu \bar{\nu}_{\mu}\gamma$ with photon energies ($x$).}
	    \label{dgvsdx}
	\end{figure}

 When the photon is hard enough, the total decay width ($\Gamma_\gamma$) for the radiative mode is expected to be suppressed by $\mathcal{O}(\alpha)$ compared to non-radiative one. In order to check if this expectation is met, we evaluate the ratio of the radiative decay width ($\Gamma_\gamma$) to the non-radiative decay width ($\Gamma$) and put $m_q\to 0$. We find that this ratio is approximately $0.01$ for hard photons with energy around $1$GeV ($x_{min}=0.5$), thus confirming the expectation.  
 
 \subsection{Non-perturbative parameters}
 \label{nppara}
After determining the decay width, we now propose a simple yet efficient method for calculating the non-perturbative parameters $\lambda_1$ and $\lambda_2$. We observe that using the ratio of the decay widths, instead of the widths themselves, is more suitable since the latter may contain uncertainties due to the CKM element $V_{ub}$. Moreover, such ratios yield ratios of simple functions of $\lambda_1$ and $\lambda_2$. By knowing the experimentally measured values of $R_1$ and $R_2$, we can simultaneously solve these two linear equations to obtain $\lambda_1$ and $\lambda_2$. The two ratios that we propose to employ are as in Eqn.(\ref{btoxur}). Both numerator and denominator have the form $A+B\lambda_1 +C \lambda_2$. Thus, each of the ratios has the form 
\begin{eqnarray}
     \frac{A+B\lambda_1 +C \lambda_2}{A'+B'\lambda_1 +C' \lambda_2}=R_1 \ \text{(say)}, 
     \label{btoxur1}
 \end{eqnarray}
 and same for $R_2$. These are still two linear equations in $\lambda_1$ and $\lambda_2$.
 To further illustrate the practicality of our proposed method, we perform a sample calculation for non-perturbative parameters $\lambda_1$ and $\lambda_2$. At this point, there is no data available to directly determine $\lambda_1$ and $\lambda_2$. To proceed further in order to demonstrate the case of obtaining $\lambda_1$ and $\lambda_2$ once the suggested ratios are experimentally available, we obtain a value for $R_2$ using the known values of $\lambda_1$ and $\lambda_2$, and for $R_1$ we assume that the decay rate for $B\to X_c\mu\bar{\nu}\gamma$ is $\alpha_{em}$ times the decay rate for $B\to X_c\mu\bar{\nu}$. With these values, we can then numerically calculate the non-perturbative parameters $\lambda_1$ and $\lambda_2$. Our results yield $\lambda_1=-0.29$ $GeV^2$ and $\lambda_2=0.13$ $GeV^2$, which are consistent with previously values reported in the literature \cite{Gremm:1996df,DELPHI:2005mot,Finauri:2023kte}. This motivates the need for a measurement of experimental measurement of the decay width of $B\to X_c\ell\bar{\nu}_{\ell}\gamma$. Such a determination will aid in the precise determination of $\lambda_1$ and $\lambda_2$. As mentioned earlier, we have focused on $\mathcal{O}(\frac{1}{m_b})$ terms in HQE and therefore are sensitive only to $\lambda_1$ and $\lambda_2$. 
 At higher orders, the expressions develop a dependence on more such non-perturbative parameters. Measurements of $B \to X_c\ell \bar{\nu}_{\ell} \gamma$ rate and ratio $R_1$ will infact be very helpful in a simultaneous and easy determination of these parameters when combined with $B\to X_c\ell \bar{\nu}_{\ell}$ data and ratio $R_2$.

\section{Summary}
\label{discussion}
To summarize, we have proposed a method to calculate the non-perturbative parameters $\lambda_1$ and $\lambda_2$ in the context of inclusive $B$ meson decays. We first calculated the decay width for the inclusive modes $B\to X_{c(u)}\ell\bar{\nu}_{\ell}$ and $B\to X_{c(u)}\ell\bar{\nu}_{\ell}\gamma$ directly using the Cutkosky method and the heavy qaurk expansion applied to the amplitude, retaining terms to order $(\frac{\Lambda_{QCD}}{m_b})$ in HQE. Since the radiative mode, $B\to X_{c(u)}\ell\bar{\nu}_{\ell} \gamma$, involves a hard photon, the most general tensorial structure will involve four indices as opposed to two for $B\to X_{c(u)} \ell \bar{\nu}_{\ell}$. In view of this and related complications to pick out the analytic properties in order to evaluate the decay rate, we decided to compute the relevant amplitude directly (or brute force) by using the Cutkosky method. The differential rate contains two non-perturbative parameters $\lambda_1$ and $\lambda_2$.  
Further, the differential and total decay rates were then obtained by integrating the four-body phase space variables. The differential rate and forward-backward asymmetry are plotted with respect to lepton energy for different $x_{min}$ values that define the hardness of the photon. It was found that the decay rate for radiative mode ($B\to X_{c(u)}\ell\bar{\nu}_{\ell}\gamma$) is $\mathcal{O}(\alpha)$ times non-radiative mode ($B\to X_{c(u)}\ell\bar{\nu}_{\ell}$) when the photon is sufficiently hard. 

Next, two ratios, $R_1$ and $R_2$, are formed using the differential rate for radiative (for photon energy $\geq \ \text{few times}\ \Lambda_{QCD}\sim 500 MeV$) and non-radiative modes respectively, in different lepton energy ranges. These ratios are independent of the CKM element and provide two linear equations in $\lambda_1$ and $\lambda_2$, allowing an unambiguous determination of these parameters. 
A simple example is also shown in Section-\ref{nppara} to demonstrate the calculation of these parameters, $\lambda_1$ and $\lambda_2$, which are found to be consistent with existing values. It should be mentioned that, at present, the radiative mode has not been measured. The results of this exercise encourage us to have precise measurements of the radiative mode, more so since, at higher orders in HQE, there are more non-perturbative parameters that enter the amplitude. The radiative inclusive semi-leptonic mode will be very useful in aiding a clean determination of these. 

The decay rate for free quark decay in $B\to X_{c(u)} \ell\bar{\nu}_{\ell}$ $\propto 2y^2(2y-3)$, where $y$ is the normalized lepton energy in the rest frame of $B$ meson. This indicates the difference in partonic and hadronic end points, with former being $\frac{m_b}{2}$ and the latter being at $\frac{m_B}{2}$, resulting in an end point region of order $\bar{\Lambda} =m_B-m_b$. 
Now, more importantly for the $B \to X_u$ mode, in order to correctly include this region, it is required to sum over the infinite terms present in the heavy quark expansion.  
However, this expansion in $\frac{\bar{\Lambda}}{m_b}$ brings higher-order derivatives of delta function at each successive order, resulting in a failure of OPE and QCD perturbation theory in this region. Therefore in order to incorporate the end-point behaviour, the distribution function of the heavy quark, or the shape function, needs to be introduced in the decay rate of $B\to X_u \ell\bar{\nu}_{\ell}$ and $B\to X_s \gamma$ mode \cite{Bigi:1993ex, Neubert:1993mb}. 

Now, in the case of radiative decay ($B\to X_u \ell\bar{\nu}_{\ell} \gamma$) mode, it is noted that the presence of a hard photon in the final state causes the endpoint to shift in comparison to the non-radiative decay. The partonic and hadronic end points are at $\frac{m_b}{2}-x_{min}$ and $\frac{m_B}{2}-x_{min}$ respectively. However, the challenge due to the difference in partonic and hadronic endpoints remains similar to $B\to X_u \ell\bar{\nu}_{\ell}$ mode. Hence, the shape function for this process is required.
Further, one can try to use a simple form of the shape function, $\propto (1-y-x)^a e^{(1+a)(x+y)}$, similar to the one suggested for non-radiative decay mode \cite{Kagan:1998ym,DeFazio:1999ptt}. However, this may be very naive at this point, and hence, it is required to compute the shape function for this process and verify its universality before any definitive conclusions can be drawn.
 While the issue of the form of the shape function remains unsettled at this point and is left for a future study, the main idea that the use of radiative inclusive decay rate can aid in a quick determination of $\lambda_1$ and $\lambda_2$ stays unaffected.    

In conclusion, our proposed method provides a simple yet efficient way to calculate non-perturbative parameters $\lambda_1$ and $\lambda_2$ in inclusive $B$ decays once the decay rate for radiative $B\to X_{c(u)}\ell \bar{\nu}_{\ell} \gamma$ is known experimentally. 

\appendix
\section{Contributions from different diagrams to the inclusive Matrix Element }
\label{appA}
The explicit structure of $\mathcal{M}_{\mu \nu}^{(m)}$, $\mathcal{L}_{\mu \nu}^{(m)}$ and $I_m$ calculated to $\mathcal{O}(z^3)\, (z=m_c/m_b)$ are listed below\footnote{Mathematical expressions for amplitude are very long, so we have only shown the leading term in $z$ here. However, we have included terms up to $\mathcal{O}(z^3)$ in our numerical computations.}. All the relevant expressions for Fig.2(a) are listed in Section-\ref{radiative}. We now present explicit expressions for diagrams Fig.2(b) to Fig.2(i).

\textbf{1. Fig.2(b):}
\begin{eqnarray}
     \mathcal{M}_{\mu \nu}^{(2)}&=& 2(-ig^{\alpha\beta})\bar{b}(-ieQ_b)\gamma_{\alpha}i\left(\slashed{ p_b}+\slashed{\Pi}-\slashed{k}+m_b\right)\gamma^{\nu}(1-\gamma^5)i\left(\slashed{p_b}+\slashed{\Pi}-\slashed{k}-\slashed{q}+m_c\right)\gamma ^{\mu}\left(1-\gamma ^5\right)b,\nonumber\\
     \mathcal{L}_{\mu \nu}^{(2)}&=&\left(\bar{\ell}(-ieQ_{\ell})\gamma_{\beta}i\left(\slashed{ p_l}+\slashed{k}+m_l\right)\gamma^{\mu }(1-\gamma^5)\nu_{\ell}\right) \left(\bar{\nu}_{\ell} \gamma^{\nu }(1-\gamma^5)\ell\right),\\
     \label{m2l2}
       I_2 &=& \frac{1}{ k^2((p_b-q-k)^2-m_c^2)}\Big[ \frac{-1}{(p_b.k)(p_{\ell}.k)}-\frac{(p_b-k).\Pi}{(p_{\ell}.k)(p_b.k)^2}-\frac{\Pi^2}{2(p_{\ell}.k)(p_b.k)^2}-\frac{((p_b-k).\Pi)^2}{2(p_{\ell}.k)(p_b.k)^3}\Big]\nonumber\\&-& \frac{1}{ k^2((p_b-q-k)^2-m_c^2)^2}\Big[ \frac{2p_c.\Pi}{(p_{\ell}.k)(p_b.k)}+\frac{2(p_c.\Pi)(p_b-k).\Pi}{(p_{\ell}.k)(p_b.k)^2}+\frac{\Pi^2}{(p_\ell.k)(p_b.k)}\Big]\nonumber\\&-&\frac{1}{ k^2((p_b-q-k)^2-m_c^2)^3}\Big[\frac{2(p_c.\Pi)^2}{(p_{\ell}.k)(p_b.k)}\Big].
     \label{i23}
 \end{eqnarray}
\begin{eqnarray}
\mathcal{O}(\Pi^0):\nonumber\\
   \mathcal{J}_2(0;a)&=&\frac{-1}{m_b(p_b.k)(p_{\ell}.k)}16\Big(-2(p_b.p_{\nu})\Big(2(p_b.p_{\ell})^2+(p_{\ell}.k)^2+p_{\ell}.k(p_b.q+p_{\ell}.q-3p_b.p_{\ell})- (p_b.k) (p_{\ell}.k)\nonumber\\&-&p_b.p_{\ell}(q.k-2p_b.k+2p_{\ell}.q)\Big)+m_b^2\Big(p_{\ell}.k((p_b+q).p_{\nu})+p_{\ell}.p_{\nu}(p_b.k-q.k-2p_{\ell}.k)\nonumber\\&-&((q-p_b).p_{\ell})(p_{\nu}.k+2p_{\ell}.p_{\nu}) \Big)-2((p_b.p_{\ell})(p_{\nu}.k)-(p_b.k)(p_{\ell}.p_{\nu}))((q+k-p_b).p_{\ell}) \Big),\\
    \label{2pi0}
     \mathcal{O}(\Pi):\nonumber\\ \mathcal{J}_2(1;a)&=&\frac{-1}{3m_b^3(p_{b}.k)^2(p_{\ell}.k)}32(\lambda_1+3\lambda_2)\Big( 2m_b^2\Big(p_{\nu}.k\big(p_b.p_{\ell}(p_{\ell}.q-2q.k-3p_{\ell}.k)-(p_b.p_{\ell})^2\nonumber\\&+&2p_{\ell}.k(p_b.q+2p_{\ell}.k+2p_{\ell}.q)\big)+p_b.p_{\nu}\big(6(p_b.p_{\ell})^2-p_b.p_{\ell}(7p_{\ell}.k+3q.k+6p_{\ell}.q)\nonumber\\&+&p_{\ell}.k(3p_b.q+p_{\ell}.(q+k))\big)\Big)+(p_b.k)^2\big(10(p_{\ell}.p_{\nu}(q+k-p_b).p_{\ell})+p_b.p_{\nu}(p_{\ell}.q-2p_b.p_{\ell})\nonumber\\&+& m_b^2p_{\ell}.p_{\nu}\big)+p_b.k\Big(-10p_b.p_{\nu}\big(2(p_b.p_{\ell})^2-p_b.p_{\ell}(3p_{\ell}.k+q.k+2p_{\ell}.q)+p_{\ell}.k(p_b.q+(q+k).p_{\ell})\big)\nonumber\\&+&m_b^2\big(p_{\nu}.k(9p_b-5q).p_{\ell}+p_{\ell}.k(9p_b-12p_{\ell}+q)-p_{\ell}.p_{\nu}(12(q-p_b).p_{\ell}+q.k)-6p_b.p_{\nu}(q-2p_b).p_{\ell}\big)\nonumber\\&-& 10(p_b.p_{\ell})(p_{\nu}.k)((q+k-p_b).p_{\ell})-3m_b^4(p_{\ell}.p_{\nu})\Big)+m_b^4\big(3(p_b-q).p_{\ell} ((k+2p_{\ell}).p_{\nu})+3(q.k)(p_{\ell}.p_{\nu})\nonumber\\&-&3(p_{\ell}.k) ( p_{\nu}.(p_b-2p_{\ell}+q))+4p_{\nu}.k\big)\Big),
    \label{2pi1}
\end{eqnarray}
\begin{eqnarray}
    \mathcal{J}_2 (1;c)&=&\frac{1}{3m_b^3(p_b.k)(p_{\ell}.k)}128(\lambda_1+3\lambda_2)\Big(m_b^4\big(p_{\nu}.k((q-3p_b).p_{\ell})-p_{\ell}.k((4k+5q+3p_b-6p_{\ell}).p_{\nu})\nonumber\\&+&p_{\ell}.p_{\nu}(5k.q+6p_{\ell}.q-6p_b.p_{\ell})\big)+\big(2((4k+2q+p_b).p_{\nu})(p_{\ell}.k)^2+\big(2p_{\nu}.k(2q.p_b+6q.p_{\ell}-3p_b.p_{\ell})\nonumber\\&+&4p_{\nu}.q((q-2p_b).p_{\ell})+2q^2(p_b.p_{\nu})-2((q+7p_b).p_{\ell})p_b.p_{\nu}+p_b.q((5q+11p_b-6p_{\ell}).p_{\nu})\nonumber\\&-&4(q.k)(p_{\ell}.p_{\nu}) \big)p_{\ell}.k+4(p_{\nu}.q)(p_b.p_{\ell})((p_b-q).p_{\ell})+p_{\nu}.k\big(4(q.p_{\ell})^2-2(p_b.p_{\ell})^2-p_b.q((q-5p_b).p_{\ell})\nonumber\\&-&2p_b.p_{\ell}(2k.q+q^2+p_{\ell}.q)\big)-2p_b.p_{\nu}\big(2q.p_{\ell}((k+p_{\ell}).q)+p_b.p_{\ell}((3k+4p_{\ell}).q)-6(p_b.p_{\ell})^2\big)\nonumber\\&+&p_{\ell}.p_{\nu}\big(6p_b.q((p_b-q).p_{\ell})+k.q(4p_b.p_{\ell}-4q.p_{\ell}-5q.p_b)\big)\big)m_b^2+10q.p_b\big(-(p_{\nu}.k)(p_b.p_{\ell})\nonumber\\&&((k+q-p_b).p_{\ell})-p_b.p_{\nu}\big(2(p_b.p_{\ell})^2-p_b.p_{\ell}(k.q+3k.p_{\ell}+2q.p_{\ell})+p_{\ell}.k(k.p_{\ell}+q.p_b+q.p_{\ell})\big)\big)\nonumber\\&+&(p_b.k)^2\big(m_b^2 p_{\ell}.p_{\nu}+10((q-2p_b).p_{\ell})(p_b.p_{\nu})+((k+q-p_b).p_{\ell})(p_{\ell}.p_{\nu})\big)+p_b.k\big(-3(p_{\ell}.p_{\nu})m_b^4\nonumber\\&+&\big(p_{\nu}.k((9p_b-q).p_{\ell})+p_{\ell}.k((q+p_b).p_{\nu})+p_{\ell}.p_{\nu}(2q^2+q.p_b-12k.p_{\ell}-k.q)-2((2q+p_b+6p_{\ell}).p_{\nu})\nonumber\\&&(q.p_{\ell})+4p_b.p_{\ell}((q+3p_b+3p_{\ell}).p_{\nu})\big)m_b^2+10\big(-(p_{\nu}.k)(p_b.p_{\ell})((k+q-p_b).p_{\ell})-p_b.p_{\nu}\big((p_{\ell}.k)^2\nonumber\\&+&p_{\ell}.k(q.p_b+q.p_{\ell}-3p_b.p_{\ell})+2(p_b.p_{\ell})^2-(q.p_b)(q.p_{\ell})-((k-2p_b+2p_{\ell}).q)\big)+((k+q-p_b).p_{\ell})\nonumber\\&&(p_b.q)(p_{\ell}.p_{\nu})\big)\big)\Big).\\
     \mathcal{O}(\Pi^2):\nonumber\\ \mathcal{J}_2 (2;a)&=&\frac{1}{3m_b^3(p_b.k)^2(p_{\ell}.k)}16\lambda_1 \Big(\big(3(p_{\ell}.k)((q+p_b).p_{\nu})-3p_{\ell}.p_{\nu}((3q+10p_{\ell}).k)-3p_{\ell}.(q-p_b)\nonumber\\&&((k+2p_{\ell}).p_{\nu})\big)m_b^4-2\big(p_{\nu}.k((7k+5q-5p_b).p_{\ell})(p_b.p_{\ell})+p_{b}.p_{\nu}\big(6(p_b.p{\ell})^2-3p_b.p_{\ell}\nonumber\\&&(k.q+5k.p_{\ell}+2q.p_{\ell})+p_{\ell}.k(7p_{\ell}.k+3p_b.q+5p_{\ell}.q)\big)\big)m_b^2+(p_b.k)^2\big(2p_{\ell}.q((p_b-p_{\ell}).p_{\nu})\nonumber\\&-&2p_b.p_{\ell}((2p_b+p_{\ell}).p_{\nu})-p_{\ell}.p_{\nu}(3m_b^2+2p_{\ell}.k)\big)+p_b.k\big(3(p_{\ell}.p_{\nu})m_b^4-4(p_b.p_{\nu})m_{\ell}^2m_b^2\nonumber\\&+&6((q-2p_b).p_{\ell})(p_b.p_{\nu})m_b^2-(k.q-8q.p_{\ell}+8p_b.p_{\ell})(p_{\ell}.p_{\nu})m_b^2+p_b.p_{\nu}\big(2(p_{\ell}.k)^2-4(p_b.p_{\ell})^2\nonumber\\&+&2(q.k)(p_b.p_{\ell})+4(p_{\ell}.q)(p_b.p_{\ell})\big)+p_{\nu}.k\big(2(pb.p_{\ell})((q+p_b).p_{\ell})-((q-5p_b).p_{\ell})m_b^2\big)\nonumber\\&+&p_{\ell}.k\big(((q+5p_b+12p_{\ell}).p_{\nu})m_b^2+2(p_{\nu}.k)(p_b.p_{\ell})+2(p_{\ell}.q-p_b.q+3p_b.p_{\ell})(p_b.p_{\nu})\big)\big)\Big),
    \label{2pi2}
\end{eqnarray}
\begin{eqnarray}
     \mathcal{J}_2(2;c)&=&\frac{-1}{3m_b^3(p_b.k)^2(p_{\ell}.k)}128\lambda_1\big(m_b^4\big(-k.(2q+3p_b)\big(k.p_{\nu}((q-p_b).p_{\ell})-k.p_{\ell}((q+p_b).p_{\nu})\big)-p_{\ell}.p_{\nu}\big(2(k.q)^2\nonumber\\&+&q.k(5p_b.k+4(k+q-p_b).p_{\ell})+p_b.k((10k+10q-6p_b).p_{\ell}-3p_b.k)\big)\big)+2\big((p_{\ell}.p_{\nu})(p_b.k)^3\nonumber\\&+&(p_b.k)^2\big((p_b.p{\nu})(p_{\ell}.(k+5q))+p_{\ell}.p_{\nu}(7k.p_{\ell}-q^2+q.p_b+7q.p_{\ell})+p_b.p_{\ell}((k+2q-6p_b-3p_{\ell}).p_{\nu})\big)\nonumber\\&+&\big(-7(p_b.p_{\nu})(p_{\ell}.k)^2+\big((p_b.p_{\nu})(q^2-4q.p_b-9q.p_{\ell})+p_b.p_{\ell}((15p_b-2q-7k).p_{\nu})+2(p_{\ell}.p_{\nu})\nonumber\\&&((k+p_b).q)\big)p_{\ell}.k+2(q.p_{\nu})(p_b.p_{\ell})(p_b.p_{\ell}-k.q-q.p_{\ell})+(p_{\nu}.k)(p_b.p_{\ell})(q^2-q.p_b-7q.p_{\ell}+5p_b.p_{\ell})\nonumber\\&-&2(q.p_{\ell})^2(p_b.p_{\nu})-(p_b.p_{\ell})(p_b.p_{\nu})(6p_b.p_{\ell}-3k.q-12q.p_{\ell})+2(p_{\ell}.p_{\nu})((q.p_b)(q.p_{\ell})+k.q\big(q.p_b+q.p_{\ell}\nonumber\\&-&p_b.p_{\ell}\big))\big)p_b.k+2k.q\big(-(p_{\nu}.k)((k+q-p_b).p_{\ell})(p_b.p_{\ell})-p_b.p_{\nu}\big(2(p_b.p_{\ell})^2-p_b.p_{\ell}(k.q+3p_{\ell}.k+2q.p_{\ell})\nonumber\\&+&p_{\ell}.k(k.p_{\ell}+q.(p_b+p_{\ell}))\big)\big)\big)m_b^2-4k.p_b((k+q).p_b)\big(-p_b.p_{\nu}(p_{\ell}.k)^2-p_{\ell}.k\big((p_{\nu}.k)(p_b.p_{\ell})+(q.p_{\ell})\nonumber\\&&(p_b.p_{\nu})-(p_b.k)(p_{\ell}.p_{\nu})\big)+p_b.p_{\ell}\big(2p_b.p_{\nu}((k+p_{\ell}).p_b)-(k.p_{\nu})(q.p_{\ell})\big)+(k.p_b)(q.p_{\ell})(p_{\ell}.p_{\nu})\big)\Big),\\
    \label{2pi2d}
     \mathcal{J}_2(2;d)&=&\frac{-1}{3m_b^3(p_b.k)(p_{\ell}.k)}256\lambda_1 \big(((q+k).p_b)^2-m_b^2(2k.q+q^2)\big)\Big(-2p_b.p_{\nu}\big(2(p_b.p_{\ell})^2+(p_{\ell}.k)^2+p_{\ell}.k\nonumber\\&&(q.(p_{\ell}+p_b)-3p_b.p_{\ell})-(p_b.k)(q.p_{\ell})-p_b.p_{\ell}(k.q+2q.p_{\ell}-2k.p_b)\big)+m_b^2\big(k.p_{\ell}((p_b+q).p_{\nu})\nonumber\\&+&(p_{\ell}.p_{\nu})((p_b-2p_{\ell}-q).k)-((q-p_b).p_{\ell})((k+2p_{\ell}).p_{\nu})\big)-2(p_b.p_{\ell})(k.p_{\nu})(-p_b+k+q).p_{\ell}\nonumber\\&+&2(k.p_b)(p_{\ell}.p_{\nu})((k+q-p_b).p_{\ell})\Big). 
    \label{2pi2dd}
\end{eqnarray}
\textbf{2. Fig.2(c):} 
 \begin{eqnarray}
     \mathcal{M}_{\mu \nu}^{(3)}&=&2(-ig^{\alpha\beta})\bar{b}\gamma^{\nu}(1-\gamma^5)i\left(\slashed{ p_b}+\slashed{\Pi}-\slashed{q}-\slashed{k}+m_c\right)\gamma ^{\mu}\left(1-\gamma ^5\right)i\left(\slashed{p_b}+\slashed{\Pi}-\slashed{k}+m_b\right)(-ieQ_b)\gamma_{\alpha}b,\nonumber\\
     \mathcal{L}_{\mu \nu}^{(3)}&=&\left(\bar{\ell}\gamma^{\mu }(1-\gamma^5)\nu_{\ell}\right) \left(\bar{\nu}_{\ell} \gamma^{\nu }(1-\gamma^5)(-ieQ_l)\gamma_{\alpha}i\left(\slashed{ p_l}+\slashed{k}+m_l\right)\ell\right),\\
     \label{m3l3}
     I_3 &=& I_2.\\
      \mathcal{O}(\Pi^0):\nonumber\\ \mathcal{J}_3(0;a)&=&\frac{1}{m_b(p_b.k)(p_{\ell}.k)}16\Big(-(p_b.p_{\nu})(p_{\ell}.k)^2+p_{\ell}.k\big(m_b^2((p_b-2p_{\ell}+q).p_{\nu})-2p_b.p_{\nu}(-2p_b.p_{\ell}+q.(p_b+p_{\ell}))\big)\nonumber\\&-&m_b^2\big((q.p_{\ell}-p_b.p_{\ell})(k.p_{\nu}+2p_{\ell}.p_{\nu})+(k.q)(p_{\ell}.p_{\nu})\big)+k.p_b\big(p_{\ell}.p_{\nu}(2q.p_{\ell}+m_b^2)+2p_b.p_{\nu}((q-2p_b).p_{\ell})\nonumber\\&+&m_{\ell}^2k.p_{\nu}+k.p_{\ell}((p_{\ell}-2k).p_{\nu})\big)+2p_b.p_{\ell}\big(p_b.p_{\nu}(k.q+2q.p_{\ell}-2p_b.p_{\ell})-(k.p_{\nu})(q.p_{\ell})\big)\Big).
    \label{3pi0}
 \end{eqnarray}
\begin{eqnarray}
    \mathcal{O}(\Pi):\nonumber\\ \mathcal{J}_3(1;a)&=&\frac{-1}{3m_b^3(p_{b}.k)^2(p_{\ell}.k)}32(\lambda_1+3\lambda_2)\Big((p_b.k)^2\big(5p_{\ell}.k(2k.p_{\nu}-p_{\ell}.p_{\nu})-5m_{\ell}^2(k.p_{\nu})-10p_b.p_{\nu}((q-2p_b).p_{\ell})\nonumber\\&-&P_{\ell}.p_{\nu}(10q.p_{\ell}+m_b^2)\big)+p_b.k\big(5(p_b.p_{\nu})(p_{\ell}.k)^2+m_b^2\big(p_{\ell}.p_{\nu}(k.q+12q.p_{\ell}-6p_b.p_{\ell})+6p_b.p_{\nu}\big(q.p_{\ell}\nonumber\\&-&2p_b.p_{\ell}\big)\big)+k.p_{\ell}\big(m_b^2((q-9p_{\ell}+6k-9p_b).p_{\nu})+15(p_b.p_{\ell})(k.p_{\nu})+10p_b.p_{\nu}(q.(p_b+p_{\ell})-2p_b.p_{\ell})\big)\nonumber\\&+&k.p_{\nu}\big(5q.p_{\ell}(2p_b.p_{\ell}+m_b^2)+3m_b^2(m_{\ell}^2-3p_b.p_{\ell})\big)-10(p_b.p_{\ell})(p_b.p_{\nu})(-2p_b.p_{\ell}+k.q+2q.p_{\ell})+3m_b^4\nonumber\\&&(p_{\ell}.p_{\nu})\big)+m_b^2\big(-p_{\nu}.k\big(4(p_b.p_{\ell})^2+p_b.p_{\ell}(5k.p_{\ell}-4k.q+2q.p_{\ell})+4k.p_{\ell}(q.p_b+2k.p_{\ell}+2q.p_{\ell})\big)\nonumber\\&+&p_b.p_{\nu}\big((k.p_{\ell})^2-2k.p_{\ell}(3q.p_b+q.p_{\ell}-4p_b.p_{\ell})+6p_b.p_{\ell}(k.q+2q.p_{\ell}-2p_b.p_{\ell})\big)+m_b^2\big(k.p_{\ell}\nonumber\\&&((3p_b-2p_{\ell}+q).p_{\nu}+4k.p_{\nu})-3(q.p_{\ell}-p_b.p_{\ell})(k.p_{\nu}+2p_{\ell}.p_{\nu})-3(k.q)(p_{\ell}.p_{\nu})\big)\big)\Big),
    \label{3pi1}
\end{eqnarray}
\begin{eqnarray}
    \mathcal{J}_3(1;c)&=&\frac{1}{3m_b^3(p_b.k)(p_{\ell}.k)}128(\lambda_1+3\lambda_2)\Big(5(p_b.k)^2\big(k.p_{\ell}((2k+4q-4p_b-p_{\ell}).p_{\nu})-m_{\ell}^2((k+2q-2p_b).p_{\nu})\nonumber\\&+&2p_b.p_{\ell}((p_{\ell}-2(k+q-p_b)).p_{\nu})\big)+\big(5(p_b.p_{\nu})(p_{\ell}.k)^2+\big(3m_b^2((p_{\ell}+4p_b-4q).p_{\nu})+10p_b.p_{\ell}\nonumber\\&&((2q-3p_b).p_{\nu})+5q.p_b((4q-4p_b-p_{\ell}).p_{\nu})+k.p_{\nu}(-6m_b^2+10q.p_b+15p_b.p_{\ell})\big)p_{\ell}.k+\big(2(q-p_b).p_{\nu}\nonumber\\&&(6m_b^2-5q.p_b)+p_{\nu}.k(9m_b^2-5q.p_b)\big)m_{\ell}^2+2p_b.p_{\ell}\big(3m_b^2((2q-2p_b-p_{\ell}).p_{\nu})-10(q.p_b+p_b.p_{\ell})\nonumber\\&&(q.p_{\nu}-p_b.p_{\nu})+5(q.p_b)(p_{\ell}.p_{\nu})+k.p_{\nu}\big(6m_b^2-10q.p_b-15p_b.p_{\ell}\big)\big)\big)p_b.k+2m_b^2\big(k.q((k+2q-2p_b).p_{\nu})\nonumber\\&+&3((k+q-p_b).p_{\nu})(q.p_b-m_b^2)\big)m_{\ell}^2+2p_b.p_{\ell}\big(\big((2q.p_{\ell}+3p_b.p_{\ell})(3k.p_{\nu}+2q.p_{\nu}-2p_b.p_{\nu})+k.q\nonumber\\&&((4(k+q-p_b)-2p_{\ell}).p_{\nu})\big)m_b^2-5(q.p_b)(p_b.p_{\ell})(3k.p_{\nu}+2q.p_{\nu}-2p_b.p_{\nu})\big)+p_{\ell}.k\big(\big(k.p_{\nu}(7p_b.p_{\ell}-6q.p_{\ell})\nonumber\\&-&8(q.p_{\ell})(q.p_{\nu})+2(4q.p_{\ell}+5p_b.p_{\ell})p_b.p_{\nu}+2k.q\big(4p_b.p_{\nu}+p_{\ell}.p_{\nu}-4q.p_{\ell}-2k.p_{\nu}\big)\big)m_b^2+5(q.p_b)(p_b.p_{\ell})\nonumber\\&&((3k+4q-6p_b).p_{\nu})\big)+(p_{\ell}.k)^2\big(5p_b.p_{\nu}(m_b^2+q.p_b)-2(4k.p_{\nu}+5q.p_{\nu})m_b^2\big)\Big).
    \label{3pi1d}
\end{eqnarray}
\begin{eqnarray}
      \mathcal{O}(\Pi^2):\nonumber\\ \mathcal{J}_3(2;a)&=&\frac{1}{3m_b^3(p_b.k)^2(p_{\ell}.k)}16\lambda_1 \Big((p_b.k)^2\big(-m_{\ell}^2(p_{\nu}.k)+2((q-2p_b).p_{\ell})(p_b.p_{\nu})+k.p_{\ell}((2k-p_{\ell}).p_{\nu})\nonumber\\&-&p_{\ell}.p_{\nu}(9m_b^2+2q.p_{\ell}+4p_b.p_{\ell})\big)+\big(3(p_{\ell}.p_{\nu})m_b^4+(p_b.p_{\nu})m_b^2(2m_{\ell}^2+6q.p_{\ell}-12p_b.p_{\ell})-\big(k.q-8q.p_{\ell}\nonumber\\&+&2p_b.p_{\ell}\big)(p_{\ell}.p_{\nu})m_b^2+((k.p_{\ell})^2-4(p_b.p_{\ell})^2+2(k.q)(p_b.p_{\ell})+4(q.p_{\ell})(p_b.p_{\ell}))(p_b.p_{\nu})+p_{\ell}.k\nonumber\\&&\big(((q+2p_b+9p_{\ell}).p_{\nu})m_b^2+2(q.p_{\ell}+4p_b.p_{\ell}-q.p_b)p_b.p_{\nu}+3(p_{\nu}.k)(p_b.p_{\ell}-2m_b^2)\big)+p_{\nu}.k\big(3m_b^2m_{\ell}^2\nonumber\\&+&q.p_{\ell}(2p_b.p_{\ell}-m_b^2)+4p_b.p_{\ell}(2m_b^2+p_b.p_{\ell})\big)\big)p_b.k+m_b^2\big(-11(p_b.p_{\nu})(p_{\ell}.k)^2+\big(m_b^2((3q+3p_b-10p_{\ell}).p_{\nu})\nonumber\\&-&17(k.p_{\nu})(p_b.p_{\ell})-2p_b.p_{\nu}(3q.p_b+5q.p_{\ell}-12p_b.p_{\ell})\big)p_{\ell}.k+3(k.q+2q.p_{\ell}-2p_b.p_{\ell})\big(2(p_b.p_{\ell})(p_b.p_{\nu})\nonumber\\&-&(p_{\ell}.p_{\nu})m_b^2\big)+k.p_{\nu}\big(p_b.p_{\ell}(3m_b^2+4p_b.p_{\ell})-q.p_{\ell}(3m_b^2+10p_b.p_{\ell})\big)\big)\Big),\\
    \label{3pi2}
    \mathcal{J}_3(2;c)&=&\frac{-1}{3m_b^3(p_b.k)^2(p_{\ell}.k)}128\Big(2 \big(\left(2 q.p_{\nu}+k.p_{\nu}\right) m_l^2+k.p_{\ell} \left(-2 k.p_{\nu}+  p_{\ell}.p_{\nu}-4 q.p_{\nu}\right)+p_b.p_{\ell} \nonumber\\&&\left(4 p_b.p_{\nu}-2 p_{\ell}.p_{\nu}\right)\big) ( k.p_b )^3+\big(-2 \left(p_b .  p_{\nu}\right) (k.p_{\ell})^2+k.p_{\ell} \big(\left(28q.p_{\nu}-20 \left(p_b .  p_{\nu}\right)-3 \left(p_{\ell} .  p_{\nu}\right)\right) m_b^2\nonumber\\&+&\left(22 m_b^2-4 q.p_b-6 p_b.p_{\ell}\right) k.p_{\nu}+4 p_b.p_{\ell} \left(p_b .  p_{\nu}-2q.p_{\nu}\right)+2 q.p_b \left(p_{\ell} .  p_{\nu}-4q.p_{\nu}\right)\big)+m_l^2 \big(\big(-11 k.p_{\nu}\nonumber\\&+&6 p_b .  p_{\nu}-14 q.p_{\nu}\big) m_b^2+2 q.p_b \left(2 q .  p_{\nu}+  k. p_{\nu}\right)\big)+4 p_b.p_{\ell} \left(\left(p_b .  p_{\ell}\right) \left(2 p_b .  p_{\nu}+  k.p_{\nu}\right)+q.p_b \left(2 p_b.p_{\nu}-p_{\ell} .  p_{\nu}\right)\right)\nonumber\\&+&2 m_b^2 \left(2 \left(2 q.p_{\ell}-5 p_b.p_{\ell}\right) k.p_{\nu}+4 q.p_{\ell} \left(q .  p_{\nu}-p_b .  p_{\nu}\right)+p_b.p_{\ell} \left(-10 q .  p_{\nu}+6 p_b .  p_{\nu}+3 p_{\ell} .  p_{\nu}\right)\right)\big)\nonumber\\&& (k.p_b)^2+2 \left(k .  q\right) m_b^2 \big(p_b.p_{\nu} (k.p_{\ell})^2+2 m_b^2 m_l^2 \left(-p_b .  p_{\nu}+.  k p_{\nu}+.  q p_{\nu}\right)-2 (p_b.p_{\ell})^2 \big(3 k .  p_{\nu}+2 q .  p_{\nu}\nonumber\\&-&2p_b.p_{\nu}\big)+\left(k .  p_{\ell}\right) \left(p_b .  p_{\ell}\right) \left(3 k .  p_{\nu}+4 q .  p_{\nu}-6 p_b.p_{\nu}\right)\big)+k.p_b \big(\left(3 m_b^2-2 q.p_b\right) \left(p_b.p_{\nu}\right) (k.p_{\ell})^2\nonumber\\&-&2 m_b^2 \left(\left(3 p_b.p_{\nu}-5 \left(  k.p_{\nu}+ q. p_{\nu}\right)\right) m_b^2+2 q.p_b \left(k.p_{\nu}+ q.p_{\nu}\right)+k.q \left(2 q .  p_{\nu}+  k .p_{\nu}-2 p_b .  p_{\nu}\right)\right) m_l^2\nonumber\\&+&2 p_b.p_{\ell} \big(\big(2 p_b.p_{\nu} \left(3 p_b.p_{\ell}-2 q.p_{\ell}\right)+\left(4 q.p_{\ell}-13 p_b.p_{\ell}\right) k.p_{\nu}+2 \left(2 q.p_{\ell}-5 p_b.p_{\ell}\right) q.p_{\nu}+2 \left(k .  q\right) \nonumber\\&&\left(p_{\ell}.p_{\nu}-2 \left(-p_b .  p_{\nu}+.  k p_{\nu}+.  q p_{\nu}\right)\right)\big) m_b^2+2 \left(q .  p_b\right) \left(p_b .  p_{\ell}\right) \left(2 p_b .  p_{\nu}+ k.p_{\nu}\right)\big)+\left(k .  p_{\ell}\right)\nonumber\\&& \big(\left(p_b .  p_{\ell} \left(25 k .  p_{\nu}+28 q .  p_{\nu}-26 p_b .  p_{\nu}\right)+\left(k .  q\right) \left(4 k .  p_{\nu}+8 q .  p_{\nu}-8 \left(p_b .  p_{\nu}\right)-2 \left(p_{\ell} .  p_{\nu}\right)\right)\right) m_b^2\nonumber\\&-&2 \left(q .  p_b\right) \left(p_b .  p_{\ell}\right) \left(3 k .  p_{\nu}+4 q .  p_{\nu}-2 p_b .  p_{\nu}\right)\big)\big)\Big),\\
  \label{3pi2d}
   \mathcal{J}_3(2;d)&=&\frac{1}{3m_b^3(p_b.k)(p_{\ell}.k)}256\lambda_1 \big(\left(k.p_b+ q.p_b\right)^2-m_b^2 \left(2k.q+q^2\right)\big) \Big(-2 \left(p_b. p_{\ell}\right)^2 \left(-2  p_b.p_{\nu}+3k.p_{\nu}+2q.p_{\nu}\right)\nonumber\\&+&\left(  p_b.p_{\nu}\right) \left(k.p_{\ell}\right)^2+2 m_b^2 m_l^2 \left(-p_b.p_{\nu}+k.p_{\nu}+q.p_{\nu}\right)+k.p_b \big(m_l^2 \left(2 p_b.p_{\nu}- k.p_{\nu}-2q.p_{\nu}\right)+\left(k.p_{\ell}\right)\nonumber\\&& \left(-4p_b.p_{\nu}+2k.p_{\nu}-p_{\ell}.p_{\nu}+4q.p_{\nu}\right)+2 \left(p_b.p_{\ell}\right) \left(p_{\ell}.p_{\nu}-2 \left(-p_b.p_{\nu}+ k.p_{\nu}+ q. p_{\nu}\right)\right)\big)+\left(p_b.p_{\ell}\right) \left(  k.p_{\ell}\right)\nonumber\\&& \left(-6  p_b.p_{\nu}+3 k. p_{\nu}+4q.p_{\nu}\right)\Big).
    \label{3pi2dd}
\end{eqnarray}
\textbf{3. Fig.2(d):}
\begin{eqnarray}
     \mathcal{M}_{\mu \nu}^{(4)}&=& \bar{b}\gamma^{\nu}(1-\gamma^5)i\left(\slashed{ p_b}+\slashed{\Pi}-\slashed{q}-\slashed{k}+m_b\right)\gamma ^{\mu}\left(1-\gamma ^5\right) b,\nonumber\\
     \mathcal{L}_{\mu \nu}^{(4)}&=& (-ig^{\alpha\beta})\left(\bar{\ell}(-ieQ_l)\gamma_{\alpha}i\left(\slashed{ p_l}+\slashed{k}+m_l\right)\gamma^{\mu }(1-\gamma^5)\nu_{\ell}\right) \Big(\bar{\nu}_{\ell} \gamma^{\nu }(1-\gamma^5)i\left(\slashed{ p_l}+\slashed{k}+m_l\right)\nonumber\\&&(-ieQ_l)\gamma_{\beta}\ell\Big),\\
     \label{m4l4}
      I_4&=& \frac{1}{ k^2((p_b-q-k)^2-m_c^2)}\Big[ \frac{1}{(p_{\ell}.k)^2}\Big]- \frac{1}{ k^2((p_b-q-k)^2-m_c^2)^2}\Big[ \frac{2p_c.\Pi}{(p_{\ell}.k)^2}+\frac{\Pi^2}{(p_{\ell}.k)^2}\Big]\nonumber\\&+&\frac{1}{ k^2((p_b-q-k)^2-m_c^2)^3}\Big[\frac{2(p_c.\Pi)^2}{(p_{\ell}.k)^2}\Big].
     \label{i4}
 \end{eqnarray}
\begin{eqnarray}
\mathcal{O}(\Pi^0):\nonumber\\
    \mathcal{J}_4(0;a)&=&\frac{1}{m_b(p_{\ell}.k)^2}64p_b.p_{\nu}\Big(k.p_{\ell}(k.q-k.p_b)-m_{\ell}^2(k.p_{\ell}+k.q+q.p_{\ell}-k.p_b-p_b.p_{\ell})\Big).
    \label{4pi0}
\end{eqnarray}
\begin{eqnarray}
    \mathcal{O}(\Pi):\nonumber\\ \mathcal{J}_4(1;a)&=&\frac{-1}{3m_b^3(p_{\ell}.k)^2}64(\lambda_1+3\lambda_2)\Big(m_{\ell}^2(2m_b^2(k.p_{\nu}+p_{\ell}.p_{\nu})-5p_b.p_{\nu}(k.p_b+p_b.p_{\ell}))+k.p_{\ell}\big(5(k.p_b)\nonumber\\&&(p_b.p_{\nu})-2m_b^2k.p_{\nu}\big)\Big),\\
    \label{4pi1}
     \mathcal{J}_4(1;c)&=&\frac{1}{3m_b^3(p_{\ell}.k)^2}512(\lambda_1+3\lambda_2)\big(m_{\ell}^2(k.p_{\ell}+k.q+q.p_{\ell}-k.p_b-p_b.p_{\ell})+k.p_{\ell}(k.p_b-k.q)\big)\nonumber\\&&\big(2m_b^2(k.p_{\nu}+q.p_{\nu})+p_b.p_{\nu}(3m_b^2-5(k.p_b+q.p_b))\big).
    \label{4pi1d}
\end{eqnarray}
\begin{eqnarray}
   \mathcal{O}(\Pi^2):\nonumber\\ \mathcal{J}_4(2;a)&=&0,\\
    \label{4pi2}
    \mathcal{J}_4(2;c)&=&\frac{-1}{3m_b^3(p_{\ell}.k)^2}256\lambda_1\Big(4p_b.p_{\nu} \left(m_{\ell}^2-  k.p_{\ell}\right) \left( k.p_b\right)^2+2 k.p_b \big(m_{\ell}^2 \big(m_b^2 \left(k. p_{\nu}+q.p_{\nu}\right)+p_b.p_{\nu} \big(2 \left(p_b.p_{\ell}+  q.p_b\right)\nonumber\\&+&3 m_b^2\big)\big)-\left(p_b.p_{\nu}\right) \left(k.p_{\ell}\right) \left(2q.p_b+3 m_b^2\right)-m_b^2 m_{\ell}^2 \left(k.p_{\nu}+q.p_{\nu}\right)\big)+m_{\ell}^2 \big(m_b^2 \big( p_b.p_{\ell} \left(6 p_b.p_{\nu}+ k.p_{\nu}+ q. p_{\nu}\right)\nonumber\\&-&p_b.p_{\nu} \left(11 \left(k. p_{\ell}+q.p_{\ell}\right)+12 k.q\right)\big)+4 \left(p_b. p_{\ell}\right) \left(p_b.p_{\nu}\right) \left(q. p_b\right)\big)+m_b^2 \big(10 \left(k. q\right) \left(p_b. p_{\nu}\right) \left(k. p_{\ell}\right)+m_{\ell}^2 \big(\left(  p_b.p_{\nu}\right)\nonumber\\&& \left(k.p_{\ell}+2 k.q+q.p_{\ell}\right)-\left(p_b.p_{\ell}\right) \left(k.p_{\nu}+  q.p_{\nu}\right)\big)\big)\Big),\\
  \label{4pi2d}
   \mathcal{J}_4(2;d)&=&\frac{1}{3m_b^3(p_{\ell}.k)^2}1024\lambda_1 p_b.p_{nu}\big((k.p_b+q.p_b)^2-m_b^2(2k.q+q^2)\big)\Big(m_{\ell}^2(-k.p_b-p_b.p_{\ell}+k.p_{\ell}+k.q+q.p_{\ell})\nonumber\\&+&k.p_{\ell}(k.p_b-k.q)\Big). 
    \label{4pi2dd}
\end{eqnarray}
\textbf{4. Fig.2(e):}
 \begin{eqnarray}
     \mathcal{M}_{\mu \nu}^{(5)}&=& (-ig^{\alpha\beta}) \bar{b}(-ieQ_b)\gamma^{\alpha}i\left(\slashed{ p_b}+\slashed{\Pi}-\slashed{k}+m_b\right)\gamma^{\nu}(1-\gamma^5)i\left(\slashed{ p_b}+\slashed{\Pi}-\slashed{q}-\slashed{k}+m_c\right)\gamma ^{\mu}\left(1-\gamma ^5\right)i\Big(\slashed{ p_b}\nonumber\\ &+& \slashed{\Pi}-\slashed{k}+m_b\Big)(-ieQ_b)\gamma^{\beta} b, \nonumber\\
      \mathcal{L}_{\mu \nu}^{(5)}&=&\left(\bar{\ell}\gamma^{\mu }(1-\gamma^5)\nu_{\ell}\right) \left(\bar{\nu}_{\ell} \gamma^{\nu }(1-\gamma^5)\gamma_{\alpha}\ell\right),\\
      \label{m5l5}
       I_5&=& \frac{1}{ k^2((p_b-q-k)^2-m_c^2)}\Big[ \frac{1}{(p_b.k)^2}+\frac{2(p_b-k).\Pi}{(p_b.k)^3}+\frac{\Pi^2}{(p_b.k)^3}+\frac{((p_b-k).\Pi)^2}{(p_b.k)^4}\Big]\nonumber\\&+& \frac{1}{ k^2((p_b-q-k)^2-m_c^2)^2}\Big[ \frac{-2p_c.\Pi}{(p_b.k)^2}-\frac{4(p_c.\Pi)(p_b-k).\Pi}{(p_b.k)^3}-\frac{\Pi^2}{(p_b.k)^2}\Big]\nonumber\\&+&\frac{1}{ k^2((p_b-q-k)^2-m_c^2)^3}\frac{2(p_c.\Pi)^2}{(p_b.k)^2}.
     \label{i5}
 \end{eqnarray}

\begin{eqnarray}
    \mathcal{O}(\Pi^0):\nonumber\\ \mathcal{J}_5(0;a)&=&\frac{1}{m_b(p_b.k)^2}32\Big(m_b^2\big(k.p_{\nu}(p_b.p_{\ell}-2q.p_{\ell})+k.p_{\ell}(p_b.p_{\nu}-2k.p_{\nu})-p_b.p_{\ell}(q.p_{\nu})+p_b.p_{\nu}(q.p_{\ell})\big)\nonumber\\&+&2k.p_b (k.p_{\nu})((k+q-p_b).p_{\ell}) \Big).
    \label{5pi0}
\end{eqnarray}
\begin{eqnarray}  
\mathcal{O}(\Pi):\nonumber\\ \mathcal{J}_5 (1;a)&=&\frac{1}{3m_b^3(p_b.k)^3}128(\lambda_1+3\lambda_2)\Big(10k.p_{\nu}(k.p_b)^2((k+q-p_b).p_{\ell})+m_b^4\big(k.p_{\nu}(8q.p_{\ell}-5p_b.p_{\ell})+k.p_{\ell}\nonumber\\&&((2q+10k-5p_b).p_{\nu})+3p_b.p_{\ell}(q.p_{\nu})-3p_b.p_{\nu}(q.p_{\ell})\big) + m_b^2 p_b.k\big(k.p_n(11p_b.p_{\ell}-16q.p_{\ell})+k.p_{\ell}\nonumber\\&&(p_b.p_{\nu}-16k.p_{\nu})-5p_b.p_{\ell}(q.p_{\nu})+p_b.p_{\nu}(4p_b.p_{\ell}+q.p_{\ell})\big)\Big),\\
    \label{5pi1}
    \mathcal{J}_5 (1;c)&=&\frac{1}{3m_b^3(p_b.k)^2}256(\lambda_1+3\lambda_2)\Big(m_b^2\big(4p_{\ell}.p_{\nu}(k.p_b)^2+k.p_b\big(k.p_{\nu}(11p_b.p_{\ell}-16q.p_{\ell})+k.p_{\ell}\big(p_b.p_{\nu}\nonumber\\&-&4(4k.p_{\nu}+q.p_{\nu})\big)-q.p_{\nu}(p_b.p_{\ell}+4q.p_{\ell})+p_b.p_{\nu}(4p_b.p_{\ell}+q.p_{\ell})\big)-4k.q(k.p_{\nu}-p_b.p_{\nu})((k+q-p_b).p_{\ell})\nonumber\\&+&q.p_b\big(k.p_{\nu}(p_b.p_{\ell}-6q.p_{\ell})+k.p_{\ell}(p_b.p_{\nu}-6k.p_{\nu})-5p_b.p_{\ell}(q.p_{\nu})+p_b.p_{\nu}(4p_b.p_{\ell}+q.p_{\ell}) \big)\big)+m_b^4\big(5k.p_{\nu}\nonumber\\&&(2q.p_{\ell}-p_b.p_{\ell})+k.p_{\ell}((10k+44q-5p_b).p_{\nu})+q.p_{\nu}(p_b.p_{\ell}+4q.p_{\ell})-5p_b.p{\nu}(q.p_{\ell})\big)+10k.p_b(k.p_{\nu})\nonumber\\&&(k.p_b+q.p_b)(k.p_{\ell}+q.p_{\ell}-p_b.p_{\ell})\Big).
    \label{5pi1d} 
\end{eqnarray}
{\small \begin{eqnarray}
    \mathcal{O}(\Pi^2):\nonumber\\ \mathcal{J}_5(2;a)&=&\frac{1}{3m_b^3(p_b.k)^3}64\lambda_1\Big(2k.p_{\nu}(k.p_b)^2((p_b+k+q).p_{\ell})+m_b^4\big( k.p_{\nu}(10q.p_{\ell}-7p_b.p_{\ell})+7k.p_{\ell}(2k.p_{\nu}-p_b.p_{\nu})\nonumber\\&-&2k.p_b(p_{\ell}.p_{\nu})+3p_b.p_{\ell}(q.p_{\nu})-3(p_b.p_{\nu})(q.p_{\ell})\big)+m_b^2 k.p_b\big(k.p_{\nu}(7p_b.p_{\ell}-12q.p_{\ell})-8k.p_{\ell}(p_b.p_{\nu}+2k.p_{\nu})\nonumber\\&+& 2k.p_b(p_{\ell}.p_{\nu})-p_b.p_{\ell}(q.p_{\nu}-13p_b.p_{\nu})-6p_b.p_{\nu}(q.p_{\ell}) \big)\Big),\\
    \label{5pi2}
     \mathcal{J}_5 (2;c)&=&\frac{-1}{3m_b^3(p_b.k)^3}256\lambda_1\Big(4(k.p_b+q.p_b)(k.p_b)^2\big(2k.p_{\nu}(k.p_{\ell}+q.p_{\ell}-p_b.p_{\ell})-p_b.p_{\nu}(k.p_{\ell}) \big)+m_b^4 \big(4k.q\nonumber\\&&\big(p_b.p_{\ell}(q.p_{\nu})-p_b.p_{\nu}(k.p_{\ell}+q.p_{\ell})-p_b.p_{\ell}(k.p_{\nu})+2 q.p_{\ell}(k.p_{\nu})\big)+k.p_b\big(k.p_{\nu}(p_b.p_{\ell}+2q.p_{\ell})+k.p_{\ell}\nonumber\\&&(p_b.p_{\nu}-2k.p_{\nu}-8q.p_{\nu})+q.p_{\nu}(7p_b.p_{\ell}-4q.p_{\ell})+p_b.p_{\nu}(q.p_{\ell})\big)\big)+2m_b^2p_b.k\big(2q.p_b \big(p_b.p_{\nu}\big(\big(q+2k\nonumber\\&-&2p_b\big).p_{\ell}\big)-2k.p_{\nu}((q+k-p_b).p_{\ell})\big)+2k.q\big(k.p_{\nu}(p_b.p_{\ell}-2q.p_{\ell})+p_b.p_{\ell}(p_b.p_{\nu}-q.p_{\nu})-2k.p_{\ell}(k.p_{\nu})\big)\nonumber\\&+& k.p_b\big(k.p_{\nu}(7p_b.p_{\ell}-5q.p_{\ell})+k.p_{\ell}(2p_b.p_{\nu}-3k.p_{\nu}+4q.p_{\nu})-4p_b.p_{\ell}(p_b.p_{\nu})+2q.p_{\ell}(q.p_{\nu})\big)\big)\Big),\\
  \label{5pi2d}
   \mathcal{J}_5 (2;d)&=&\frac{-1}{3m_b^3(p_{\ell}.k)^2}512\lambda_1 \big((pb.k+p_b.q)^2-m_b^2(2k.q+q^2)\big)\big(m_b^2\big(p_{\nu}.k(p_b.p_{\ell}-2q.p_{\ell})+k.p_{\ell}(p_b.p_{\nu}-2k.p_{\nu})\nonumber\\&-&p_b.p_{\ell}(q.p_{\nu})+(p_b.p_{\nu})(q.p{\ell})\big)+2k.p_b(k.p_{\nu})(k.p_{\ell}+q.p_{\ell}-p_b.p_{\ell})\Big). 
    \label{5pi2dd}
\end{eqnarray}}
\textbf{5. Fig.2(f):}
  \begin{eqnarray}
     \mathcal{M}_{\mu \nu}^{(6)}&=& (-ig^{\alpha\beta}) \bar{b}\gamma^{\nu}(1-\gamma^5)i\left(\slashed{ p_b}+\slashed{\Pi}-\slashed{q}-\slashed{k}+m_c\right)(-ieQ_u)\gamma^{\alpha}i\left(\slashed{ p_b}+\slashed{\Pi}-\slashed{q}+m_c\right)\gamma ^{\mu}\left(1-\gamma ^5\right) b, \nonumber\\
     \mathcal{L}_{\mu \nu}^{(6)}&=&\left(\bar{\ell}\gamma^{\mu }(1-\gamma^5)\nu_{\ell}\right) \left(\bar{\nu}_{\ell} \gamma^{\nu }(1-\gamma^5)i\left(\slashed{ p_l}+\slashed{k}+m_l\right)(-ieQ_l)\gamma_{\beta}\ell\right),\\
     \label{m6l6}
     I_6 &=& I_7=\frac{1}{ k^2((p_b-q-k)^2-m_c^2)}\Big[ \frac{1}{(p_c.k)(p_{\ell}.k)}-\frac{(p_b-q).\Pi}{(p_{\ell}.k)(p_c.k)^2}-\frac{\Pi^2}{2(p_{\ell}.k)(p_c.k)^2}+\frac{((p_b-q).\Pi)^2}{2(p_{\ell}.k)(p_c.k)^3}\Big]\nonumber\\&+& \frac{1}{ k^2((p_b-q-k)^2-m_c^2)^2}\Big[ \frac{-2p_c.\Pi}{(p_{\ell}.k)(p_c.k)}+\frac{2(p_c.\Pi)(p_b-q).\Pi}{(p_{\ell}.k)(p_c.k)^2}-\frac{\Pi^2}{(p_\ell.k)(p_c.k)}\Big]\nonumber\\&+&\frac{1}{ k^2((p_b-q-k)^2-m_c^2)^3}\frac{2(p_c.\Pi)^2}{(p_{\ell}.k)(p_c.k)}.
     \label{i67}
 \end{eqnarray}
\begin{eqnarray}
     \mathcal{O}(\Pi^0):\nonumber\\ \mathcal{J}_6(0;a)&=&\frac{1}{m_b(p_{\ell}.k)(p_c.k)}32 p_b.p_{\nu}\Big(2(q.p_{\ell}-p_b.p_{\ell})\big(k.p_{\ell}+k.q+q.p_{\ell}-k.p_b-p_b.p_{\ell}\big)-m_{\ell}^2\big(k.q+m_b^2+q^2\nonumber\\&-&k.p_b-2q.p_b\big)\Big).\\
    \label{6pi0}   
\mathcal{O}(\Pi):\nonumber\\ \mathcal{J}_6 (1;a)&=&\frac{1}{3m_b^3(p_{\ell}.k)(k.p_c)^2}64(\lambda_1+3\lambda_2)\big(2m_b^2q.p_{\nu}+p_b.p_{\nu}(3m_b^2-5q.p_b)\big)\Big(2(q.p_{\ell}-p_b.p_{\ell})\big(k.p_{\ell}+k.q+q.p_{\ell}\nonumber\\&-&p_b.p_{\ell}-k.p_b\big)-m_{\ell}^2(k.q+q^2+m_b^2-k.p_b-2q.p_b)\Big),\\
    \label{6pi1}
\mathcal{J}_6 (1;c)&=&\frac{1}{3m_b(p_{\ell}.k)(p_c.k)}256(\lambda_1+3\lambda_2)\Big(m_{\ell}^2(k.q+q^2+m_b^2-k.p_b-2q.p_b)-2(q.p_{\ell}-p_b.p_{\ell})\big(k.p_{\ell}+k.q+q.p_{\ell}\nonumber\\&-&k.p_b-p_b.p_{\ell}\big)\Big)\big(p_b.p_{\nu}(5(k.q+q.p_b)-3m_b^2)-2m_b^2(k.p_{\nu}+q.p_{\nu})\big).
    \label{6pi1d}
\end{eqnarray}

{\small
\begin{eqnarray}
   \mathcal{O}(\Pi^2):\nonumber\\  \mathcal{J}_6 (2;a)&=&\frac{1}{3m_b^3(p_{\ell}.k)(p_c.k)^3}32\lambda_1 p_b.p_{\nu}\Big(2k.p_c\big(m_{\ell}^2(m_b^2(k.q+2q^2)-2(q.p_b)^2)+2(k.p_{\ell}+2q.p_{\ell}-2p_b.p_{\ell})\nonumber\\&&\big((p_b.p_{\ell})(q.p_b)-m_b^2(q.p_{\ell})\big)+2k.q\big(p_b.p_{\ell}(q.p_b+m_b^2)-2m_b^2q.p_{\ell}\big)+k.p_b\big(2m_b^2q.p_{\ell} +q.p_b(2q.p_{\ell}\nonumber\\&-&4p_b.p_{\ell}-m_{\ell}^2)\big)\big)-\big(2(q.p_{\ell}-p_b.p_{\ell})(k.p_{\ell}+k.q+q.p_{\ell}-k.p_b-p_b.p_{\ell})-m_{\ell}^2\big(k.q+q^2+m_b^2-k.p_b\nonumber\\&-&2q.p_b\big)\big)\big((q.p_b)^2+m_b^2(3k.p_c-q^2)\big)-2(k.p_c)^2\big(m_b^2(m_{\ell}^2-2k.p_{\ell})+2p_b.p_{\ell}(k.p_b+p_b.p_{\ell})\big)\Big),\\
    \label{6pi2}
    \mathcal{J}_6(2;c)&=&\frac{-1}{3m_b^3(p_{\ell}.k)(p_c.k)^2}128\lambda_1\Big(m_b^2(k.p_c)(q.p_{\nu})(2(p_b.p_{\ell})^2-m_b^2p_{\ell}.k+k.p_{\ell}(q.p_b)-2(k.q)(p_b.p_{\ell})-3(k.p_{\ell})(p_b.p_{\ell})\nonumber\\&-&4(q.p_{\ell})(p_b.p_{\ell}))+m_b^2(q.p_{\ell})^2(p_b.p_{\nu})\big(8k.q-24k.p_c+8q^2\big)+m_b^2(p_b.p_{\ell})^2(p_b.p_{\nu})\big(8k.q-12k.p_c+8q^2\big)\nonumber\\&+&m_b^2p_b.p_{\nu}\big((p_c.k)(p_{\ell}.k)(q.p_b-6p_{\ell}.k-8k.q-q^2)+\big)\Big),\\
  \label{6pi2d}
  \mathcal{J}_6 (2;d)&=&-\frac{1}{3 \left(k . p_{\ell}\right) ( k . p_c )^2}128  \lambda _1 \Big(\left(q . p_{\nu}\right) \left(k . p_c\right)\big(-\left(k . p_{\ell}\right) m_b^4-4\left(q . p_{\ell}\right) \left(p_b . p_{\ell}\right) m_b^2-2 \left(k . q\right) \left(p_b . p_{\ell}\right) m_b^2\nonumber\\&-&3 \left(k . p_{\ell}\right) \left(p_b . p_{\ell}\right) m_b^2\big)+ \left(k . p_c\right)\left(p_b . p_{\nu}\right) \big(-24 \left(q . p_{\ell}\right)^2  m_b^2-12 \left(p_b . p_{\ell}\right)^2 m_b^2\big)+q^2 \left(p_b . p_{\nu}\right) m_b^2\big( 8 \left(q . p_{\ell}\right)^2\nonumber\\&+&8 \left(p_b . p_{\ell}\right)^2 -\left(k . p_{\ell}\right) \left(k . p_c\right)\big)+  8\left(k . q\right) \left(p_b . p_{\nu}\right) m_b^2\big( \left(q . p_{\ell}\right)^2+ \left(p_b . p_{\ell}\right)^2 + (k . q + q^2) q . p_{\ell}\big)+ \left(q . p_{\ell}\right) \left(p_b . p_{\nu}\right) \nonumber\\&&m_b^2\big(8 (q^2 + k . q) \left(k . p_{\ell}\right) -6(4 k . q +5 k . p_{\ell}) k . p_c\big)- 8\left(p_b . p_{\nu}\right) m_b^2 \big(p_b . p_{\ell}\big( k . q + k . p_{\ell} +2q.p_{\ell}\big)(q^2 + k . q)\nonumber\\& -&(6 k . p_{\ell} + 8 k . q)\left(k . p_{\ell}\right)\left(k . p_c\right)  \big)+\left(k . p_c\right) m_b^2\big(q . p_{\nu} \big(2 \left(p_b . p_{\ell}\right){}^2  +\left(q . p_b\right) \left(k . p_{\ell}\right)\big) +\left(q . p_b\right) \left(k . p_{\ell}\right) \left(p_b . p_{\nu}\right)\nonumber\\& +&\big( 19 k . q +26 k . p_{\ell}  +38 q . p_{\ell} +2 q^2  +3 k . q \big) \left(p_b . p_{\ell}\right) \left(p_{\ell} . p_{\nu}\right) \big)+\big(4 \big(\big(\left(q^2+ k .q+ p_b. q\right) m_b^2+q . p_b \big(q^2+ k.q\nonumber\\&-&3 q . p_b\big)\big) m_l^2-2 \left(q . p_{\ell}-p_b . p_{\ell}\right) \big(\left(q^2+ k. q\right) m_b^2+q . p_b \big( k.q- p_b. q+ p_{\ell}.q-p_b . p_{\ell}+ k. p_{\ell}\big)\big)\big) \left(p_b . p_{\nu}\right)\nonumber\\&+&\big(-\left(p_{\ell} . p_{\nu}\right) \left(q . p_b\right)^2-m_b^2\big((k . p_{\ell} +2 q . p_{\ell}) \left(q . p_{\nu}\right) +6(2 \left(p_b . p_{\ell}\right)  + m_l^2)\left(p_b . p_{\nu}\right) \big)+\big(\left(p_b . p_{\ell}\right)^2\nonumber\\&-&\left(2 q . p_{\ell}+ k. p_{\ell}-2 p_b . p_{\ell}\right) m_b^2\big) \left(k . p_{\nu}\right)+2 (p_b . p_{\ell} + m_b^2) \left(p_b . p_{\ell}\right) \left(q . p_{\nu}\right)+\left(p_b . p_{\nu}\right)\big(-16 \left(p_b . p_{\ell}\right)^2 +8 m_b^2 \left(k . p_{\ell}\right) \nonumber\\&+&19 m_b^2 \left(q . p_{\ell}\right)+7\left(p_b . p_{\ell}\right)\big(k . q  +k . p_{\ell} +2 q . p_{\ell} \big)\big)+(q^2 +2 k . q) m_b^2 \left(p_{\ell} . p_{\nu}\right)+\left(q . p_b\right) \big(-9 \left(p_b . p_{\nu}\right) m_l^2\nonumber\\&+&\left(7 q . p_{\ell}+. k p_{\ell}\right) \left(p_b . p_{\nu}\right)+\left(p_b . p_{\ell}\right) \left(2 q . p_{\nu}-16 \left(p_b . p_{\nu}\right)-3 \left(p_{\ell} . p_{\nu}\right)\right)\big)\big)\left(k . p_c\right)\big) \left(k . p_b\right) -8( \left(q . p_b\right)^2 \left(q . p_{\ell}\right)^2\nonumber\\& +&( q . p_b +2 k . p_c)\left(q . p_b\right) \left(p_b . p_{\ell}\right){}^2 )\left(p_b . p_{\nu}\right)+\big(-4 \left(k . q\right)^2 m_b^2-4 \left(q^2+m_b^2-2 \left(q . p_b\right)\right) \left(q^2 m_b^2-\left(q . p_b\right){}^2\right)\nonumber\\&+&\big(-4 m_b^4+\big(-8 q^2+8 . p_b q+15 k . p_c\big) m_b^2+4 \left(q . p_b\right)^2\big) \left(k . q\right)+6 \left(m_b^4+2 \left(q^2-q . p_b\right) m_b^2-\left(q . p_b\right)^2\right) \nonumber\\&&\left(k . p_c\right)\big) m_l^2 \left(p_b . p_{\nu}\right)-8\left(q . p_b\right){}^2 \left(q . p_{\ell}\right) \left(p_b . p_{\nu}\right)\big( k . q + k . p_{\ell}\big) +8 \left(q . p_b\right)^2 \left(p_b . p_{\ell}\right) \left(p_b . p_{\nu}\right)(k . q + k . p_{\ell} +2q . p_{\ell} )\nonumber\\&-&\left(k . p_c\right)\left(p_{\ell} . p_{\nu}\right)(2  \left(q . p_b\right){}^2 \left(p_b . p_{\ell}\right) +\left(k . p_b\right){}^3 )+\left(k . p_b\right){}^2 \big(4 \left(q . p_b\right) \left(-m_l^2+2 q . p_{\ell}-2 \left(p_b . p_{\ell}\right)\right) \left(p_b . p_{\nu}\right)\nonumber\\&+&\left(p_b . p_{\ell}\right) \left(k . p_{\nu}\right) \left(k . p_c\right)+\big(-3 \left(p_b . p_{\nu}\right) m_l^2+\left(7 q . p_{\ell}+ k. p_{\ell}\right) \left(p_b . p_{\nu}\right)-2 \left(q . p_b\right) \left(p_{\ell} . p_{\nu}\right)+\left(p_b . p_{\ell}\right) \nonumber\\&&\left(2 q . p_{\nu}-p_{\ell} . p_{\nu}-16 p_b . p_{\nu}\right)\big) k . p_c\big)+\big(-\left(k . p_{\ell}\right) m_b^4+p_b . p_{\ell}\left(q^2-2 k . p_{\ell}+2 p_b . p_{\ell}-3 q . p_{\ell}\right) m_b^2 \nonumber\\&+&\left(\left(p_b . p_{\ell}\right){}^2+m_b^2 k . p_{\ell}\right) q . p_b-\left(q . p_b\right){}^2 \left(p_b . p_{\ell}\right)\big) \left(k . p_{\nu}\right) \left(k . p_c\right)+(2 \left(p_b . p_{\ell}\right) \left(q . p_{\nu}\right) +7\big( k . q + k . p_{\ell} +2 q . p_{\ell})   \nonumber\\&& p_b . p_{\nu} \big)\left(q . p_b\right)\left(p_b . p_{\ell}\right)\left(k . p_c\right)\Big). 
    \label{6pi2dd}
\end{eqnarray}
}

\textbf{6. Fig.2(g):}
 \begin{eqnarray}
     \mathcal{M}_{\mu \nu}^{(7)}&=& (-ig^{\alpha\beta}) \bar{b}\gamma^{\nu}(1-\gamma^5)i\left(\slashed{ p_b}+\slashed{\Pi}-\slashed{q}+m_c\right)(-ieQ_u)\gamma^{\alpha}i\left(\slashed{ p_b}+\slashed{\Pi}-\slashed{q}-\slashed{k}+m_c\right)\gamma ^{\mu}\left(1-\gamma ^5\right) b,\nonumber\\
     \mathcal{L}_{\mu \nu}^{(7)}&=&\left(\bar{\ell}(-ieQ_l)\gamma_{\beta}i\left(\slashed{ p_l}+\slashed{k}+m_l\right)\gamma^{\mu }(1-\gamma^5)\nu_{\ell}\right) \left(\bar{\nu}_{\ell} \gamma^{\nu }(1-\gamma^5)\ell\right),\\
     \label{m7l7}
      I_7&=&I_6.
 \end{eqnarray}
 \begin{eqnarray}
  \mathcal{O}(\Pi^0):\  \mathcal{J}_7(0;a)&=& \mathcal{J}_6 (0;a),\\
  \mathcal{O}(\Pi):\ \mathcal{J}_7(1;a)&=& \mathcal{J}_6 (1;a),\mathcal{J}_7 (1;c)= \mathcal{J}_6 (1;c),\\ 
  \mathcal{O}(\Pi^2):\ \mathcal{J}_7(2;a)&=& \mathcal{J}_6 (2;a), \ \mathcal{J}_7(2;c)=\mathcal{J}_6 (2;c),\ 
     \mathcal{J}_7(2;d)=\mathcal{J}_6 (2;d).
 \end{eqnarray}

\textbf{7. Fig.2(h):}
  \begin{eqnarray}
     \mathcal{M}_{\mu \nu}^{(8)}&=&(-ig^{\alpha\beta}) \bar{b}\gamma^{\nu}(1-\gamma^5)i\left(\slashed{ p_b}+\slashed{\Pi}-\slashed{q}+m_c\right)(-ieQ_u)\gamma^{\alpha}i\left(\slashed{ p_b}+\slashed{\Pi}-\slashed{q}-\slashed{k}+m_c\right)\gamma ^{\mu}\left(1-\gamma ^5\right)i\Big(\slashed{ p_b}\nonumber\\ &+&\slashed{\Pi}-\slashed{k}+m_b\Big)(-ieQ_b)\gamma^{\alpha} b,\nonumber\\
     \mathcal{L}_{\mu \nu}^{(8)}&=&\left(\bar{\ell}\gamma^{\mu }(1-\gamma^5)\nu_{\ell}\right) \left(\bar{\nu}_{\ell} \gamma^{\nu }(1-\gamma^5)\ell\right),\\
     \label{m8l8}
      I_8 &=&\frac{1}{ k^2((p_b-q-k)^2-m_c^2)}\Big[ \frac{-1}{(p_c.k)(p_b.k)}+\frac{(p_b-q).\Pi}{(p_b.k)(p_c.k)^2}-\frac{(p_b-k).\Pi}{(p_b.k)^2(p_c.k)}+\frac{\Pi^2}{2(p_b.k)(p_c.k)^2}\nonumber\\&-&\frac{\Pi^2}{2(p_b.k)^2(p_c.k)}-\frac{((p_b-q).\Pi)^2}{2(p_b.k)(p_c.k)^3}-\frac{((p_b-k).\Pi)^2}{2(p_b.k)^3(p_c.k)}+\frac{((p_b-q).\Pi)((p_b-k).\Pi)}{(p_c.k)^2(p_b.k)^2}\Big]\nonumber\\&+& \frac{1}{ k^2((p_b-q-k)^2-m_c^2)^2}\Big[ \frac{2p_c.\Pi}{(p_b.k)(p_c.k)}-\frac{2(p_c.\Pi)(p_b-q).\Pi}{(p_b.k)(p_c.k)^2}+\frac{2(p_c.\Pi)(p_b-k).\Pi}{(p_b.k)^2(p_c.k)}+\frac{\Pi^2}{(p_b.k)(p_c.k)}\Big]\nonumber\\&+&\frac{1}{ k^2((p_b-q-k)^2-m_c^2)^3}\frac{-2(p_c.\Pi)^2}{(p_b.k)(p_c.k)}.\\
     \label{i89}
      \mathcal{O}(\Pi^0):\nonumber\\ \mathcal{J}_8(0;a)&=&\frac{1}{m_b \left( k. p_b\right) \left(  k. p_c\right)}16 \Big(-2 \left( p_{\ell}. p_{\nu}\right) \left( k. p_b\right){}^2+m_b^2 \big(-\left(  p_{\ell}. p_{\nu}\right) \left(-2 \left(  k.p_b\right)-2 \left(  q.p_b\right)+  k. q+q^2\right)\nonumber\\&-&2 \left( p_b. p_{\ell}\right) \left( q. p_{\nu}\right)+2 \left(q. p_{\ell}\right) \left(  p_b.p_{\nu}+  q. p_{\nu}\right)\big)+\left( k. p_{\nu}\right) \big(\left(  q. p_{\ell}\right) \left(2 \left(  q.p_b\right)-m_b^2\right)-\left(  p_b. p_{\ell}\right) \big(-2 k.p_b\nonumber\\&+&2 k .q+q^2\big)\big)+\left(k.p_{\ell}\right) \left(2 \left( k. p_{\nu}\right) \left(q. p_b-m_b^2\right)-2 \left(  k. p_b\right) \left(q. p_{\nu}\right)+\left(p_b .p_{\nu}\right) \left(2 \left(  k .p_b\right)-q^2\right)+m_b^2 \left(  q. p_{\nu}\right)\right)\nonumber\\&-&2 \left( k .p_b\right) \left(  p_b. p_{\nu}\right) \left(  q. p_{\ell}\right)-2 \left( k.p_b\right) \left( q. p_{\ell}\right) \left( q. p_{\nu}\right)+2 \left(  k. p_b\right) \left(  p_b .p_{\ell}\right) \left(  q. p_{\nu}\right)+2 \left(  k .q\right) \left(  p_b. p_{\nu}\right) \left(  q. p_{\ell}\right)\nonumber\\&+&q^2 \left(  k. p_b\right) \left(p_{\ell}.p_{\nu}\right)+2 \left(  k. q\right) \left(  k. p_b\right) \left(  p_{\ell}. p_{\nu}\right)-2 \left(  k. p_b\right) \left( q. p_b\right) \left(p_{\ell}. p_{\nu}\right)+m_b^4 \left(-\left(  p_{\ell}. p_{\nu}\right)\right)-4 \left(  p_b .p_{\nu}\right) \left(q. p_b\right) \nonumber\\&&\left(q. p_{\ell}\right)+2 q^2 \left( p_b .p_{\ell}\right) \left(  p_b. p_{\nu}\right)\Big).
    \label{8pi0}
 \end{eqnarray}

{\footnotesize
\begin{eqnarray}
   \mathcal{O}(\Pi):\nonumber\\  \mathcal{J}_8(1;a)&=&\frac{1}{3m_b^3 \left( k. p_b\right)^2 \left(  k. p_c\right)^2}16(\lambda_1+3\lambda_2)\big(2p_{\ell}.p_{\nu}(k.p_b)^3(5q.p_b-3m_b^2)+(p_b.k)^2\big(6m_b^4(p_{\ell}.p_{\nu})+m_b^2\big(-2q.p_{\nu}\nonumber\\&&(3k.p_{\ell}+7q.p_{\ell}-5p_b.p_{\ell})+k.p_{\nu}(4q.p_{\ell}+6p_b.p_{\ell})+6p_b.p_{\nu}(k.p_{\ell}-q.p_{\ell})+(2k.q+8k.p_c+5q^2-16q.p_b)\nonumber\\&&p_{\ell}.p_{\nu}\big)+10\big(-(k.p_c-q.p_b)(k.p_{\ell}+q.p_{\ell})q.p_{\nu}-q.p_b(q.p_{\ell}-k.p_{\ell})(p_b.p_{\nu})\big)+5\big(2(q.p_b)^2-(2k.q+q^2)q.p_b\nonumber\\&+&k.p_c(2k.q+q^2)\big)p_{\ell}.p_{\nu}\big)+p_b.k\big(-3(p_{\ell}.p_{\nu})m_b^6+m_b^4\big(-8(q.p_{\nu})(p_b.p_{\ell})+2q.p_{\ell}(5q.p_{\nu}+4p_b.p_{\nu})\nonumber\\&+&p_{\ell}.p_{\nu}\big(k.q-6k.p_c-3q^2+9q.p_b\big)\big)+m_b^2\big(2q.p_{\nu}\big(5(k.p_c-q.p_b)q.p_{\ell}+p_b.p_{\ell}(2k.q-4k.p_c+q^2+3q.p_b)\big)\nonumber\\&+&2p_b.p_{\nu}\big(q.p_{\ell}(5k.q+2k.p_c-q^2-13q.p_b)+(5q^2-2k.q)p_b.p_{\ell}\big)+p_{\ell}.p_{\nu}\big(4(k.q)^2+k.q\big(2q^2-15k.p_c\nonumber\\&-&3q.p_b\big)+3q.p_b(q^2-2q.p_b)+k.p_c(8q.p_b-5q^2)\big)\big)+10p_b.p_{\nu}\big(q.p_{\ell}(2(q.p_b)^2+k.q(k.p_c-q.p_b))-q^2\nonumber\\&&(q.p_b)(p_b.p_{\ell})\big)+k.p_{\nu}\big(-7(q.p_{\ell})m_b^4-m_b^2\big(q.p_{\ell}(4k.q+k.p_c-2q^2-15q.p_b)+p_b.p_{\ell}\big(6k.q+2k.p_c+9q^2\nonumber\\&-&4q.p_b\big)\big)+5(k.p_c-q.p_b)\big(2(q.p_b)(q.p_{\ell})-(2k.q+q^2)p_b.p_{\ell}\big)\big)+k.p_{\ell}\big(3m_b^4(q.p_{\nu}-2k.p_{\nu})-\big(4k.p_{\nu}\nonumber\\&&(2k.p_c-q^2-3q.p_b)+q.p_{\nu}\big(4k.q-15k.p_c+2q^2+q.p_b\big)+p_b.p_{\nu}(5(2k.p_c+q^2)-4k.q)\big)m_b^2+5\big(k.p_c\nonumber\\&-&q.p_b\big)(2k.p_{\nu}(q.p_b)-q^2(p_b.p_{\nu}))\big)\big)+(k.p_c)m_b^2\big(3(p_b.p_{\ell})m_b^4-m_b^2\big(k.p_{\nu}(q.p_{\ell}-2p_b.p_{\ell})-6(q.p_{\nu})(p_b.p_{\ell})\nonumber\\&+&6q.p_{\ell}(q.p_{\nu}+p_b.p_{\nu})-3(q^2-2q.p_b)p_{\ell}.p_{\nu}\big)-(2(q.p_b)(q.p_{\ell})-q^2(p_b.p_{\ell}))(k.p_{\nu}-6p_b.p_{\nu})+k.q\big(3m_b^2\nonumber\\&&(p_{\ell}.p_{\nu})+k.p_{\nu}(14p_b.p_{\ell}-8q.p_{\ell})-2p_b.p_{\nu}(q.p_{\ell}+2p_b.p_{\ell})\big)+k.p_{\ell}\big(-3m_b^2(q.p_{\nu})+p_b.p_{\nu}\big(-2m_b^2+q^2\nonumber\\&+&4q.p_b\big)+2k.p_{\nu}(5m_b^2+2q^2-7q.p_b)\big)\big)\Big),\\
    \label{8pi1}
     \mathcal{J}_8(1;c)&=&\frac{1}{3 \big(k .  p_b\big) \big(k .  p_c\big) m_b}128 \big(-3 \big(p_{\ell} .  p_{\nu}\big) m_b^6+\big(\big(p_{\ell} .  p_{\nu}\big) \big(-3 q^2+  k. q+9  p_b. q\big)+\big(3 q .  p_{\ell}+2 p_b .  p_{\ell}\big) \big(k .  p_{\nu}\big)\nonumber\\&+&2 \big(5 q .  p_{\ell}+4 p_b .  p_{\ell}\big) \big(q .  p_{\nu}\big)-8 \big(q .  p_{\ell}\big) \big(p_b .  p_{\nu}\big)-\big(k .  p_{\ell}\big) \big(10 k .  p_{\nu}+3 q .  p_{\nu}+2 p_b .  p_{\nu}\big)\big) m_b^4+\big(-2 \big(k .  p_{\nu}\big)\nonumber\\&& \big(q .  p_{\ell}\big) q^2-3 \big(k .  p_{\nu}\big) \big(p_b .  p_{\ell}\big) q^2-2 \big(q .  p_{\nu}\big) \big(p_b .  p_{\ell}\big) q^2+2 \big(q .  p_{\ell}\big) \big(p_b .  p_{\nu}\big) q^2+10 \big(p_b .  p_{\ell}\big) \big(p_b .  p_{\nu}\big) q^2-\big(k .  p_{\nu}\big)\nonumber\\&& \big(q .  p_b\big) \big(q .  p_{\ell}\big)-10 \big(q .  p_b\big) \big(q .  p_{\nu}\big) \big(q .  p_{\ell}\big)-4 \big(k .  p_{\nu}\big) \big(q .  p_b\big) \big(p_b .  p_{\ell}\big)-26 \big(q .  p_b\big) \big(q .  p_{\nu}\big) \big(p_b .  p_{\ell}\big)-4 \big(k .  q\big) \big(q .  p_{\ell}\big)\nonumber\\&& \big(k .  p_{\nu}\big)+4 \big(k .  q\big) \big(p_b .  p_{\ell}\big) \big(k .  p_{\nu}\big)+6 \big(k .  q\big) \big(p_b .  p_{\ell}\big) \big(q .  p_{\nu}\big)+4 \big(k .  q\big) \big(q .  p_{\ell}\big) \big(p_b .  p_{\nu}\big)+6 \big(q .  p_b\big) \big(q .  p_{\ell}\big) \big(p_b .  p_{\nu}\big)\nonumber\\&+&\big(-6 \big(q .  p_b\big){}^2+2 \big(k .  q\big) \big(q^2+2 k .  q\big)+3 \big(q^2-k .  q\big) \big(q .  p_b\big)\big) \big(p_{\ell} .  p_{\nu}\big)+\big(k .  p_{\ell}\big) \big(\big(q .  p_b\big) \big(20 k .  p_{\nu}+11 q .  p_{\nu}\nonumber\\&+&4 p_b .  p_{\nu}\big)+\big(q^2+2 k .  q\big) \big(2 \big(q .  p_{\nu}\big)-7 \big(p_b .  p_{\nu}\big)\big)\big)\big) m_b^2+\big(9 \big(p_{\ell} .  p_{\nu}\big) m_b^4+\big(\big(p_{\ell} .  p_{\nu}\big) \big(8 q^2+7 k .  q-22 \big(q .  p_b\big)\big)\nonumber\\&-&4 \big(5 q .  p_{\ell}+4 p_b .  p_{\ell}\big) \big(q .  p_{\nu}\big)+16 \big(q .  p_{\ell}\big) \big(p_b .  p_{\nu}\big)\big) m_b^2+\big(-11 \big(q .  p_{\ell}\big) m_b^2+\big(5 q^2+6 m_b^2\big) \big(p_b .  p_{\ell}\big)+10 \big(q .  p_b\big)\nonumber\\&& \big(q .  p_{\ell}-p_b .  p_{\ell}\big)\big) \big(k .  p_{\nu}\big)+10 \big(\big(q .  p_b\big) \big(q .  p_{\ell}\big)-\big(k .  q-3 \big(q .  p_b\big)\big) \big(p_b .  p_{\ell}\big)\big) \big(q .  p_{\nu}\big)-10 \big(\big(p_b .  p_{\ell}\big) q^2+\big(q .  p_b\big) \nonumber\\&&\big(q .  p_{\ell}\big)\big) \big(p_b .  p_{\nu}\big)-5 \big(q^2+2 k .  q-2 \big(q .  p_b\big)\big) \big(q .  p_b\big) \big(p_{\ell} .  p_{\nu}\big)+\big(k .  p_{\ell}\big) \big(\big(10 k .  p_{\nu}+14 p_b .  p_{\nu}+  q. p_{\nu}\big) m_b^2+5 \nonumber\\&&\big(q^2+2 k .  q\big) \big(p_b .  p_{\nu}\big)-10 \big(q .  p_b\big) \big(.  k p_{\nu}+.  q p_{\nu}+.  p_b p_{\nu}\big)\big)\big) \big(k .  p_b\big)+10 \big(k .  p_b\big){}^3 \big(p_{\ell} .  p_{\nu}\big)+\big(k .  p_b\big){}^2 \big(-16\nonumber\\&& \big(p_{\ell} .  p_{\nu}\big) m_b^2-5 \big(q^2+2 k .  q-4 \big(q .  p_b\big)\big) \big(p_{\ell} .  p_{\nu}\big)+10 \big(-\big(p_b .  p_{\nu}\big) \big(  k .p_{\ell}+  q. p_{\ell}\big)+\big(q .  p_{\ell}-p_b .  p_{\ell}\big) \big(k .  p_{\nu}\big)\nonumber\\&+&\big(  q .p_{\ell}+  p_b .p_{\ell}\big) \big(q .  p_{\nu}\big)\big)\big)+5 \big(q .  p_b\big) \big(\big(p_b .  p_{\ell}\big) \big(q^2 \big(k .  p_{\nu}-2 \big(p_b .  p_{\nu}\big)\big)-2 \big(k .  q-2 \big(q .  p_b\big)\big) \big(q .  p_{\nu}\big)\big)+\big(k .  p_{\ell}\big)\nonumber\\&& \big(\big(q^2+2 k .  q\big) \big(p_b .  p_{\nu}\big)-2 \big(q .  p_b\big) \big(  k. p_{\nu}+ q. p_{\nu}\big)\big)\big)\big) \big(\lambda _1+3 \lambda _2\big).
    \label{8pi1d}
\end{eqnarray}

\begin{eqnarray}
   \mathcal{O}(\Pi^2):\nonumber\\  \mathcal{J}_8(2;a)&=&\frac{1}{3 \big(k .  p_b\big){}^2 \big(k .  p_c\big){}^3 m_b^3}16 \Big(2 \big(k .  p_c\big){}^2 \big(k .  p_b\big) \big(2 \big(3 q .  p_{\ell}+ k. p_{\ell}\big) \big(p_b .  p_{\nu}\big)+\big(-3 m_b^2+4 k .  p_b+2 q .  p_b\big) \big(p_{\ell} .  p_{\nu}\big)\nonumber\\&-& 2 \big(p_b .  p_{\ell}\big) \big(3 k .  p_{\nu}+ q. p_{\nu}\big)\big) m_b^2+\big(\big(k .  p_c\big) \big(2 \big(k .  q\big)-3 \big(k .  p_c\big)\big) m_b^2+\big(\big(3 \big(k .  p_c\big)-q^2\big) m_b^2+\big(k .  p_c-q .  p_b\big){}^2\big)\nonumber\\&& \big(k .  p_b\big)\big) \big(\big(p_{\ell} .  p_{\nu}\big) m_b^4+\big(\big(p_{\ell} .  p_{\nu}\big) \big(q^2+ k.q-2 \big(k .  p_b\big)-2 \big(q .  p_b\big)\big)+2 \big(p_b .  p_{\ell}\big) \big(q .  p_{\nu}\big)-2 \big(q .  p_{\ell}\big) \nonumber\\&&\big(  q. p_{\nu}+ p_b. p_{\nu}\big)\big) m_b^2+\big(-\big(q .  p_{\nu}\big) m_b^2+2 \big(m_b^2-q .  p_b\big) \big(k .  p_{\nu}\big)+2 \big(k .  p_b\big) \big(q .  p_{\nu}\big)+\big(q^2-2 \big(k .  p_b\big)\big) \big(p_b .  p_{\nu}\big)\big) \nonumber\\&&\big(k .  p_{\ell}\big)-2 \big(k .  p_b\big) \big(q .  p_{\nu}\big) \big(p_b .  p_{\ell}\big)+\big(\big(p_b .  p_{\ell}\big) \big(q^2+2 k .  q-2 \big(k .  p_b\big)\big)+\big(m_b^2-2 \big(q .  p_b\big)\big) \big(q .  p_{\ell}\big)\big) \big(k .  p_{\nu}\big)\nonumber\\&+&2 \big(k .  p_b\big) \big(q .  p_{\ell}\big) \big(q .  p_{\nu}\big)-2 \big(k .  q\big) \big(q .  p_{\ell}\big) \big(p_b .  p_{\nu}\big)+2 \big(k .  p_b\big) \big(q .  p_{\ell}\big) \big(p_b .  p_{\nu}\big)+4 \big(q .  p_b\big) \big(q .  p_{\ell}\big) \big(p_b .  p_{\nu}\big)\nonumber\\&-&2 q^2 \big(p_b .  p_{\ell}\big) \big(p_b .  p_{\nu}\big)+2 \big(( \big(k .  p_b\big) )\big){}^2 \big(p_{\ell} .  p_{\nu}\big)-\big(k .  p_b\big) q^2 \big(p_{\ell} .  p_{\nu}\big)-2 \big(k .  q\big) \big(k .  p_b\big) \big(p_{\ell} .  p_{\nu}\big)+2 \big(k .  p_b\big) \nonumber\\&& \big(q .  p_b\big) \big(p_{\ell} .  p_{\nu}\big)\big)-2 \big(k .  p_c\big)\big(\big(\big(k .  p_{\ell}\big) \big(k .  p_c\big) \big(2 k .  p_{\nu}-p_b .  p_{\nu}-2 q .  p_{\nu}\big)+\big(k .  p_b\big) \big(-2 \big(p_{\ell} .  p_{\nu}\big) q^2\nonumber\\&-&\big(q .  p_{\nu}\big) \big(p_b .  p_{\ell}\big)-2 \big(q .  p_{\ell}\big) \big(k .  p_{\nu}\big)+\big(q .  p_{\ell}\big) \big(2 q .  p_{\nu}+ p_b. p_{\nu}\big)\big)+\big(\big(p_b .  p_{\ell}\big) \big(k .  p_{\nu}\big)+2 \big(k .  q\big) \big(p_{\ell} .  p_{\nu}\big)\big) \big(k .  p_c\big)\big) m_b^4\nonumber\\&+&\big(2 \big(\big(q^2+  k .q\big) \big(p_{\ell} .  p_{\nu}\big)-\big(q .  p_{\nu}\big) \big(-p_b .  p_{\ell}+  k. p_{\ell}+  q. p_{\ell}\big)\big) \big(k .  p_b\big){}^2+\big(k .  p_b\big) \big(-\big(p_b .  p_{\ell}\big) \big(q .  p_{\nu}\big) q^2+\big(q .  p_{\ell}\big)\nonumber\\&& \big(p_b .  p_{\nu}\big) q^2+2 \big(p_b .  p_{\ell}\big) \big(p_b .  p_{\nu}\big) q^2+2 \big(\big(q .  p_b\big) \big(  q. p_{\ell}+p_b. p_{\ell}\big)-\big(q^2+  k .q+2 k .  p_c\big) \big(p_b .  p_{\ell}\big)\big) \big(k .  p_{\nu}\big)\nonumber\\&+&2 \big(q .  p_b\big) \big(q .  p_{\ell}\big) \big(q .  p_{\nu}\big)- 2 \big(q .  p_b\big) \big(p_b .  p_{\ell}\big) \big(q .  p_{\nu}\big)+2 \big(k .  q\big) \big(q .  p_{\ell}\big) \big(p_b .  p_{\nu}\big)-2 \big(q .  p_b\big) \big(q .  p_{\ell}\big) \big(p_b .  p_{\nu}\big)-2 \big(k .  q\big)\nonumber\\&& \big(p_b .  p_{\ell}\big) \big(p_b .  p_{\nu}\big)-2 \big(\big(q .  p_b\big) \big(k .  p_c-q .  p_b\big)+\big(k .  q\big) \big(k .  p_c\big)\big) \big(p_{\ell} .  p_{\nu}\big)+2 \big(p_b .  p_{\ell}\big) \big(q .  p_{\nu}\big) \big(k .  p_c\big)+2 \big(k .  p_{\ell}\big) \nonumber\\&&\big(  k. p_{\nu}+  q. p_{\nu}\big) \big(k .  p_c\big)\big)+\big(\big(p_b .  p_{\nu}\big) \big(\big(k .  p_{\ell}\big) \big(q^2+2p_b. q\big)-2 \big(k .  q\big) \big(  q. p_{\ell}+  p_b. p_{\ell}\big)\big)+\big(\big(q^2+2 k .  q\big) \big(p_b .  p_{\ell}\big)\nonumber\\&-&2 \big(q .  p_b\big) \big(  k. p_{\ell}+ q .p_{\ell}\big)\big) \big(k .  p_{\nu}\big)\big) \big(k .  p_c\big)\big) m_b^2+2 \big(k .  p_b\big) \big(k .  p_c-q .  p_b\big) \big(\big(p_{\ell} .  p_{\nu}\big) \big(( \big(k .  p_b\big) )\big){}^2-\big(\big(k .  p_{\ell}-q .  p_{\ell}\big)\nonumber\\&& \big(p_b .  p_{\nu}\big)-\big(q .  p_b\big) \big(p_{\ell} .  p_{\nu}\big)+\big(p_b .  p_{\ell}\big) \big(.  k p_{\nu}+.  q p_{\nu}\big)\big) \big(k .  p_b\big)+\big(2 \big(q .  p_b\big) \big(q .  p_{\ell}\big)-q^2 \big(p_b .  p_{\ell}\big)\big) \big(p_b .  p_{\nu}\big)\big)\big)\Big) \lambda _1\nonumber\\&&,
    \label{8pi2}
\end{eqnarray}
\begin{eqnarray}
    \mathcal{J}_8(2;c)&=&-\frac{1}{3 \big(k .  p_b\big){}^2 \big(k .  p_c\big){}^2}128 \Big(4 \big(q .  p_b\big) \big(p_{\ell} .  p_{\nu}\big) \big(k .  p_b\big){}^4+2 \big(-2 \big(q^2+  k. q+  p_b. q\big) \big(p_{\ell} .  p_{\nu}\big) m_b^2-2 \big(k .  p_{\nu}\big) \big(\big(q .  p_b\big)\nonumber\\&& \big(p_b .  p_{\ell}-q .  p_{\ell}\big)+\big(q .  p_{\ell}\big) \big(k .  p_c\big)\big)+\big(q .  p_b\big) \big(-\big(p_{\ell} .  p_{\nu}\big) \big(q^2+2 k .  q-4 \big(q .  p_b\big)\big)+2 \big( q .p_{\ell}+  p_b .p_{\ell}\big) \big(q .  p_{\nu}\big)\nonumber\\&-&2 \big(  k. p_{\ell}+ q .p_{\ell}\big) \big(p_b .  p_{\nu}\big)\big)+\big(\big(p_{\ell} .  p_{\nu}\big) \big(q^2+2 k .  q+5 m_b^2\big)-2 \big(q .  p_{\ell}\big) \big(q .  p_{\nu}\big)\big) \big(k .  p_c\big)\big) \big(k .  p_b\big){}^3+\big(2 \big(2 \big(q^2\nonumber\\&+&  k. q-2 \big(k .  p_c\big)\big)+ q .p_b\big) \big(p_{\ell} .  p_{\nu}\big) m_b^4+\big(\big(p_{\ell} .  p_{\nu}\big) \big(4 \big(k .  q\big)^2-8 \big(q .  p_b\big){}^2-2 \big(-3 q^2+  p_b. q+6 k .  p_c\big) \big(k .  q\big)\nonumber\\&-&2 \big(q^2-5 \big(k .  p_c\big)\big) \big(q .  p_b\big)+q^2 \big(2 q^2-9 \big(k .  p_c\big)\big)\big)+2 \big(\big(q .  p_{\ell}\big) \big(p_b .  p_{\nu}\big) \big(2 k .  q+2 \big(q^2+  p_b. q\big)-5 \big(k .  p_c\big)\big)\nonumber\\&+&\big(p_b .  p_{\ell}\big) \big(\big(k .  p_{\nu}\big) \big(2 q^2+2 k .  q-3 \big(k .  p_c\big)\big)-\big(2 k .  q+2 \big(q^2+  p_b. q\big)-5 \big(k .  p_c\big)\big) \big(q .  p_{\nu}\big)\big)+\big(q .  p_{\ell}\big) \big(-\big(q .  p_{\nu}\big)\nonumber\\&& \big(2 k .  q+2 \big(q^2+  p_b. q\big)-7 \big(k .  p_c\big)\big)+\big(2 \big(q^2+  k. q-4 \big(k .  p_c\big)\big)+  q. p_b\big) \big(-k .  p_{\nu}\big)\big)\big)\big) m_b^2+2 \big(\big(2 \big(p_b .  p_{\nu}\big) \big(q^2\nonumber\\&+&  k. q\big)+\big(q .  p_b\big) \big(q .  p_{\nu}\big)\big) m_b^2+2 \big(q .  p_b\big) \big(m_b^2-q .  p_b+ k. p_c\big) \big(k .  p_{\nu}\big)+\big(q .  p_b\big) \big(\big(q^2+2 k .  q\big) \big(p_b .  p_{\nu}\big)-2 \big(q .  p_b\big)\nonumber\\&& \big(  q. p_{\nu}+ p_b. p_{\nu}\big)\big)+\big(-\big(q .  p_{\nu}\big) m_b^2-\big(p_b .  p_{\nu}\big) \big(q^2+2 k .  q+7 m_b^2\big)+2 \big(q .  p_b\big) \big(q .  p_{\nu}\big)\big) \big(k .  p_c\big)\big) \big(k .  p_{\ell}\big)\nonumber\\&+&2 \big(\big(\big(p_b .  p_{\ell}\big) \big(\big(q^2-2 \big(q .  p_b\big)\big) \big(q .  p_b\big)-\big(k .  p_c\big) q^2\big)+2 \big(q .  p_b\big) \big(q .  p_{\ell}\big) \big(q .  p_b-k .  p_c\big)\big) \big(k .  p_{\nu}\big)+2 \big(\big(p_b .  p_{\ell}\big) \nonumber\\&&\big(3 \big(q .  p_b\big){}^2+\big(k .  q\big) \big(k .  p_c-q .  p_b\big)\big)+\big(q .  p_b\big) \big(q .  p_{\ell}\big) \big(q .  p_b-k .  p_c\big)\big) \big(q .  p_{\nu}\big)-2 \big(q .  p_b\big) \big(\big(p_b .  p_{\ell}\big) q^2+\big(q .  p_b\big)\nonumber\\&& \big(q .  p_{\ell}\big)\big) \big(p_b .  p_{\nu}\big)+\big(2 \big(q .  p_b\big){}^2-\big(q^2+2 k .  q\big) \big(q .  p_b\big)+\big(q^2+2 k .  q\big) \big(k .  p_c\big)\big) \big(q .  p_b\big) \big(p_{\ell} .  p_{\nu}\big)\big)\big) \big(k .  p_b\big){}^2\nonumber\\&+&\big(\big(-2 q^2-2 k .  q+3 k .  p_c\big) \big(p_{\ell} .  p_{\nu}\big) m_b^6+\big(-2 \big(p_{\ell} .  p_{\nu}\big) \big(k .  q\big)^2+\big(\big(2 q^2-7 \big(k .  p_c\big)\big) \big(q .  p_{\ell}\big)-2 \big(k .  p_c\big) \big(p_b .  p_{\ell}\big)\big) \nonumber\\&&\big(k .  p_{\nu}\big)+2 \big(2 \big(p_b .  p_{\ell}\big) \big(q^2-2 \big(k .  p_c\big)\big)+\big(q .  p_{\ell}\big) \big(2 q^2-5 \big(k .  p_c\big)\big)\big) \big(q .  p_{\nu}\big)+4 \big(2 \big(k .  p_c\big)-q^2\big) \big(q .  p_{\ell}\big) \big(p_b .  p_{\nu}\big)\nonumber\\&+&\big(2 \big(q .  p_b\big){}^2+q^2 \big(7 k .  p_c-2 q^2\big)+4 \big(q .  p_b\big) \big(q^2-2 k .  p_c\big)\big) \big(p_{\ell} .  p_{\nu}\big)+\big(k .  q\big) \big(\big(p_{\ell} .  p_{\nu}\big) \big(-4 q^2+4p_b .q+3 k .  p_c\big)\nonumber\\&-&4 \big(p_b .  p_{\nu}\big) \big(q .  p_{\ell}\big)+2 \big(q .  p_{\ell}\big) \big(k .  p_{\nu}\big)+4 \big( q. p_{\ell}+  p_b. p_{\ell}\big) \big(q .  p_{\nu}\big)\big)\big) m_b^4+\big(4 \big(\big(q .  p_{\nu}\big) \big(p_b .  p_{\ell}\big)-\big(k .  p_c\big) \big(p_{\ell} .  p_{\nu}\big)\big)\nonumber\\&& \big(k .  q\big)^2+2 \big(-2 \big(p_{\ell} .  p_{\nu}\big) \big(q .  p_b\big){}^3+\big(\big(p_{\ell} .  p_{\nu}\big) q^2-2 \big(  q. p_{\ell}+  p_b .p_{\ell}\big) \big(q .  p_{\nu}\big)+2 \big(q .  p_{\ell}\big) \big(p_b .  p_{\nu}\big)\big) \big(q .  p_b\big){}^2+q^2 \nonumber\\&&\big(\big(\big(q .  p_{\ell}\big) \big(k .  p_c\big)+\big(p_b .  p_{\ell}\big) \big(2 q^2-5 \big(k .  p_c\big)\big)\big) \big(p_b .  p_{\nu}\big)-\big(k .  p_c\big) \big(q .  p_{\nu}\big) \big(p_b .  p_{\ell}\big)\big)+\big(q .  p_b\big) \big(2 \big(5 \big(k .  p_c\big)-2 q^2\big)\nonumber\\&& \big(q .  p_{\nu}\big) \big(p_b .  p_{\ell}\big)-\big(k .  p_c\big) q^2 \big(p_{\ell} .  p_{\nu}\big)\big)\big)+\big(\big(p_b .  p_{\ell}\big) \big(\big(k .  p_c\big) \big(5 q^2+4 .  p_b q\big)-2 q^4\big)+2 \big(q .  p_b\big) \big(k .  p_c-q .  p_b\big) \nonumber\\&&\big(q .  p_{\ell}\big)\big) \big(k .  p_{\nu}\big)+2 \big(k .  q\big) \big(\big(p_{\ell} .  p_{\nu}\big) \big(( \big(q .  p_b\big) )\big){}^2+\big(2 \big(k .  p_c\big) \big(q .  p_{\ell}\big)-\big(q^2+2 k .  p_c\big) \big(p_b .  p_{\ell}\big)\big) \big(k .  p_{\nu}\big)\nonumber\\&+&2 \big(p_b .  p_{\ell}\big) \big(\big(p_b .  p_{\nu}\big) q^2+\big(q^2-2 \big(q .  p_b\big)\big) \big(q .  p_{\nu}\big)\big)+\big(\big(p_{\ell} .  p_{\nu}\big) \big(q .  p_b-q^2\big)-5 \big(p_b .  p_{\ell}\big) \big(q .  p_{\nu}\big)+2 \big(q .  p_{\ell}\big) \big(q .  p_{\nu}\nonumber\\&-& p_b .  p_{\nu}\big)\big) \big(k .  p_c\big)\big)\big) m_b^2+\big(2 \big(q .  p_b-m_b^2\big) \big(-2 \big(( \big(q .  p_b\big) )\big){}^2+2 \big(q^2+  k. q\big) m_b^2+\big(2 \big(q .  p_b\big)-5 m_b^2\big) \big(k .  p_c\big)\big)\nonumber\\&& \big(k .  p_{\nu}\big)-2 \big(\big(q .  p_b\big){}^2-\big(q^2+  k .q\big) m_b^2\big) \big(\big(q .  p_{\nu}\big) \big(2 \big(q .  p_b\big)-m_b^2\big)-\big(q^2+2 k .  q\big) \big(p_b .  p_{\nu}\big)\big)+\big(\big(7 q .  p_{\nu}+2 p_b .  p_{\nu}\big)\nonumber\\&& m_b^4+\big(9 q^2+10 k .  q\big) \big(p_b .  p_{\nu}\big) m_b^2+2 \big(-8 \big(q .  p_{\nu}\big) m_b^2-\big(p_b .  p_{\nu}\big) \big(q^2+2 k .  q+2 m_b^2\big)\big) \big(q .  p_b\big)+4 \big(q .  p_b\big){}^2 \big(q .  p_{\nu}\big)\big)\nonumber\\&& \big(k .  p_c\big)\big) \big(k .  p_{\ell}\big)+2 \big(q .  p_b\big) \big(p_b .  p_{\ell}\big) \big(-2 \big(q .  p_b\big) \big(p_b .  p_{\nu}\big) q^2+\big(k .  p_{\nu}\big) \big(q .  p_b-k .  p_c\big) q^2+2 \big(2 \big(q .  p_b\big){}^2+\big(k .  q\big)\nonumber\\&& \big(k .  p_c-q .  p_b\big)\big) \big(q .  p_{\nu}\big)\big)\big) \big(k .  p_b\big)+2 \big(k .  q\big) m_b^2 \big(\big(p_{\ell} .  p_{\nu}\big) m_b^4+\big(\big(p_{\ell} .  p_{\nu}\big) \big(q^2+  k. q-2 \big(q .  p_b\big)\big)-2 \big(q .  p_{\nu}\big) \big( q. p_{\ell}\nonumber\\&+& p_b. p_{\ell}\big)+2 \big(q .  p_{\ell}\big) \big(p_b .  p_{\nu}\big)+\big(q .  p_{\ell}\big) \big(-k .  p_{\nu}\big)\big) m_b^2+\big(k .  p_{\ell}\big) \big(\big(2 k .  p_{\nu}+ q. p_{\nu}\big) m_b^2+\big(q^2+2 k .  q\big) \big(p_b .  p_{\nu}\big)\nonumber\\&-&2 \big(q .  p_b\big) \big(  k. p_{\nu}+  q. p_{\nu}\big)\big)+\big(p_b .  p_{\ell}\big) \big(q^2 \big(k .  p_{\nu}-2 \big(p_b .  p_{\nu}\big)\big)-2 \big(k .  q-2 \big(q .  p_b\big)\big) \big(q .  p_{\nu}\big)\big)\big) \big(k .  p_c\big)\Big) \lambda _1,
  \label{8pi2d}
\end{eqnarray}
\begin{eqnarray}
     \mathcal{J}_8(2;d)&=&\frac{1}{3 m_b^3 \big(  k. p_b\big) \big( k. p_c\big)}256 \lambda _1 \big(\big(  k.p_b+ q. p_b\big){}^2-m_b^2 \big(2  k. q+q^2\big)\big) \big(2 \big(p_{\ell}. p_{\nu}\big) \big( k. p_b\big){}^2+m_b^2 \big(\big(  p_{\ell} .p_{\nu}\big) \nonumber\\&&\big(-2 \big(  q. p_b\big)+ k. q+q^2\big)-2 \big(  q .p_{\nu}\big) \big(p_b. p_{\ell}+q .p_{\ell}\big)+2 \big(p_b.p_{\nu}\big) \big(q. p_{\ell}\big)-\big(k. p_{\nu}\big) \big(q .p_{\ell}\big)\big)+\big( k .p_{\ell}\big)\nonumber\\&& \big(m_b^2 \big(2 k .p_{\nu}+ q.p_{\nu}\big)+\big(p_b. p_{\nu}\big) \big(-2 \big( k. p_b\big)+2 k. q+q^2\big)-2 \big(  q. p_b\big) \big( k. p_{\nu}+  q. p_{\nu}\big)\big)+\big(  k. p_b\big) \nonumber\\&&\big(-\big(p_{\ell} .p_{\nu}\big) \big(-2 \big(  q .p_b\big)+2 k .q+q^2+2 m_b^2\big)+2 \big(k. p_{\nu}\big) \big(q. p_{\ell}- p_b. p_{\ell}\big)-2 \big(  p_b. p_{\nu}\big) \big( q. p_{\ell}\big)+2 \big( q. p_{\nu}\big)\nonumber\\&& \big( p_b .p_{\ell}+ q. p_{\ell}\big)\big)+\big(p_b. p_{\ell}\big) \big(q^2 \big(  k.p_{\nu}-2 \big(p_b. p_{\nu}\big)\big)-2 \big( q. p_{\nu}\big) \big(k. q-2 \big(q. p_b\big)\big)\big)+m_b^4 \big(p_{\ell} .p_{\nu}\big)\big).
    \label{8pi2dd}
\end{eqnarray}
}
 \textbf{8. Fig.2(i):}
\begin{eqnarray}
     \mathcal{M}_{\mu \nu}^{(9)}&=& (-ig^{\alpha\beta}) \bar{b}(-ieQ_b)\gamma^{\alpha}i\left(\slashed{ p_b}+\slashed{\Pi}-\slashed{k}+m_b\right)\gamma^{\nu}(1-\gamma^5)i\left(\slashed{ p_b}+\slashed{\Pi}-\slashed{q}-\slashed{k}+m_c\right)(-ieQ_u)\gamma^{\beta}i\Big(\slashed{ p_b}\nonumber\\ &+&\slashed{\Pi}-\slashed{q}+m_c\Big)\gamma ^{\mu}\left(1-\gamma ^5\right) b, \nonumber\\
     \mathcal{L}_{\mu \nu}^{(9)}&=&\left(\bar{\ell}\gamma^{\mu }(1-\gamma^5)\nu_{\ell}\right) \left(\bar{\nu}_{\ell} \gamma^{\nu }(1-\gamma^5)\ell\right),\\
     \label{m9l9}
      I_9&=&I_8.
 \end{eqnarray}
\begin{eqnarray}
    \mathcal{O}(\Pi^0):\nonumber\\ \mathcal{J}_9(0;a)&=&\frac{1}{m_b \big( k. p_b\big) \big(k. p_c\big)}16 \big(-2 \big(p_{\ell}. p_{\nu}\big) \big( k. p_b\big){}^2+m_b^2 \big(\big(p_{\ell} .p_{\nu}\big) \big(2 k. p_b+2 q. p_b- k. q-q^2\big)-2 \big(p_b. p_{\ell}\big) \big(q. p_{\nu}\big)\nonumber\\&+&2 \big(q. p_{\ell}\big) \big(p_b. p_{\nu}+q. p_{\nu}\big)\big)+\big(k. p_{\nu}\big) \big(\big(q. p_{\ell}\big) \big(2 \big(q. p_b\big)-m_b^2\big)-\big(p_b .p_{\ell}\big) \big(-2k. p_b+2 k. q+q^2\big)\big)+\big(k. p_{\ell}\big)\nonumber\\&& \big(2 \big(k. p_{\nu}\big) \big(q. p_b-m_b^2\big)-2 \big(k. p_b\big) \big(q. p_{\nu}\big)+\big(p_b. p_{\nu}\big) \big(2 \big(k. p_b\big)-q^2\big)+m_b^2 \big( q. p_{\nu}\big)\big)-2 \big(k. p_b\big) \big(p._b p_{\nu}\big) \big(q. p_{\ell}\big)\nonumber\\&-&2 \big( k. p_b\big) \big(q. p_{\ell}\big) \big(q. p_{\nu}\big)+2 \big(k. p_b\big) \big(p_b. p_{\ell}\big) \big(q. p_{\nu}\big)+2 \big(k. q\big) \big(p_b. p_{\nu}\big) \big(q. p_{\ell}\big)+q^2 \big(k. p_b\big) \big(p_{\ell}. p_{\nu}\big)+2 \big(k. q\big) \big(k. p_b\big)\nonumber\\&& \big(p_{\ell}. p_{\nu}\big)-2 \big(k. p_b\big) \big(q. p_b\big) \big(  p_{\ell} .p_{\nu}\big)-m_b^4 p_{\ell}. p_{\nu}-4 \big(p_b .p_{\nu}\big) \big(q. p_b\big) \big(q. p_{\ell}\big)+2 q^2 \big(p_b. p_{\ell}\big) \big(p_b. p_{\nu}\big)\big).
     \label{9pi0}
\end{eqnarray}
{\footnotesize
\begin{eqnarray}
   \mathcal{O}(\Pi):\nonumber\\ \mathcal{J}_9(1;a)&=&\frac{1}{3 \big(k .  p_b\big){}^2 \big(k .  p_c\big){}^2 m_b^3}16 \big(2 \big(5 \big(q .  p_b\big)-3 m_b^2\big) \big(p_{\ell} .  p_{\nu}\big) \big(k .  p_b\big){}^3+\big(6 \big(p_{\ell} .  p_{\nu}\big) m_b^4+\big(\big(p_{\ell} .  p_{\nu}\big) \big(5 q^2+2 k .  q\nonumber\\&-&16 \big(q .  p_b\big)+8 k .  p_c\big)+\big(4 q .  p_{\ell}+6 p_b .  p_{\ell}\big) \big(k .  p_{\nu}\big)-2 \big(3 k .  p_{\ell}+7 q .  p_{\ell}-5 \big(p_b .  p_{\ell}\big)\big) \big(q .  p_{\nu}\big)+6 \big(k .  p_{\ell}-q .  p_{\ell}\big)\nonumber\\&& \big(p_b .  p_{\nu}\big)\big) m_b^2+5 \big(2 \big(q .  p_b\big){}^2-\big(q^2+2 k .  q\big) \big(q .  p_b\big)+\big(q^2+2 k .  q\big) \big(k .  p_c\big)\big) \big(p_{\ell} .  p_{\nu}\big)+10 \big(-\big(q .  p_b\big) \big(k. p_{\nu}\nonumber\\&+& q. p_{\nu}\big) \big(p_b .  p_{\ell}\big)-\big(k .  p_c-q .  p_b\big) \big(k. p_{\ell}+ q. p_{\ell}\big) \big(q .  p_{\nu}\big)+\big(q .  p_b\big) \big(q .  p_{\ell}-k .  p_{\ell}\big) \big(p_b .  p_{\nu}\big)\big)\big) \big(k .  p_b\big){}^2+\big(k .  p_b\big) \nonumber\\&&\big(-3 \big(p_{\ell} .  p_{\nu}\big) m_b^6+\big(\big(p_{\ell} .  p_{\nu}\big) \big(-3 q^2+.  k q+9 .  p_b q-6 \big(k .  p_c\big)\big)-8 \big(p_b .  p_{\ell}\big) \big(q .  p_{\nu}\big)+2 \big(q .  p_{\ell}\big) \big(5 q .  p_{\nu}\nonumber\\&+&4 p_b .  p_{\nu}\big)\big) m_b^4+\big(\big(p_{\ell} .  p_{\nu}\big) \big(4 \big(k .  q\big)^2+3 \big(q .  p_b\big) \big(q^2-2 \big(q .  p_b\big)\big)+\big(k .  q\big) \big(2 q^2-3 \big(q .  p_b\big)-18 k .  p_c\big)\nonumber\\&+&\big(8 \big(q .  p_b\big)-5 q^2\big) \big(k .  p_c\big)\big)+2 \big(5 \big(q .  p_{\ell}\big) \big(k .  p_c-q .  p_b\big)+\big(q^2+2 k .  q+3 .  p_b q-4 \big(k .  p_c\big)\big) \big(p_b .  p_{\ell}\big)\big) \big(q .  p_{\nu}\big)\nonumber\\&+&2 \big(\big(p_b .  p_{\ell}\big) \big(5 q^2-2 \big(k .  q\big)\big)+\big(-q^2+5 k .  q-13 \big(q .  p_b\big)+2 k .  p_c\big) \big(q .  p_{\ell}\big)\big) \big(p_b .  p_{\nu}\big)\big) m_b^2+\big(-7 \big(q .  p_{\ell}\big) m_b^4\nonumber\\&-&\big(\big(p_b .  p_{\ell}\big) \big(9 q^2+6 k .  q-4 \big(q .  p_b\big)+2 k .  p_c\big)+\big(-2 q^2+4 k .  q-15 \big(q .  p_b\big)+4 k .  p_c\big) \big(q .  p_{\ell}\big)\big) m_b^2+5 \big(k .  p_c\nonumber\\&-&q .  p_b\big) \big(2 \big(q .  p_b\big) \big(q .  p_{\ell}\big)-\big(q^2+2 k .  q\big) \big(p_b .  p_{\ell}\big)\big)\big) \big(k .  p_{\nu}\big)+10 \big(\big(2 \big(q .  p_b\big){}^2+\big(k .  q\big) \big(k .  p_c-q .  p_b\big)\big) \big(q .  p_{\ell}\big)\nonumber\\&-&q^2 \big(q .  p_b\big) \big(p_b .  p_{\ell}\big)\big) \big(p_b .  p_{\nu}\big)+\big(k .  p_{\ell}\big) \big(3 \big(q .  p_{\nu}-2 \big(k .  p_{\nu}\big)\big) m_b^4-\big(\big(p_b .  p_{\nu}\big) \big(5 \big(q^2+2 k .  p_c\big)-4 \big(k .  q\big)\big)+4 \big(-q^2\nonumber\\&-&3 \big(q .  p_b\big)+2 k .  p_c\big) \big(k .  p_{\nu}\big)+\big(2 q^2+4 k .  q+.  p_b q-18 \big(k .  p_c\big)\big) \big(q .  p_{\nu}\big)\big) m_b^2+5 \big(k .  p_c-q .  p_b\big) \big(2 \big(k .  p_{\nu}\big) \big(q .  p_b\big)\nonumber\\&-&q^2 \big(p_b .  p_{\nu}\big)\big)\big)\big)+m_b^2 \big(3 \big(p_{\ell} .  p_{\nu}\big) m_b^4-\big(-3 \big(p_{\ell} .  p_{\nu}\big) \big(q^2-2 \big(q .  p_b\big)\big)-6 \big(q .  p_{\nu}\big) \big(p_b .  p_{\ell}\big)+\big(q .  p_{\ell}-2 \big(p_b .  p_{\ell}\big)\big) \nonumber\\&&\big(k .  p_{\nu}\big)+6 \big(q .  p_{\ell}\big) \big(.  q p_{\nu}+.  p_b p_{\nu}\big)\big) m_b^2+\big(-3 \big(q .  p_{\nu}\big) m_b^2+2 \big(2 q^2+5 m_b^2-7 \big(q .  p_b\big)\big) \big(k .  p_{\nu}\big)+\big(q^2+4 .  p_b q\nonumber\\&-& 2 m_b^2\big) \big(p_b .  p_{\nu}\big)\big) \big(k .  p_{\ell}\big)-\big(2 \big(q .  p_b\big) \big(q .  p_{\ell}\big)-q^2 \big(p_b .  p_{\ell}\big)\big) \big(k .  p_{\nu}-6 \big(p_b .  p_{\nu}\big)\big)+\big(k .  q\big) \big(3 \big(p_{\ell} .  p_{\nu}\big) m_b^2+\big(14\nonumber\\&& \big(p_b .  p_{\ell}\big)-8 \big(q .  p_{\ell}\big)\big) \big(k .  p_{\nu}\big)-2 \big(2 p_b .  p_{\ell}+.  q p_{\ell}\big) \big(p_b .  p_{\nu}\big)\big)\big) \big(k .  p_c\big)\big) \big(\lambda _1+3 \lambda _2\big),\\
    \label{9pi1}
     \mathcal{J}_9(1;c)&=&-\frac{1}{3 \big(k .  p_b\big) \big(k .  p_c\big) m_b^3}128 \big(-3 \big(p_{\ell} .  p_{\nu}\big) m_b^6+\big(\big(p_{\ell} .  p_{\nu}\big) \big(-3 q^2+ k. q+9 p_b.q\big)-\big(k .  p_{\nu}\big) \big(3 q .  p_{\ell}+2 p_b .  p_{\ell}\big)\nonumber\\&+&2 \big(5 \big(q .  p_{\ell}\big)-4 \big(p_b .  p_{\ell}\big)\big) \big(q .  p_{\nu}\big)+8 \big(q .  p_{\ell}\big) \big(p_b .  p_{\nu}\big)+\big(k .  p_{\ell}\big) \big(-10 k .  p_{\nu}+3 q .  p_{\nu}+2 p_b .  p_{\nu}\big)\big) m_b^4+\big(-2 \nonumber\\&&\big(p_b .  p_{\nu}\big) \big(q .  p_{\ell}\big) q^2-7 \big(k .  p_{\nu}\big) \big(p_b .  p_{\ell}\big) q^2+2 \big(q .  p_{\ell}\big) \big(k .  p_{\nu}\big) q^2+2 \big(p_b .  p_{\ell}\big) \big(q .  p_{\nu}\big) q^2+10 \big(p_b .  p_{\ell}\big) \big(p_b .  p_{\nu}\big) q^2\nonumber\\&-&10 \big(q .  p_b\big) \big(q .  p_{\nu}\big) \big(q .  p_{\ell}\big)-26 \big(q .  p_b\big) \big(p_b .  p_{\nu}\big) \big(q .  p_{\ell}\big)-14 \big(k .  q\big) \big(k .  p_{\nu}\big) \big(p_b .  p_{\ell}\big)+4 \big(k .  q\big) \big(q .  p_{\ell}\big) \big(k .  p_{\nu}\big)+11 \big(q .  p_b\big)\nonumber\\&& \big(q .  p_{\ell}\big) \big(k .  p_{\nu}\big)+4 \big(q .  p_b\big) \big(p_b .  p_{\ell}\big) \big(k .  p_{\nu}\big)+4 \big(k .  q\big) \big(p_b .  p_{\ell}\big) \big(q .  p_{\nu}\big)+6 \big(q .  p_b\big) \big(p_b .  p_{\ell}\big) \big(q .  p_{\nu}\big)+6 \big(k .  q\big) \big(q .  p_{\ell}\big) \nonumber\\&&\big(p_b .  p_{\nu}\big)+\big(-6 \big(q .  p_b\big){}^2+2 \big(k .  q\big) \big(q^2+2 k .  q\big)+3 \big(q^2-k .  q\big) \big(q .  p_b\big)\big) \big(p_{\ell} .  p_{\nu}\big)+\big(k .  p_{\ell}\big) \big(20 \big(q .  p_b\big) \big(k .  p_{\nu}\big)\nonumber\\&-&\big(2 q^2+4 k .  q+.  p_b q\big) \big(q .  p_{\nu}\big)+\big(-3 q^2+4 k .  q-4 \big(q .  p_b\big)\big) \big(p_b .  p_{\nu}\big)\big)\big) m_b^2+\big(9 \big(p_{\ell} .  p_{\nu}\big) m_b^4+\big(\big(p_{\ell} .  p_{\nu}\big) \big(8 q^2\nonumber\\&+&7 k .  q-22 \big(q .  p_b\big)\big)+16 \big(p_b .  p_{\ell}\big) \big(q .  p_{\nu}\big)-4 \big(q .  p_{\ell}\big) \big(5 q .  p_{\nu}+4 p_b .  p_{\nu}\big)\big) m_b^2+\big(\big(q .  p_{\ell}\big) m_b^2+\big(5 q^2+10 k .  q\nonumber\\&+&14 m_b^2\big) \big(p_b .  p_{\ell}\big)-10 \big(q .  p_b\big) \big(.  q p_{\ell}+.  p_b p_{\ell}\big)\big) \big(k .  p_{\nu}\big)+10 \big(q .  p_b\big) \big(q .  p_{\ell}-p_b .  p_{\ell}\big) \big(q .  p_{\nu}\big)-10 \big(\big(p_b .  p_{\ell}\big) q^2+\big(k .  q\nonumber\\&-&3 \big(q .  p_b\big)\big) \big(q .  p_{\ell}\big)\big) \big(p_b .  p_{\nu}\big)-5 \big(q^2+2 k .  q-2 \big(q .  p_b\big)\big) \big(q .  p_b\big) \big(p_{\ell} .  p_{\nu}\big)+\big(k .  p_{\ell}\big) \big(5 \big(p_b .  p_{\nu}\big) q^2-10 \big(q .  p_b\big) \big(-q .  p_{\nu}\nonumber\\&+&  k. p_{\nu}+  p_b. p_{\nu}\big)+m_b^2 \big(10 k .  p_{\nu}+6 p_b .  p_{\nu}-11 q .  p_{\nu}\big)\big)\big) \big(k .  p_b\big)+\big(k .  p_b\big){}^2 \big(-\big(p_{\ell} .  p_{\nu}\big) \big(16 m_b^2+5 \big(q^2+2 k .  q\nonumber\\&-&4 \big(q .  p_b\big)\big)\big)-10 \big(p_b .  p_{\ell}\big) \big( k. p_{\nu}+  q. p_{\nu}\big)+10 \big(q .  p_{\ell}\big) \big(  q. p_{\nu}+  p_b .p_{\nu}\big)+10 \big(k .  p_{\ell}\big) \big(q .  p_{\nu}-p_b .  p_{\nu}\big)\big)+10 \big(k .  p_b\big){}^3 \nonumber\\&&\big(p_{\ell} .  p_{\nu}\big)+5 \big(q .  p_b\big) \big(\big(p_b .  p_{\nu}\big) \big(q^2 \big(k .  p_{\ell}-2 p_b .  p_{\ell}\big)-2 \big(k .  q-2 \big(q .  p_b\big)\big) \big(q .  p_{\ell}\big)\big)+\big(\big(q^2+2 k .  q\big) \big(p_b .  p_{\ell}\big)\nonumber\\&-&2 \big(q .  p_b\big) \big(  k. p_{\ell}+  q. p_{\ell}\big)\big) \big(k .  p_{\nu}\big)\big)\big) \big(\lambda _1+3 \lambda _2\big).
    \label{9pi1d}
\end{eqnarray}
}

{\footnotesize
\begin{eqnarray}
   \mathcal{O}(\Pi^2):\nonumber\\ \mathcal{J}_9 (2;a)&=&-\frac{1}{3 \big(k .  p_b\big){}^2 \big(k .  p_c\big){}^3 m_b^3}16 \big(2 \big(-\big(q .  p_b\big){}^2+\big(k .  p_c\big){}^2+q^2 m_b^2-3 m_b^2 \big(k .  p_c\big)\big) \big(p_{\ell} .  p_{\nu}\big) \big(k .  p_b\big){}^3+\big(2 \big(3 k .  p_c\nonumber\\&-&q^2\big) \big(p_{\ell} .  p_{\nu}\big) m_b^4+\big(\big(p_{\ell} .  p_{\nu}\big) \big(2 \big(q .  p_b\big){}^2+q^2 \big(9 k.p_c-q^2\big)+\big(k .  q\big) \big(6 k .  p_c-2 q^2\big)+2 \big(q .  p_b\big) \big(q^2-5 k .  p_c\big)\big)\nonumber\\&+&2 \big(\big(p_b .  p_{\ell}\big) \big(5 \big(k .  p_c\big)-q^2\big)+\big(q .  p_{\ell}\big) \big(q^2-7 k .  p_c\big)\big) \big(q .  p_{\nu}\big)+2 \big(q .  p_{\ell}\big) \big(p_b .  p_{\nu}\big) \big(q^2-3 \big(k .  p_c\big)\big)\big) m_b^2+\big(-\big(2 \nonumber\\&& p_b .  p_{\nu}+ q. p_{\nu}\big) \big(( \big(k .  p_c\big) )\big){}^2-2 \big(q .  p_{\nu}-p_b .  p_{\nu}\big) \big(\big(( \big(q .  p_b\big) )\big){}^2-q^2 m_b^2\big)+2 \big(\big(3 \big(p_b .  p_{\nu}\big)-5 \big(q .  p_{\nu}\big)\big) m_b^2\nonumber\\&+&2 \big(q .  p_b\big) \big(q .  p_{\nu}\big)\big) \big(k .  p_c\big)\big) \big(k .  p_{\ell}\big)-2 \big(k .  p_c\big){}^2 \big(q .  p_{\nu}\big) \big(p_b .  p_{\ell}\big)+\big(-\big(2 p_b .  p_{\ell}+  q. p_{\ell}\big) \big(k .  p_c\big){}^2+2 \big(\big(q .  p_b\big){}^2\nonumber\\&-&q^2 m_b^2\big) \big(p_b .  p_{\ell}\big)+6 m_b^2 \big(p_b .  p_{\ell}\big) \big(k .  p_c\big)\big) \big(k .  p_{\nu}\big)-2 \big(q .  p_b\big){}^2 \big(q .  p_{\ell}\big) \big(q .  p_{\nu}\big)-2 \big(k .  p_c\big){}^2 \big(q .  p_{\ell}\big) \big(q .  p_{\nu}\big)+2 \big(q .  p_b\big){}^2\nonumber\\&& \big(p_b .  p_{\ell}\big) \big(q .  p_{\nu}\big)-2 \big(q .  p_b\big){}^2 \big(q .  p_{\ell}\big) \big(p_b .  p_{\nu}\big)+2 \big(k .  p_c\big){}^2 \big(q .  p_{\ell}\big) \big(p_b .  p_{\nu}\big)-2 \big(q .  p_b\big){}^3 \big(p_{\ell} .  p_{\nu}\big)+\big(q .  p_b\big){}^2 q^2 \big(p_{\ell} .  p_{\nu}\big)\nonumber\\&+&\big(k .  p_c\big){}^2 q^2 \big(p_{\ell} .  p_{\nu}\big)+2 \big(q .  p_b\big){}^2 \big(k .  q\big) \big(p_{\ell} .  p_{\nu}\big)+\big(k .  p_c\big){}^2 \big(k .  q\big) \big(p_{\ell} .  p_{\nu}\big)+2 \big(k .  p_c\big){}^2 \big(q .  p_b\big) \big(p_{\ell} .  p_{\nu}\big)-2 \big(k .  p_c\big)\nonumber\\&& q^2 \big(q .  p_b\big) \big(p_{\ell} .  p_{\nu}\big)-4 \big(k .  q\big) \big(k .  p_c\big) \big(q .  p_b\big) \big(p_{\ell} .  p_{\nu}\big)+4 \big(q .  p_b\big) \big(q .  p_{\ell}\big) \big(q .  p_{\nu}\big) \big(k .  p_c\big)\big) \big(k .  p_b\big){}^2+\big(k .  p_b\big) \big(\big(p_{\ell} .  p_{\nu}\big)\nonumber\\&& \big(q^2-3 \big(k .  p_c\big)\big) m_b^6+\big(2 \big(q .  p_{\ell}\big) \big(p_b .  p_{\nu}\big) \big(4 \big(k .  p_c\big)-q^2\big)+2 \big(\big(q .  p_{\ell}\big) \big(5 \big(k .  p_c\big)-q^2\big)+\big(p_b .  p_{\ell}\big) \big(q^2-4 k .  p_c\big)\big)\nonumber\\&& \big(q .  p_{\nu}\big)+\big(q^4-2 \big(q .  p_b\big) q^2-\big(q .  p_b\big){}^2-\big(k .  p_c\big){}^2+\big(k .  q\big) \big(q^2+.  k p_c\big)+\big(8 \big(q .  p_b\big)-7 q^2\big) \big(k .  p_c\big)\big) \big(p_{\ell} .  p_{\nu}\big)\big)\nonumber\\&& m_b^4+\big(\big(p_{\ell} .  p_{\nu}\big) \big(2 \big(q .  p_b\big){}^3-\big(q^2+  k. q\big) \big(q .  p_b\big){}^2+\big(2 \big(k .  q-2 \big(k .  p_c\big)\big) q^2+\big(k .  q\big) \big(4 \big(k .  q\big)-13 \big(k .  p_c\big)\big)\big)\nonumber\\&& \big(k .  p_c\big)+2 \big(q^2-2 \big(k .  q\big)\big) \big(q .  p_b\big) \big(k .  p_c\big)\big)+2 \big(\big(p_b .  p_{\ell}\big) \big(\big(k .  p_c\big) \big(2 \big(k .  q\big)-q^2\big)-\big(q .  p_b\big){}^2\big)+\big(\big(q .  p_b\big){}^2\nonumber\\&+&4 \big(k .  p_c\big){}^2-2 \big(k .  q\big) \big(k .  p_c\big)\big) \big(q .  p_{\ell}\big)\big) \big(q .  p_{\nu}\big)+2 \big(\big(\big(q .  p_b\big){}^2+\big(k .  q\big) \big(5 \big(k .  p_c\big)-q^2\big)+\big(k .  p_c\big) \big(q^2-2 \big(k .  p_c\big)\big)\nonumber\\&+&2 \big(q .  p_b\big) \big(q^2-5 \big(k .  p_c\big)\big)\big) \big(q .  p_{\ell}\big)-\big(\big(q^2-5 k .  p_c\big) q^2+2 \big(k .  q\big) \big(k .  p_c\big)\big) \big(p_b .  p_{\ell}\big)\big) \big(p_b .  p_{\nu}\big)\big) m_b^2+\big(-\big(-13 \nonumber\\&&\big(q .  p_{\nu}\big) m_b^2+\big(q .  p_b\big) \big(q .  p_{\nu}\big)+\big(q^2+10 m_b^2\big) \big(p_b .  p_{\nu}\big)\big) \big(k .  p_c\big){}^2+\big(q^2 m_b^2-\big(q .  p_b\big)^2\big) \big(q^2 \big(p_b .  p_{\nu}\big)-\big(q .  p_{\nu}\big) m_b^2\big)\nonumber\\&+&2 \big(\big(q^2-3 k .  p_c\big) m_b^4+\big(-\big(q .  p_b\big){}^2+\big(k .  p_c\big){}^2+\big(5 k .  p_c-q^2\big) \big(q .  p_b\big)\big) m_b^2+\big(q .  p_b\big) \big(k .  p_c-q .  p_b\big){}^2\big) \big(k .  p_{\nu}\big)\nonumber\\&+&\big(3 \big(q .  p_{\nu}\big) m_b^4+4 \big(k .  q\big) \big(p_b .  p_{\nu}-q .  p_{\nu}\big) m_b^2+q^2 \big(2 \big(q .  p_b\big)-5 m_b^2\big) \big(p_b .  p_{\nu}\big)\big) \big(k .  p_c\big)\big) \big(k .  p_{\ell}\big)+\big(\big(\big(q .  p_{\ell}\big) m_b^2\nonumber\\&-&\big(p_b .  p_{\ell}\big) \big(q^2+2 k .  q+2 m_b^2\big)+3 \big(q .  p_b\big) \big(q .  p_{\ell}\big)\big) \big(k .  p_c\big){}^2+\big(\big(q .  p_b\big){}^2-q^2 m_b^2\big) \big(\big(q .  p_{\ell}\big) \big(2 \big(q .  p_b\big)-m_b^2\big)-\big(q^2\nonumber\\&+&2 k .  q\big) \big(p_b .  p_{\ell}\big)\big)+\big(-7 \big(q .  p_{\ell}\big) m_b^4-3 \big(3 q^2+2 k .  q\big) \big(p_b .  p_{\ell}\big) m_b^2+2 \big(7 \big(q .  p_{\ell}\big) m_b^2+\big(q^2+2 k .  q+2 m_b^2\big) \big(p_b .  p_{\ell}\big)\big)\nonumber\\&& \big(q .  p_b\big)-4 \big(q .  p_b\big){}^2 \big(q .  p_{\ell}\big)\big) \big(k .  p_c\big)\big) \big(k .  p_{\nu}\big)-4 \big(q .  p_b\big){}^3 \big(q .  p_{\ell}\big) \big(p_b .  p_{\nu}\big)+2 \big(q .  p_b\big){}^2 \big(k .  q\big) \big(q .  p_{\ell}\big) \big(p_b .  p_{\nu}\big)\nonumber\\&+&2 \big(k .  p_c\big){}^2 \big(k .  q\big) \big(q .  p_{\ell}\big) \big(p_b .  p_{\nu}\big)+4 \big(k .  p_c\big){}^2 \big(q .  p_b\big) \big(q .  p_{\ell}\big) \big(p_b .  p_{\nu}\big)-4 \big(k .  q\big) \big(k .  p_c\big) \big(q .  p_b\big) \big(q .  p_{\ell}\big) \big(p_b .  p_{\nu}\big)+2 \nonumber\\&&\big(q .  p_b\big){}^2 q^2 \big(p_b .  p_{\ell}\big) \big(p_b .  p_{\nu}\big)-2 \big(k .  p_c\big)^2 q^2 \big(p_b .  p_{\ell}\big) \big(p_b .  p_{\nu}\big)+\big(k .  p_c\big){}^2 \big(k .  q\big) \big(q .  p_b\big) \big(p_{\ell} .  p_{\nu}\big)\big)+m_b^2 \big(-2 \big(\big(p_{\ell} .  p_{\nu}\big)\nonumber\\&& m_b^2+2 \big(p_b .  p_{\ell}\big) \big(k .  p_{\nu}\big)-2 \big(q .  p_{\ell}\big) \big(p_b .  p_{\nu}\big)\big) \big(k .  q\big)^2+\big(k .  q\big) \big(-2 \big(p_{\ell} .  p_{\nu}\big) m_b^4+\big(\big(p_{\ell} .  p_{\nu}\big) \big(-2 q^2+4 .  p_b q\nonumber\\&+&7 k .  p_c\big)-2 \big(q .  p_{\ell}\big) \big(k .  p_{\nu}\big)+4 \big(q .  p_{\ell}-p_b .  p_{\ell}\big) \big(q .  p_{\nu}\big)+4 \big(q .  p_{\ell}\big) \big(p_b .  p_{\nu}\big)+2 \big(k .  p_{\ell}\big) \big(q .  p_{\nu}-2 \big(k .  p_{\nu}\big)\big)\big) m_b^2\nonumber\\&+&2 \big(\big(p_b .  p_{\ell}\big) \big(7 \big(k .  p_c\big)-q^2\big)+2 \big(q .  p_b\big) \big(.  k p_{\ell}+.  q p_{\ell}\big)\big) \big(k .  p_{\nu}\big)-2 \big(\big(k .  p_{\ell}\big) q^2+\big(4 q .  p_b+5 k .  p_c\big) \big(q .  p_{\ell}\big)\nonumber\\&+&2 \big(k .  p_c-q^2\big) \big(p_b .  p_{\ell}\big)\big) \big(p_b .  p_{\nu}\big)\big)+\big(3 \big(p_{\ell} .  p_{\nu}\big) m_b^4+\big(3 \big(p_{\ell} .  p_{\nu}\big) \big(q^2-2 \big(q .  p_b\big)\big)+\big(3 q .  p_{\ell}+2 p_b .  p_{\ell}\big) \big(k .  p_{\nu}\big)\nonumber\\&+&6 \big(p_b .  p_{\ell}\big) \big(q .  p_{\nu}\big)-6 \big(q .  p_{\ell}\big) \big(.  q p_{\nu}+.  p_b p_{\nu}\big)\big) m_b^2+\big(-7 \big(q .  p_{\nu}\big) m_b^2+2 \big(5 m_b^2-7 \big(q .  p_b\big)\big) \big(k .  p_{\nu}\big)+\big(5 q^2\nonumber\\&+&4 .  p_b q-2 m_b^2\big) \big(p_b .  p_{\nu}\big)\big) \big(k .  p_{\ell}\big)-\big(2 \big(q .  p_b\big) \big(q .  p_{\ell}\big)-q^2 \big(p_b .  p_{\ell}\big)\big) \big(5 \big(k .  p_{\nu}\big)-6 \big(p_b .  p_{\nu}\big)\big)\big) \big(k .  p_c\big)\big) \big(k .  p_c\big)\big) \lambda _1,
    \label{9pi2}
\end{eqnarray}
\begin{eqnarray}
     \mathcal{J}_9(2;c)&=&-\frac{1}{3 \big(k .  p_b\big){}^2 \big(k .  p_c\big){}^2}128 \big(4 \big(q .  p_b\big) \big(p_{\ell} .  p_{\nu}\big) \big(k .  p_b\big){}^4+2 \big(-2 \big(q^2+ k. q+p_b .q\big) \big(p_{\ell} .  p_{\nu}\big) m_b^2+\big(q .  p_b\big) \nonumber\\&&\big(-\big(p_{\ell} .  p_{\nu}\big) \big(q^2+2 k .  q-4 \big(q .  p_b\big)\big)-2 \big(p_b .  p_{\ell}\big) \big( k. p_{\nu}+  q. p_{\nu}\big)+2 \big(q .  p_{\ell}\big) \big(q. p_{\nu}+  p_b. p_{\nu}\big)\big)-2 \big(k .  p_{\ell}\big) \nonumber\\&&\big(\big(q .  p_b\big) \big(p_b .  p_{\nu}-q .  p_{\nu}\big)+\big(q .  p_{\nu}\big) \big(k .  p_c\big)\big)+\big(\big(p_{\ell} .  p_{\nu}\big) \big(q^2+2 k .  q+3 m_b^2\big)-2 \big(q .  p_{\ell}\big) \big(q .  p_{\nu}\big)\big) \big(k .  p_c\big)\big) \nonumber\\&&\big(k .  p_b\big){}^3+\big(2 \big(2 \big(q^2+  k .q-2 \big(k .  p_c\big)\big)+  q. p_b\big) \big(p_{\ell} .  p_{\nu}\big) m_b^4+\big(\big(p_{\ell} .  p_{\nu}\big) \big(4 \big(k .  q\big)^2-8 \big(q .  p_b\big){}^2-2 \big(-3 q^2\nonumber\\&+&  p_b. q+7 k .  p_c\big) \big(k .  q\big)-2 \big(q^2-3 \big(k .  p_c\big)\big) \big(q .  p_b\big)+q^2 \big(2 q^2-11 k .  p_c\big)\big)+2 \big(\big(p_b .  p_{\ell}\big) \big(2 q^2+2 k .  q-9 \big(k .  p_c\big)\big)\nonumber\\&+&\big(q .  p_b\big) \big(q .  p_{\ell}\big)\big) \big(k .  p_{\nu}\big)+2 \big(\big(2 k .  q+2 \big(q^2+.  p_b q\big)-3 \big(k .  p_c\big)\big) \big(p_b .  p_{\ell}\big)-\big(2 k .  q+2 \big(q^2+  p_b. q\big)-9 \big(k .  p_c\big)\big)\nonumber \\&&\big(q .  p_{\ell}\big)\big) \big(q .  p_{\nu}\big)-2 \big(2 k .  q+2 \big(q^2+.  p_b q\big)-3 \big(k .  p_c\big)\big) \big(q .  p_{\ell}\big) \big(p_b .  p_{\nu}\big)\big) m_b^2+2 \big(-\big(\big(2 \big(q^2+  k. q\big)+  q .p_b\big)\nonumber\\&& \big(q .  p_{\nu}\big)-2 \big(q^2+  k. q\big) \big(p_b .  p_{\nu}\big)\big) m_b^2-\big(k .  p_c\big) \big(-9 \big(q .  p_{\nu}\big) m_b^2+2 \big(q .  p_b\big) \big(q .  p_{\nu}\big)+\big(q^2+m_b^2\big) \big(p_b .  p_{\nu}\big)\big)\nonumber\\&+&2 \big(q .  p_b\big) \big(m_b^2-q .  p_b+.  k p_c\big) \big(k .  p_{\nu}\big)+\big(q .  p_b\big) \big(2 \big(q .  p_b\big) \big(q .  p_{\nu}\big)+\big(q^2-2 \big(q .  p_b\big)\big) \big(p_b .  p_{\nu}\big)\big)\big) \big(k .  p_{\ell}\big)+2 \big(-2 \nonumber\\&&\big(q .  p_b\big) \big(q .  p_{\nu}\big) \big(\big(q .  p_b\big) \big(p_b .  p_{\ell}-q .  p_{\ell}\big)+\big(q .  p_{\ell}\big) \big(k .  p_c\big)\big)+\big(\big(p_b .  p_{\ell}\big) \big(-2 \big(q .  p_b\big){}^2-\big(k .  p_c\big) \big(q^2+2 k .  q\big)+\big(q^2\nonumber\\&+&2 k .  q\big) \big(q .  p_b\big)\big)+2 \big(q .  p_b\big) \big(k .  p_c-q .  p_b\big) \big(q .  p_{\ell}\big)\big) \big(k .  p_{\nu}\big)+2 \big(\big(3 \big(q .  p_b\big){}^2+\big(k .  q\big) \big(k .  p_c-q .  p_b\big)\big) \big(q .  p_{\ell}\big)\nonumber\\&-&q^2 \big(q .  p_b\big) \big(p_b .  p_{\ell}\big)\big) \big(p_b .  p_{\nu}\big)+\big(2 \big(q .  p_b\big){}^2-\big(q^2+2 k .  q\big) \big(q .  p_b\big)+\big(q^2+2 k .  q\big) \big(k .  p_c\big)\big) \big(q .  p_b\big) \big(p_{\ell} .  p_{\nu}\big)\big)\big) \big(k .  p_b\big)^2\nonumber\\&+&\big(\big(-2 q^2-2 k .  q+3 k .  p_c\big) \big(p_{\ell} .  p_{\nu}\big) m_b^6+\big(-2 \big(p_{\ell} .  p_{\nu}\big) \big(k .  q\big)^2+\big(\big(k .  p_c\big) \big(7 q .  p_{\ell}+2 p_b .  p_{\ell}\big)-2 q^2 \big(q .  p_{\ell}\big)\big)\nonumber\\&& \big(k .  p_{\nu}\big)+2 \big(2 \big(p_b .  p_{\ell}\big) \big(2 \big(k .  p_c\big)-q^2\big)+\big(q .  p_{\ell}\big) \big(2 q^2-5 k .  p_c\big)\big) \big(q .  p_{\nu}\big)+\big(2 \big(q .  p_b\big){}^2+q^2 \big(7 \big(k .  p_c\big)-2 q^2\big)\nonumber\\&+&4 \big(q .  p_b\big) \big(q^2-2 \big(k .  p_c\big)\big)\big) \big(p_{\ell} .  p_{\nu}\big)+\big(k .  q\big) \big(\big(p_{\ell} .  p_{\nu}\big) \big(-4 q^2+4 .  p_b q+3 k .  p_c\big)-4 \big(q .  p_{\nu}\big) \big(p_b .  p_{\ell}\big)-2 \big(q .  p_{\ell}\big)\nonumber\\&& \big(k .  p_{\nu}\big)+4 \big(q .  p_{\ell}\big) \big(.  q p_{\nu}+.  p_b p_{\nu}\big)\big)+4 \big(q .  p_{\ell}\big) \big(p_b .  p_{\nu}\big) \big(q^2-2 \big(k .  p_c\big)\big)\big) m_b^4+\big(-4 \big(\big(p_b .  p_{\ell}\big) \big(k .  p_{\nu}\big)-\big(q .  p_{\ell}\big)\nonumber\\&& \big(p_b .  p_{\nu}\big)+\big(p_{\ell} .  p_{\nu}\big) \big(k .  p_c\big)\big) \big(k .  q\big)^2+\big(\big(p_b .  p_{\ell}\big) \big(\big(k .  p_c\big) \big(11 q^2-4 \big(q .  p_b\big)\big)-2 q^4\big)+2 \big(q .  p_b\big) \big(2 q^2+p_b. q\nonumber\\&-&9 k .  p_c\big) \big(q .  p_{\ell}\big)\big) \big(k .  p_{\nu}\big)+2 \big(-2 \big(p_{\ell} .  p_{\nu}\big) \big(q .  p_b\big){}^3+\big(\big(p_{\ell} .  p_{\nu}\big) q^2+2 \big(p_b .  p_{\ell}\big) \big(q .  p_{\nu}\big)-2 \big(q .  p_{\ell}\big) \big(q. p_{\nu}+  p_b. p_{\nu}\big)\big)\nonumber\\&& \big(q .  p_b\big){}^2+q^2 \big(\big(k .  p_c\big) \big(q .  p_{\nu}\big) \big(p_b .  p_{\ell}\big)-\big(\big(p_b .  p_{\ell}\big) \big(5 \big(k .  p_c\big)-2 q^2\big)+\big(q .  p_{\ell}\big) \big(k .  p_c\big)\big) \big(p_b .  p_{\nu}\big)\big)+\big(q .  p_b\big) \big(2 \big(5\nonumber\\&& \big(k .  p_c\big)-2 q^2\big) \big(q .  p_{\ell}\big) \big(p_b .  p_{\nu}\big)-\big(k .  p_c\big) q^2 \big(p_{\ell} .  p_{\nu}\big)\big)\big)+2 \big(k .  q\big) \big(\big(p_{\ell} .  p_{\nu}\big) \big(q .  p_b\big){}^2+\big(\big(p_b .  p_{\ell}\big) \big(7 \big(k .  p_c\big)-3 q^2\big)\nonumber\\&+&2 \big(q .  p_b\big) \big(q .  p_{\ell}\big)\big) \big(k .  p_{\nu}\big)+2 \big(\big(p_b .  p_{\ell}\big) q^2+\big(q^2-2 \big(q .  p_b\big)\big) \big(q .  p_{\ell}\big)\big) \big(p_b .  p_{\nu}\big)+\big(-\big(p_{\ell} .  p_{\nu}\big) \big(q^2-2 \big(q .  p_b\big)\big)+2 \nonumber\\&&\big(q .  p_{\ell}-p_b .  p_{\ell}\big) \big(q .  p_{\nu}\big)-7 \big(q .  p_{\ell}\big) \big(p_b .  p_{\nu}\big)\big) \big(k .  p_c\big)\big)\big) m_b^2+\big(-2 \big(\big(q^2+.  k q\big) m_b^2-\big(q .  p_b\big){}^2\big) \big(q^2 \big(p_b .  p_{\nu}\big)\nonumber\\&-&\big(q .  p_{\nu}\big) m_b^2\big)+2 \big(\big(k .  p_c\big) \big(5 m_b^4-9 \big(q .  p_b\big) m_b^2+2 \big(q .  p_b\big){}^2\big)-2 \big(q .  p_b-m_b^2\big) \big(\big(q .  p_b\big){}^2-\big(q^2+  k. q\big) m_b^2\big)\big) \big(k .  p_{\nu}\big)\nonumber\\&+&\big(-\big(7 q .  p_{\nu}+2 p_b .  p_{\nu}\big) m_b^4+\big(4 \big(k .  q\big) \big(q .  p_{\nu}\big)+\big(7 q^2-4 k .  q+4 .  p_b q\big) \big(p_b .  p_{\nu}\big)\big) m_b^2-2 q^2 \big(q .  p_b\big) \big(p_b .  p_{\nu}\big)\big) \big(k .  p_c\big)\big)\nonumber\\&&\big(k .  p_{\ell}\big)+2 \big(q .  p_b\big) \big(2 \big(p_b .  p_{\nu}\big) \big(\big(2 \big(q .  p_b\big){}^2+\big(k .  q\big) \big(k .  p_c-q .  p_b\big)\big) \big(q .  p_{\ell}\big)-q^2 \big(q .  p_b\big) \big(p_b .  p_{\ell}\big)\big)+\big(k .  p_c-q .  p_b\big)\nonumber\\&& \big(2 \big(q .  p_b\big) \big(q .  p_{\ell}\big)-\big(q^2+2 k .  q\big) \big(p_b .  p_{\ell}\big)\big) \big(k .  p_{\nu}\big)\big)\big) \big(k .  p_b\big)+2 \big(k .  q\big) m_b^2 \big(\big(p_{\ell} .  p_{\nu}\big) m_b^4+\big(\big(p_{\ell} .  p_{\nu}\big) \big(q^2+  k. q\nonumber\\&-&2 \big(q .  p_b\big)\big)+2 \big(p_b .  p_{\ell}\big) \big(q .  p_{\nu}\big)-2 \big(q .  p_{\ell}\big) \big( q .p_{\nu}+  p_b. p_{\nu}\big)\big) m_b^2+\big(\big(p_b .  p_{\nu}\big) q^2+2 \big(m_b^2-q .  p_b\big) \big(k .  p_{\nu}\big)+m_b^2 \big(\nonumber\\&&-\big(q .  p_{\nu}\big)\big)\big) \big(k .  p_{\ell}\big)+\big(\big(p_b .  p_{\ell}\big) \big(q^2+2 k .  q\big)+\big(m_b^2-2 \big(q .  p_b\big)\big) \big(q .  p_{\ell}\big)\big) \big(k .  p_{\nu}\big)-2 \big(\big(p_b .  p_{\ell}\big) q^2+\big(k .  q-2 \big(q .  p_b\big)\big)\nonumber\\&& \big(q .  p_{\ell}\big)\big) \big(p_b .  p_{\nu}\big)\big) \big(k .  p_c\big)\big) \lambda _1,
  \label{9pi2d}
\end{eqnarray}
}
\begin{eqnarray}
    \mathcal{J}_9(2;d)&=&\frac{1}{3 m_b^3 \big(k. p_b\big) \big(k. p_c\big)}256 \lambda _1 \big(\big(k. p_b+  q. p_b\big){}^2-m_b^2 \big(2 k. q+q^2\big)\big) \big(2 \big(p_{\ell} .p_{\nu}\big) \big(  k .p_b\big){}^2+m_b^2 \big(\big(  p_{\ell}. p_{\nu}\big) \big(-2 \nonumber\\&&\big( k. p_b\big)-2 \big(  q. p_b\big)+  k. q+q^2\big)+2 \big(  p_b. p_{\ell}\big) \big( q. p_{\nu}\big)-2 \big(q. p_{\ell}\big) \big(p_b .p_{\nu}+ q. p_{\nu}\big)\big)+\big(  k .p_{\nu}\big) \big(\big( p_b. p_{\ell}\big) \big(-2 \big(  k .p_b\big)\nonumber\\&+&2   k. q+q^2\big)+\big(  q. p_{\ell}\big) \big(m_b^2-2 q .p_b\big)\big)+\big(k. p_{\ell}\big) \big(2 \big(  k. p_{\nu}\big) \big(m_b^2-  q. p_b\big)+2 \big(  k .p_b\big) \big(  q .p_{\nu}\big)+\big( p_b. p_{\nu}\big) \big(q^2\nonumber\\&-&2 \big( k. p_b\big)\big)+m_b^2 \big(- q. p_{\nu}\big)\big)-2 \big(k. q\big) \big(  p_b. p_{\nu}\big) \big(  q.p_{\ell}\big)+2 \big(  k. p_b\big) \big(  q. p_{\ell}\big) \big(  q .p_{\nu}\big)-2 \big(  k. p_b\big) \big(  p_b. p_{\ell}\big) \big( q. p_{\nu}\big)\nonumber\\&+&2 \big(  k. p_b\big) \big(  p_b. p_{\nu}\big) \big(  q.p_{\ell}\big)-q^2 \big(  k. p_b\big) \big(  p_{\ell}. p_{\nu}\big)-2 \big(  k .q\big) \big(  k. p_b\big) \big(  p_{\ell}. p_{\nu}\big)+2 \big( k .p_b\big) \big(q. p_b\big) \big(p_{\ell} .p_{\nu}\big)+m_b^4 \big(  p_{\ell}. p_{\nu}\big)\nonumber\\&+&4 \big(p_b. p_{\nu}\big) \big(  q. p_b\big) \big(  q. p_{\ell}\big)-2 q^2 \big( p_b. p_{\ell}\big) \big(  p_b. p_{\nu}\big)\big).
    \label{9pi2dd}
\end{eqnarray}

 \section{Kinematics}
 \label{appB}
 In this section, we describe the kinematics involve in the decay. However four body decay generally consist of five independent kinematical variables. The inclusive four body decay consist of six independent variable where one extra variable is due to invariant mass squared for decayed hadron ($p_X^2$). Here, we have traded $p_X^2$ with $q'^2(=(p_{\ell}+p_{\nu}+k)^2)$. Further we define two Lorentz invariant variables as
 \begin{eqnarray}
     y=\frac{2p_B.p_{\ell}}{m_B^2}\ \text{and} \  x=\frac{2p_B.k}{m_B^2}.
 \end{eqnarray} 
  The variables $y$ and $x$ are normalized lepton and photon energy, respectively in $B$ meson rest frame. Other three variables are neutrino energy ($E_{\nu}$), and two angles: 
  (a) $\theta_{X\gamma}$ is the angle between the recoiling hadron and hard photon, (b) $\theta_{X\ell}$ is the angle between the final state recoiling hadron ($X$) and charged lepton. The general form of triple differential decay   
\begin{eqnarray}
    \frac{d^3\Gamma}{dq'^2 dE_{\ell} dE_{\nu}}&=&\int \frac{d^4 p_{\ell}}{(2\pi)^4} 2\pi \delta(p_{\ell}^2 - m_{\ell}^2)\theta(p_{\ell}^0) \int \frac{d^4 p_{\nu}}{(2\pi)^4} 2\pi \delta(p_{\nu}^2 )\theta(p_{\nu}^0) \delta(E_{\ell}-p_{\ell}^0) \delta(E_{\nu}-p_{\nu}^0) \delta(q'^2-(q+k)^2) \nonumber\\ &&\int \frac{d^4 k}{(2\pi)^4} \frac{1}{k^2((p_b+\Pi-q-k)^2-m_c^2)}|\mathcal{M}|^2 (2\pi)^4 \delta^4({p_b -q'-p_X}). 
\end{eqnarray}
Cutcosky method implies that 
\begin{eqnarray}
    \int \frac{d^4 k}{(2\pi)^4}\to \int \frac{d^4 k}{(2\pi)^4}\int\frac{d^4p_X}{(2\pi)^4}(2\pi)^4 \delta^4(p_b-q-p_X-k),
\end{eqnarray}
and the propagator is replaced with delta functions. For example, in the Fig.2(a), propagators are  
\begin{eqnarray}
    \frac{1}{k^2}&\to & -2\pi i \delta(k^2)\theta(k^0),\,\,\, \text{and}\\
    \frac{1}{((p_b+\Pi-q-k)^2-m_c^2)}&\to & -2\pi i \delta(((p_b+\Pi-q-k)^2-m_c^2))\theta((p_b+\Pi-q-k)^0).
\end{eqnarray}
Incorporating the Cutkosky method, the differential decay width is 
{\small
\begin{eqnarray}
    \frac{d^3\Gamma}{ dE_{\ell} dq'^2 dE_{\nu}}&=&\int \frac{d^3 p_{\ell}}{(2\pi)^3} \delta(p_{\ell}^2 )\theta(p_{\ell}^0) \int \frac{d^3 p_{\nu}}{(2\pi)^3} \delta(p_{\nu}^2 )\theta(p_{\nu}^0) \delta(q'^2-(q+k)^2) \int \frac{d^4 k}{(2\pi)^4} \frac{d^4 p_X}{(2\pi)^4} (-2\pi i) \delta(k^2) \theta(k^0) \nonumber\\ &(-2\pi i)& \delta((p_b+\Pi-q-k)^2-m_c^2)  (2\pi)^4 \delta^4({p_b -q'-p_X}) (2\pi)^4 |\mathcal{M}|^2  \delta^4(p_B-q-p_X-k), \nonumber\\
    &=&-\frac{1}{8\pi^2} E_{\ell} E_{\nu} \int d (cos\theta_{\ell \nu}) \delta(q'^2-(q+k)^2) \int \frac{d^3 k}{2E_{\gamma}}\delta ((p_B +\Pi -q-k)^2-m_c^2) |\mathcal{M}|^2 \nonumber\\&&\delta^4(p_B-q-p_X-k),\nonumber\\
    &=& -\frac{1}{8\pi^2}\int \frac{d^3 k}{2E_{\gamma}}\delta ((p_B +\Pi -q-k)^2-m_c^2) |\mathcal{M}|^2 \delta^4(p_B-q-p_X-k),\nonumber\\
    &=& -\frac{1}{16\pi^2}\int |k| dE_\gamma\int d\Omega_k\delta ((p_B +\Pi -q-k)^2-m_c^2) |\mathcal{M}|^2 \delta^4(p_B-q-p_X-k),
    \label{app_b_diff_decay}
\end{eqnarray}}
where $\theta_{\ell \nu}$ is the angle between the lepton and neutrino. Eqn. (\ref{app_b_diff_decay}) can be translated in terms of variables $x$ and $y$. Hence, one can write Eqn. (\ref{app_b_diff_decay}) as 
\begin{eqnarray}
    \frac{d^2\Gamma}{ dy\, dx}&=&-\frac{m_B^2}{64\pi^2}\int dq'^2 \int d\Omega_k |k| \int dE_\nu\delta ((p_B +\Pi -q-k)^2-m_c^2) |\mathcal{M}|^2 \nonumber\\&&\delta^4(p_B-q-p_X-k).
\end{eqnarray}
Further, the differential rate with respect to lepton energy is  
\begin{eqnarray}
    \frac{d\Gamma}{ dy}=\int_{x_{min}}^{x_{max}} dx \frac{d^2\Gamma}{ dy\, dx},
\end{eqnarray}
where $m_\gamma^2\leq q'^2\leq \frac{y\,m_B^2(1-y-z^2)}{1-y}$, $m_\gamma \leq x\leq 1-y-z^2$ and $0\leq y\leq 1-z^2$. The variable $z=m_{u/c}/m_b$ is defined in Sec-\ref{m_diffdecay}.

Next, the delta function with $\Pi$ can be expanded in the power of $\Pi$. Explicitly, it is given by
\begin{eqnarray}
    \delta\big(\big(p_b+\Pi-q-k\big)^2\big)&=&\delta\big(\big(p_b-q-k\big)^2\big)+2\Pi\cdot (p_b-q-k)\delta'\big(\big(p_b-q-k\big)^2\big)+\Pi^2 \Big(\delta'\big(\big(p_b-q-k\big)^2\big)\nonumber\\&&+2(p_b-q-k)^2 \delta''\big(\big(p_b-q-k\big)^2\big)\Big)+...
\end{eqnarray}
 In the rest frame of the $B$ meson, momentas involved are given by 
\begin{eqnarray}
    p_B=(m_B,{\bf{0}}),\,\,\,
    q'=(q^0,|{\bf{q'}}|),\,\,\, p_X=(E_X,-|{\bf q'}|),
\end{eqnarray}
where, 
\begin{eqnarray}
    E_X=\frac{m_B^2-q'^2+m_X^2}{2m_B},\,\, q^0=\frac{m_B^2+q'^2-m_X^2}{2m_B},\,\, \text{and} \,\,|{\bf q'}|=\frac{\lambda^{1/2}(m_B^2,q'^2,m_X^2)}{2m_B}.
\end{eqnarray}
Another important point to note is that the integration domain in $E_{\nu}$ has a boundary from below:
\begin{eqnarray}
    E_{\nu} \geq \frac{q'^2-2q.k-m_{\ell}^2}{4E_{\ell}}.
\end{eqnarray}
Therefore, it should be ensured that $E_{\nu}$ does not cross the boundary. This is enforced by introducing an appropriate theta function in the integral. This plays an important role in the integration of delta functions and their derivatives present in the differential rate in Eqn. (\ref{app_b_diff_decay}).

\bibliographystyle{ieeetr}
\bibliography{inclusivebtolnu}

\end{document}